\documentclass[a4paper,11pt]{article}
\pdfoutput=1

\usepackage{jcappub}

\usepackage{amsmath}
\usepackage{amssymb}
\usepackage{amsfonts}
\usepackage{mathtools}
\usepackage{soul}
\usepackage{lscape}
\usepackage{subfigure}
\usepackage{graphicx}
\usepackage{color}
\usepackage{bm}
\usepackage{epstopdf}
\usepackage{hyperref}
\usepackage{url}
\usepackage{booktabs}
\usepackage{mdframed}
\newcommand{\dragon}{{\tt DRAGON2}}
\newcommand{\dragonold}{{\tt DRAGON}}
\newcommand{\galprop}{{\tt GALPROP}}

\begin{document}

\title{Cosmic-ray propagation with {\tt DRAGON2}: I. numerical solver and astrophysical ingredients}

\author[a]{Carmelo Evoli}
\affiliation[a]{Gran Sasso Science Institute, viale Francesco Crispi 7, 67100 L'Aquila (AQ), Italy}
\emailAdd{carmelo.evoli@gssi.infn.it}

\author[b]{Daniele Gaggero}
\affiliation[b]{GRAPPA Institute, University of Amsterdam, Science Park 904, 1090 GL Amsterdam, The Netherlands}
\emailAdd{d.gaggero@uva.nl}

\author[c]{Andrea Vittino}
\affiliation[c]{Physik-Department T30d, Technische Universit\"{a}t M\"{u}nchen, James-Franck-Stra{\ss}e 1, D-85748 Garching, Germany}
\emailAdd{andrea.vittino@tum.de}

\author[d]{Giuseppe Di Bernardo}
\affiliation[d]{Max-Planck-Institut f{\"u}r Astrophysik, Karl-Schwarzschild-Stra{\ss}e 1, 85740 Garching bei M\"{u}nchen, Germany}
\emailAdd{bernardo@mpa-garching.mpg.de}

\author[e]{Mattia Di Mauro}
\affiliation[e]{W.~W.~Hansen Experimental Physics Laboratory, Kavli Institute for Particle Astrophysics and Cosmology, Department of Physics and SLAC National Accelerator Laboratory, Stanford University, Stanford, CA 94305, USA}
\emailAdd{mdimauro@slac.stanford.edu}

\author[f]{Arianna Ligorini}
\affiliation[f]{Instytut Fizyki J\c{a}drowej - PAN,  ul. Radzikowskiego 152, 31-342 Krak\'{o}w, Poland}
\emailAdd{arianna.ligorini@ifj.edu.pl}

\author[g]{Piero Ullio}
\affiliation[g]{Scuola Internazionale di Studi Superiori Avanzati, via Bonomea 265, 34136  Trieste, Italy}
\emailAdd{piero.ullio@sissa.it}

\author[h]{Dario Grasso}
\affiliation[h]{INFN and Dipartimento di Fisica ``E. Fermi'', Pisa University, Largo B. Pontecorvo 3, I-56127 Pisa, Italy}
\emailAdd{dario.grasso@pi.infn.it}

\date{\today}

\begin{abstract}
{We present version 2 of the {\tt DRAGON} code designed for computing realistic predictions of the CR densities in the Galaxy.
The code numerically solves the interstellar CR transport equation (including inhomogeneous and anisotropic diffusion, either in space and momentum, advective transport and energy losses), under realistic conditions.

The new version includes an updated numerical solver and several models for the astrophysical ingredients involved in the transport equation.
Improvements in the accuracy of the numerical solution are proved against analytical solutions and in reference diffusion scenarios. 

The novel features implemented in the code allow to simulate the diverse scenarios proposed to reproduce the most recent measurements of local and diffuse CR fluxes, going beyond the limitations of the homogeneous galactic transport paradigm. 
To this end, several applications using \dragon~are presented as well.

This new version facilitates the users to include their own physical models by means of a modular C++ structure.}
\end{abstract}


\maketitle

\section{Introduction}

An impressive experimental effort has been devoted during the last years to provide very precise measurements of the high-energy cosmic radiation.

Since 2006 the PAMELA \cite{Picozza:2006nm} orbital observatory has measured the spectra of many charged cosmic-ray (CR) species and discovered several intriguing anomalies in the proton, helium and positron spectrum \cite{2013AdSpR..51..219A,2013PhRvL.111h1102A,2014ApJ...791...93A}; AMS-02 \cite{Kounine:2012ega}, on board of the International Space Station since 2011, confirmed some of those results with higher accuracy extending PAMELA measurements up to the TeV \cite{2014PhRvL.113l1102A,2014PhRvL.113l1101A,2014PhRvL.113v1102A,2015PhRvL.114q1103A}. 
At even larger energies CALET \cite{Torii:2008zzb} and ISS-CREAM \cite{Seo:2014tka} should soon bridge direct measurements with those of ground based air-shower experiments, like KASCADE-Grande~\cite{Antoni:2003gd}, and ARGO-YBG \cite{Aielli:2006cj}. 

Concerning gamma-rays, the Large Area Telescope (LAT) on board of the {\it Fermi} Gamma-Ray Space Telescope \cite{2009ApJ...697.1071A} has been releasing refined all-sky maps -- as well as detailed imaging and spectral information of very interesting regions like the Galactic Center (GC) \cite{TheFermi-LAT:2015kwa} and of extended sources like Supernova Remnants (SNRs) \cite{Acero:2015prw} -- up to TeV energies \cite{TheFermi-LAT:2015ykq}, tracing the CR emission through the whole Galaxy. HAWC \cite{2015arXiv150805826S} and CTA \cite{CTA} should soon extend those measurements to higher energy.   
At the opposite edge of the electromagnetic spectrum, in the radio and microwave bands, PLANCK \cite{Planck}, LOFAR \cite{2013A&A...556A...2V} -- and in the next future SKA \cite{2015aska.confE.147Q} -- observatories are probing the synchrotron emission of CR electrons and positrons.

Moreover, the era of neutrino astronomy was recently opened by the detection of $54$ high-energy astrophysical events by the IceCube collaboration in four years of data taking \cite{Aartsen:2015zva}. Some of them are likely to be of Galactic origin as will be further investigated when the KM3NeT experiment \cite{Margiotta:2014gza} will be operating, providing a better coverage of both hemispheres. 

Such amount of recent and upcoming experiments has also another driver: the search for the first non-gravitational signal from Dark Matter (DM) particle interactions amid the detected fluxes. 
DM indirect searches are strictly affected by uncertainties on the propagation of CRs in the Galaxy. Therefore, reducing and {quantifying} these astrophysical uncertainties is crucial in order to be able to disentangle a possible DM signature from astrophysical backgrounds and to determine the experimental sensitivity to DM properties.

In the wake of the plethora of novel experimental achievements described above, a parallel effort on the CR-transport modelling in the Galaxy is needed~\cite{2015ARA&A..53..199G}. 

The \dragonold~project has been pursued in order to meet the urgent demand to model CR propagation under the most realistic and general conditions. 

Within this context, we introduce here the code version 2 (\dragon) as a general tool to simulate all {relevant} processes regarding CR transport from very low ($\sim 10$ MeV and below) to extremely high ($\sim 1$ PeV) energies.\footnote{At those energies, however, extrapolations of the spallation cross-sections must be performed and the source stochasticity may become relevant.}

In particular, it computes the solution of the diffusion-advection-loss equation describing CR transport for most CR species, from heavier ones down to protons, antiprotons, and leptons, both of astrophysical and exotic origin (i.e., coming from DM annihilations/decays).
The transport equation features fully position- and energy-dependent transport coefficients (spatial and momentum diffusion, energy losses and advection) in both spatial two-dimensional (assuming cylindrical symmetry) and three-dimensional mode.

\dragon~allows a detailed study of both small-scale and large-scale structures (e.g., the spiral structure of the Galaxy) in steady-state and transient mode, refining the spatial resolution on the regions of interest (e.g., local bubble, GC, or Galactic Plane).  

In this paper we introduce the code having in mind the specific case of CR propagation in the Galaxy; however, the code is written in a general way and can be easily used in many other different contexts and on different scales (e.g., CR transport in Galaxy clusters or {in a star forming region}).  

Moreover, the new modular structure of the code make it possible for the user to implement additional spatial distributions for all relevant astrophysical quantities in a straightforward way. 

Old versions of {\tt DRAGON}~\cite{Evoli:2008dv} have been used in several contexts. For example, to provide a solution to the CR gradient and isotropy problem in terms of inhomogeneous diffusion~\cite{Evoli:2012ha}; to compute CR electron and positron spectra in the presence of a spiral arm structure for sources~\cite{Gaggero:2013rya}; 
to model CR antiprotons as CR secondaries and from DM annihilations~\cite{Evoli:2015vaa,Evoli:2011id}; to study the synchrotron emission from galactic leptons~\cite{2013JCAP...03..036D}; and to reproduce $\gamma$-ray and neutrino diffuse emissions above the TeV~\cite{Gaggero:2015xza}. 
Most of these original results were obtained thanks to the innovative features already present in the first version of the code: these features are included also in the new version of the code and, where possible, extended to more general cases (see the discussion in Section~\ref{sec:newfeatures}).
   
With respect to the previous versions, the new code \dragon~was also largely reworked in order to optimally profit from modern programming design and computing techniques.
In this paper, we provide a detailed description of the transport equation solver (see Section~\ref{sec:discretization}) and provide in the appendixes details about different models for the relevant astrophysical ingredients (e.g., source and gas distribution, magnetic field models, spiral arm patterns) adopted (see Appendix~\ref{sec:astroingredients}).
In Section~\ref{sec:solver}, we also provide a comprehensive set of numerical tests to assess the code performances, and to study the accuracy of the solution and the time needed to reach convergence in different conditions (e.g., different grid sizes, constant and variable time step).
The main new features for galactic propagation are demonstrated in Section~\ref{sec:newfeatures} in a few example applications.

The first large project of this kind, \galprop\footnote{See \url{http://galprop.stanford.edu} and \url{http://sourceforge.net/projects/galprop}.}, is a widely used code in the community~\cite{1995ICRC....3...48S,Strong1998,1998ApJ...493..694M}.
\galprop~is designed to make predictions of direct CR measurements as well as gamma rays and synchrotron radiation consistently.
It includes realistic models for nuclear spallation processes~\cite{Moska2001ICRC,Moska2003ICRC,Moskalenko2002,Moska2003,Mashnik2004} and energy losses, but basic assumptions for the CR transport\footnote{For a detailed comparison between the two codes we refer to the \dragon~wiki-page: \url{https://github.com/cosmicrays/DRAGON2/wiki}}.
Semi-analytical solutions of the propagation equation are implemented in the USINE code developed since 2010~\cite{Maurin2001}. 
Taking advantage of much faster computation methods than numerical models, the semianalytical approach allows for an efficient scan of a wider transport parameter space~\cite{2010A&A...516A..66P,2010A&A...516A..67M,2011A&A...526A.101P}.

Recently, the {\tt PICARD} numerical code have been developed~\cite{Kissmann2014,Werner2015}. {\tt PICARD} is fully 3D in concept and implements modern numerical techniques for the numerical solver, handling high resolutions with reasonable computer resources.

\dragon~is part of a complete suite of numerical tools designed to cover most of the relevant processes involving Galactic CRs and their secondary products over a very wide energy range. 
With the help of these tools -- in particular the {\tt HeSky}\footnote{A technical documentation will be released during 2017.} package -- it is possible to compute spectra and sky-maps of radiation emitted by CRs interactions in a huge energy range, from the synchrotron radio waves up to the PeV neutrinos. 
On the low-energy side, the solar modulation can be treated either with auxiliary analytical routines implementing the force-field approximation~\cite{Gleeson:1968zza, Cholis:2015gna}, or with the {\tt HelioProp} numerical code featuring a detailed model of CR charge-dependent interaction with the Heliosphere, including diffusion, advection and energy losses due to the solar wind \cite{Maccione:2012cu}.

This paper does not contain a description of spallation processes and of off-diagonal anisotropic diffusion, which will be covered in forthcoming publications and in the evolving {\tt DRAGON} manual (see {\tt  www.dragonproject.org}).

\section{Transport of CRs in the Galaxy}
\label{sec:propagation}

\dragon~features all relevant processes for CR transport from Galactic acceleration sites to Earth: in particular, spatial and momentum diffusion, energy losses, advection, nuclear spallations and decays.

The combination of all these processes can be described by the following equation~\cite{1964ocr..book.....G,Berezinskii1990}:
\begin{equation}\label{eq:prop}
\begin{split}
{\bf \nabla} \cdot (\vec{J}_i - {\vec{v}_w} N_i) + \frac{\partial}{\partial p} \left[ p^2 D_{pp} \, \frac{\partial}{\partial p} \left( \frac{N_i}{p^2} \right) \right] - \frac{\partial}{\partial p} \left[ \dot{p} N_i - \frac{p}{3} \left(\vec{\nabla} \cdot \vec{v}_w \right) N_i \right] = \\
Q + \sum_{i<j} \left( c \beta n_{\rm gas} \, \sigma_{j \rightarrow i} + \frac{1}{\gamma \tau_{j \rightarrow i}} \right) N_j - \left( c \beta n_{\rm gas} \, \sigma_i + \frac{1}{\gamma \tau_i} \right) N_i
\end{split}
\end{equation}
where $N_i(\vec{r},p)$ is the density per total momentum $p$ of the CR species $i$, $D_{pp}(\vec{r},p)$ is the momentum diffusion coefficient, $Q(\vec{r},p)$ describes the distribution and the energy spectra of sources, $\vec{v}_w (\vec{r})$ is the Galactic wind velocity responsible for CR advection, $\dot{p}(\vec{r},p)$ accounts for the momentum losses.
The timescale for radioactive decay at rest is given by $\tau_i$, while $\sigma_i$ is the spallation cross-section with the interstellar gas. In this paper we do not consider these latter nuclear processes, and we postpone a detailed description to a forthcoming publication. 
The CR macroscopic current $\vec{J}(\vec{r},p)$ is determined by the spatial diffusion tensor $D_{ij}$, as $J_i = - D_{ij} \nabla_j N$.

These quantities can be either inferred from independent observations (e.g. the gas distribution, the magnetic field entering the loss term) or fitted to the data (e.g. the diffusion coefficient, the Galactic wind velocity). 
For all of them, different parameterizations are provided in literature and can be used to estimate the systematic uncertainty affecting the corresponding process. We therefore implement in \dragon~several options for the relevant transport quantities, as extensively described in Appendix~\ref{sec:astroingredients}; in most cases, the quantities are position-dependent. 
 
As discussed in the Introduction, one of the main novelty of our code with respect to other existing codes is the possibility to implement inhomogeneous transport\footnote{Not necessarily separable in a spatial and an energy term.} (e.g., advection, momentum and spatial diffusion).

In particular, assuming diffusion as inhomogeneous and anisotropic has a very natural motivation. 
In fact, the presence of a large scale Galactic magnetic field (GMF) clearly breaks isotropy and introduces a preferred direction, so that charged-particle diffusion should be expressed in terms of a diffusion tensor with components given by:
\begin{equation}
D_{ij} = \left(D_\parallel - D_\perp \right) b_i b_j + D_\perp \delta_{ij} + \epsilon_{ijk}~D_A b_k \, ,
\label{eq:diff_tens}
\end{equation}
where $\vec{b}$ is a unit vector along the mean (large scale) GMF. With this choice of versors, $D_\parallel$ and $D_\perp$ are the components of the diffusion tensor parallel and perpendicular to the mean magnetic field and describe diffusion due to small-scale turbulent fluctuations.
The coefficient $D_A$ gauges the anti-symmetric component of the diffusion tensor: It is usually identified as the drift coefficient since it describes a macroscopic drift orthogonal to both $\vec{b}$ and the gradient of the CR density, $\vec{\nabla} N$ \cite{Burger1985,Minnie2007}.
In this paper we always assume $D_A = 0$ since the associated drifts are negligible up to $\sim$PeV energies as shown, e.g., in~\cite{Evoli2007}.

Although the physics behind CR diffusion is far from being understood (see e.g. \cite{Schlickeiser2002} for a comprehensive review),
some basic aspects may however be clarified starting from the {\it weak-turbulence} approximation where magnetic perturbations are well-developed in k-space and small compared with the regular background component. 
Under this assumption it is possible to treat analytically the problem of resonant CR interactions with the random-phase MHD wavemodes. This framework is known as quasi-linear theory (QLT)~\cite{Morrison1957,Jokipii1966}. The classical result for QLT gives that diffusion coefficients are described by a power-law in rigidity with different slopes for the parallel and perpendicular components (see also~\cite{DeMarco2007a}). 
Moreover, these coefficients are spatially inhomogeneous since they are determined by local properties of the turbulent and regular fields. In this perspective, for the diffusion coefficients $D_\parallel$ and $D_\perp$ we adopt several phenomenological parameterizations as proposed in recent works based on local fluxes and gamma-ray data (see Appendix~\ref{sec:spatialcoefficient}).

\dragon ~can work either in a ($2+1$)-dimensional (2D) or in a ($3+1$)-dimensional (3D) configuration. 
In the 2D case we use cylindrical coordinates defined by the radial distance $r$ and the height form the Galactic disk $z$ and we assume azimuthally symmetry. 
For the 3D case we consider Cartesian coordinates $x,y,z$.
The quantities defined as function of cylindrical coordinates are consistently mapped in Cartesian coordinates by the relation $r = \sqrt{x^2 + y^2}$.

In the next Sections, we will specify the transport equation in these two configurations. 

\subsection*{2D in cylindrical coordinates and azimuthal symmetry}

In 2D mode, $D_{\parallel}$ plays no role and the derivative of the particle density along the azimuthal coordinate vanishes ($\partial_{\phi} N= 0$), then Eq.~\ref{eq:prop} can be written by substituting:
\begin{equation}
\vec{\nabla} \cdot \vec{J} \rightarrow D_{rr} (r,z,p) \, \frac{\partial^2 N}{\partial r^2} \, + \, D_{zz}(r,z,p) \, \frac{\partial^2 N}{\partial z^2} \, + \, \chi (r,z,p) \, \frac{\partial N}{\partial r}  \, + \, \psi(r,z,p) \, \frac{\partial N}{\partial z} 
\label{eq:CRdivCylCoord}
\end{equation}
where:
\begin{eqnarray*}
\chi(r,z,p) & = & \frac{D_{rr}(r,z,p)}{r} + \frac{\partial D_{rr}(r,z)}{\partial r}\\
\psi(r,z,p) & = & \frac{\partial D_{zz}(r,z,p)}{\partial z}
\end{eqnarray*}

At Galactic scale, it is common to assume an azimuthal mean field, $\vec{B} = B {\hat \phi}$, such that, as it follows from Eq.~\ref{eq:diff_tens}, $D_{rr} = D_{zz} = D_\perp$. 
In this configuration, the problem reduces to that of isotropic diffusion.

In presence of a vertical component of the mean field, as the case of the GC, with $\vec{B} = B {\hat z}$, we obtain $D_{rr} = D_\perp$ and $D_{zz} = D_\parallel$ and, in general, $D_\perp \neq  D_\parallel$.

\subsection*{3D in Cartesian coordinates}

In this configuration Eq.~\ref{eq:prop} can be written by substituting:
\begin{eqnarray}
\vec{\nabla} \cdot \vec{J} & \rightarrow & D_{xx} \frac{\partial^2 N}{\partial x^2} + D_{yy} \frac{\partial^2 N}{\partial y^2} + D_{zz} \frac{\partial^2 N}{\partial z^2} + \nonumber\\ 
& + & 2 D_{xy} \frac{\partial N}{\partial x \partial y}  + 2 D_{xz} \frac{\partial N}{\partial x \partial z} + 2 D_{yz} \frac{\partial N}{\partial y \partial z}  + \nonumber\\
& + & u_x \frac{\partial N}{\partial x} + u_y \frac{\partial N}{\partial y}  + u_z \frac{\partial N}{\partial z} 
\end{eqnarray}
where $u_i = \nabla_j~D_{ij}$.

In the present work, we consider only the case in which off-diagonal components of the diffusion tensor, $D_{i \ne j}$, are null. Under this condition, the previous equation can be simplified as this:
\begin{eqnarray}
\vec{\nabla} \cdot \vec{J} & \rightarrow & D_{xx} \frac{\partial^2 N}{\partial x^2} + D_{yy} \frac{\partial^2 N}{\partial y^2} + D_{zz} \frac{\partial^2 N}{\partial z^2} + 
\nonumber\\ 
& + & \frac{\partial D_{xx}}{\partial x}  \frac{\partial N}{\partial x} +   \frac{\partial D_{yy}}{\partial y}  \frac{\partial N}{\partial y} +  \frac{\partial D_{zz}}{\partial z}  \frac{\partial N}{\partial z} ~.
\end{eqnarray}

\section{Numerical solution of the transport equation}
\label{sec:discretization}

\subsection{Discretization over grid}
\label{2Ddiscretization}

In order to solve the transport equation numerically it is necessary to discretise the equation, i.e. to write it on a discrete grid and transform derivative operators into finite differences. 
In cylindrical coordinates (2D) we consider a grid with two spatial coordinates ($r_i$, $z_j$) and one momentum coordinate ($p_q$); the grid spacing is arbitrary and it may be irregular. 
In Cartesian coordinates (3D) the spatial grid is instead obtained with three coordinates: $x_i$, $y_j$, $z_k$.

CR density at a given position, momentum, and time can be written on this lattice as
\begin{equation*}
N_{i,j,q}^n \quad \text{or} \quad N_{i,j,k,q}^n
\end{equation*}
where $n$ is the time step index.

In order to replace the derivatives in the transport equation by their finite difference approximations, we mainly adopt the {\it centred difference} scheme for an irregular spaced grid:
\begin{eqnarray*}
\frac{\partial N}{\partial x} & \rightarrow & \frac{N_{i+1} - N_{i-1}}{x_{i+1} - x_{i-1}} \\
\frac{\partial^2 N}{\partial x^2} & \rightarrow & \frac{2}{x_{i+1} - x_{i-1}} \, \left( \frac{N_{i+1} - N_i}{x_{i+1} - x_i} - \frac{N_i - N_{i-1}}{x_{i} - x_{i-1}} \right)
\end{eqnarray*}
which gives a truncation error $O(\Delta x^2)$ and $O(\Delta x)$ for uniform and non-uniform grid respectively.

\subsection{Iteration scheme}
\label{sect:iteration}

We rewrite~\ref{eq:prop} as a time-dependent equation and we find the stationary solution by evolving an initial condition (IC), $N^0_{ijk}$, until it relaxes to an equilibrium solution, $N^\infty_{ijk}$, for which the time derivative vanishes.   

Schematically the transport equation can be now written as:
\begin{equation}\label{Eq:timedep}
\frac{\partial N}{\partial t} = \mathcal{L}(N) + Q
\end{equation}
where $\mathcal{L}$ is the operator which defines the transport equation. 

In its discretized version, Eq.~\ref{Eq:timedep} becomes:
\begin{equation}
\frac{N^{n+1}_{i} - N^{n}_{i}}{\Delta t} = \hat{\mathcal{L}}_{i} + Q_{i}
\end{equation}
where $\Delta t$ is the time step, and $i$ is now a unique index over the spatial-energy grid.

The algorithm we adopt to evolve the solution of the transport equation~\ref{eq:prop} at each time step is described in Sec.~\ref{Sec:LOD}.
{The convergence criterion is introduced in Sec.~\ref{convergence}.

\subsection{Local One Dimensional (LOD) operator splitting method}
\label{Sec:LOD}

A well-known approach to find the solution of a diffusive-advection equation is the {\it operator splitting} method.

The basic idea of this algorithm is to consider the transport equation~\ref{Eq:timedep} as a linear sum of different evolution operators (e.g., radial diffusion, vertical advection, energy loss, ...):
\begin{equation}
\frac{\partial N}{\partial t} = \sum_{l} \mathcal{L}_l (N) + Q 
\end{equation}
and for each of them to find a valid differencing scheme for updating $N$ from timestep $n$ to timestep $n + 1$, as the operator were the only one on the right-hand side of~\ref{Eq:timedep}. 
The overall evolution in the time step $\Delta t$ is obtained by using separately all the operators in sequence.

This specific implementation of the method is known as {\it Local One Dimensional} (LOD) operator splitting. 
Clear advantages of this algorithm are that one can discretise independently the different operators using different methods and different boundary conditions. {It also allows to have different time steps for the the different subproblems}.

The transport equation in cylindrical coordinates (which is obtained by substituting~\ref{eq:CRdivCylCoord} for the CR flux divergence in~\ref{eq:prop}), can be conveniently written as the sum of 5 operators:
\begin{eqnarray*}
\mathcal{L}_{r} & = & D^{rr} \, \frac{\partial^2 N}{\partial r^2} + \left[ \frac{D^{rr}}{r} + \frac{\partial D^{rr}}{\partial r} \right] \frac{\partial N}{\partial r} \\
\mathcal{L}_{z} & = & D^{zz} \, \frac{\partial^2 N}{\partial z^2} + \left[ \frac{\partial D^{zz}}{\partial z} \right] \frac{\partial N}{\partial z} \\
\mathcal{L}_{a} & = & - \frac{\partial (v_w N)}{\partial z} \\
\mathcal{L}_{p} & = & \frac{\partial}{\partial p} \left[ p^2 D^{pp} \, \frac{\partial}{\partial p} \left( \frac{N}{p^2} \right) \right] \\
\mathcal{L}_{l} & = &  \frac{\partial}{\partial p} \left[ \dot{p} N - \frac{p}{3} \left( \frac{\partial v_w}{\partial z} \right) N \right] \equiv \frac{\partial}{\partial p} \left[ \dot{P} N \right]
\end{eqnarray*}
where we assumed $\vec{v}_w = v_w(z) \vec{z}$ and $v_w(z) > 0$, and $\dot{P} \equiv \dot{p} - \frac{p}{3} \left( \frac{\partial v_w}{\partial z} \right)$.

\subsection{Crank-Nicolson coefficients}
\label{Sec:CN}

The Crank-Nicolson (CN) method is a convenient discretisation scheme for $\hat{\mathcal{L}}_i$ since it is second-order accurate in time and unconditionally stable. 
According to this scheme, the time derivative is obtained by taking the average of the right-hand side at $t$ and $t+\Delta t$, giving:
\begin{equation}\label{Eq:CN}
\frac{\partial N}{\partial t} = \mathcal{L}_l (N) + \frac{Q}{n_{\rm os}} \, \rightarrow \, \frac{N_{i}^{n+1} - N_{i}^{n}}{\Delta t} = \frac{1}{2} \, \left[ \hat{\mathcal{L}}_{i}^{n+1} + \hat{\mathcal{L}}_{i}^{n} \right] + \frac{Q_{i}}{n_{\rm os}}
\end{equation}
where $n_{\rm os}$ is the number of active operators. 

The most general expression of $\hat{\mathcal{L}}$ can be written as  
\begin{equation}
\hat{\mathcal{L}}_i^n = L_i N_{i-1}^n - C_i N_i^n + U_i N_{i+1}^n 
\end{equation}
and consequently~\ref{Eq:CN} becomes: 
\begin{equation}
\begin{aligned}
- \frac{\Delta t}{2} L_i N_{i-1}^{n+1} + \left( 1 + \frac{\Delta t}{2} C_i \right) N_i^{n+1} - \frac{\Delta t}{2} U_i N_{i+1}^{n+1} =  \\ 
\frac{\Delta t}{2} L_i N_{i-1}^{n} + \left( 1 - \frac{\Delta t}{2} C_i \right) N_i^{n} + \frac{\Delta t}{2} U_i N_{i+1}^{n} + \frac{\Delta t}{n_{\rm os}} Q_i 
\end{aligned}
\label{eq:CN_extended}
\end{equation}%
which is a tridiagonal set of simultaneous linear equations that we solve at each timestep to compute $N^{n+1}_i$ once $N^n_i$ and $Q_i$ are given.
Tridiagonal systems of linear equations can easily be solved by standard methods like Cholesky decomposition (see Sec.~2.9 in \cite{Press2002}) or LU decomposition (see Sec.~2.4 in \cite{Press2002}).

CN coefficients $L_i$, $C_i$, and $U_i$, for the transport operators in the cylindrical symmetric version of the transport equation are reported in Table~\ref{Tab:CN2D}.
Boundary conditions (b.c.) are also reported in the same table: We assume $N(r=R,|z| = H) = 0$, and $N$ to be symmetrical around $r = 0$. %
Boundary conditions in momentum are given by: $N(p = p_{\rm max}) = 0$ and $\left( \frac{d}{dp} \frac{N}{p^2} \right)_{p_{\rm min}} = 0$ (see also~\ref{App:analyticalsol_reacc}).

The three-dimensional anisotropic version in Cartesian coordinates is easily obtained by discretizing $\mathcal{L}_x$ and $\mathcal{L}_y$ similarly to $\mathcal{L}_z$ with different diffusion coefficients ($D_{xx}$, $D_{yy}$, $D_{zz}$) and by imposing as boundary conditions: $N(|x|=R,|y|=R,|z|=H) = 0$.
The spatial operator CN coefficients in Cartesian three-dimensional grid are listed in Table~\ref{Tab:CN3D}. 

Differently than other operators, we adopt for the energy loss a 2-nd order accurate method which is obtained by integrating in momentum the energy loss term and evaluating the integral with the trapezoidal rule.
This approach is described in detail in~\cite{Kissmann2014} and implemented in the {\tt PICARD} code.

Following this idea, we end up with a tridiagonal set of linear equations:
\\
\begin{equation}
\left( 1 + C_i \right) N_i^{n+1} + \left( 1 - U_i \right) N_{i+1}^{n+1} =
\left( 1 - C_i \right) N_i^{n}   + \left( 1 + U_i \right) N_{i+1}^{n} + \Delta t \left(Q_{i+1} + Q_i \right)
\end{equation}
where
\begin{eqnarray*}
\frac{{\rm C_i}}{\Delta t} & = & -\frac{\dot{P}_i}{p_{i+1} - p_i} \\
\frac{{\rm U_i}}{\Delta t} & = & -\frac{\dot{P}_{i+1}}{p_{i+1} - p_i} 
\end{eqnarray*}
that we invert with standard methods to find $N^{n+1}_i$.

In Sec.~\ref{sec:elosstest}, we compare the performances of this scheme with the 1-st order upwind discretisation combined with the CN scheme (see Table~\ref{Tab:CN2D}), as implemented in the previous version of {\tt DRAGON}.

\section{Validation of the numerical solver}
\label{sec:solver}

In this section we discuss a complete set of numerical tests aimed at assessing the performances of the evolution equation components. In particular, our intent is to investigate both the convergence and the accuracy of the numerical approach that we use. 

In order to test the convergence of the numerical solution, we introduce the concept of {\it residuals}. 
In a steady-state condition, the transport equation can be written as: 
\begin{equation}
\mathcal{L}(\vec{x},p) + Q(\vec{x},p) = 0
\end {equation}

One can then introduce the {\it normalised residual} as:
\begin{equation}
\mathcal{R}(\vec{x},p)  = \frac{\mathcal{L}(\vec{x},p)  + Q(\vec{x},p) }{Q(\vec{x},p) }
\end{equation}

When solving the transport equation on a discretized grid, the quantity $\mathcal{R}$ corresponds to a matrix over the spatial-energy grid. 
To identify this matrix with one single value we compute the normalized l2-residual:
\begin{equation}
\|\mathcal{R}\|^2 = \frac{1}{N_Q} \sum_{i} \left(\frac{\hat{\mathcal{L}}_{i} + Q_{i}}{Q_{i}} \right)^2
\end{equation} 
where $N_Q$ is the number of grid points where $Q \ne 0$.
During the iteration procedure, the residual decreases and finally converges to a constant value which depends on the grid resolution and on the adopted discretisation scheme.

The tests that we discuss in this section are performed under simplified assumptions regarding both the geometry of the Galaxy and the astrophysical quantities. In doing so, we are able to compare the numerical scheme of each transport operator against an analytical solution and this allows us to evaluate the accuracy of the numerical solution.  

We quantify such accuracy by means of the {\it relative error} $\epsilon_{\mathrm{rel}}$. 
This quantity is defined as follows: 
\begin{equation}
\epsilon_{\mathrm{rel}} = \mathrm{max} \left( \frac{N^a_{i} - N^m_{i}}{N^a_{i}} \right)
\end{equation}  
where $N^a_{i}$ is the analytical solution evaluated at the grid point corresponding to the index $i$, and $N^m_{i}$ is the corresponding numerical solution as obtained after convergence has been attained (i.e., after the residual has reached the plateau value).

In the various tests, we evaluate the accuracy of the numerical solution for different values of the time step $\Delta t$ and of the grid spacings and we check if the scaling of $\epsilon_{\mathrm{rel}}$ with the grid step is consistent with the scaling expected from the discretisation order. 

We derive the analytical solutions for the test cases used in this section in Appendix~\ref{App:A}.

\subsection{Spatial diffusion}
\label{sec:spatialdiff}

We first describe the case of a two-dimensional anisotropic and spatially dependent diffusion coefficient with a steady-state source term.  
The possibility to simulate inhomogeneous diffusion has been a relevant feature of the {\tt DRAGON} code since its first version.

The transport equation adopted for this particular test, together with its analytical solution, is described in~\ref{App:analyticalsol_diff}. 
In particular, we study the case of anisotropic diffusion by setting $f \equiv D_{zz} / D_{xx} = 0.1$.

The left panel of Fig.~\ref{fig:diff_residual} shows the evolution with time of the squared norm of the residual $\|\mathcal{R}\|^2$, for a fixed value of the timestep $\Delta t$ and for different spatial resolutions of the grid. 
The grid resolution is given in terms of the number of bins along $r$ and $z$ (we take $n_r=n_z=n$). 

The right panel of Fig.~\ref{fig:diff_residual} illustrates the impact of the time-step on the residual, for a fixed spatial resolution.  

\begin{figure}[!t]
\centering
\includegraphics[width=0.49\textwidth]{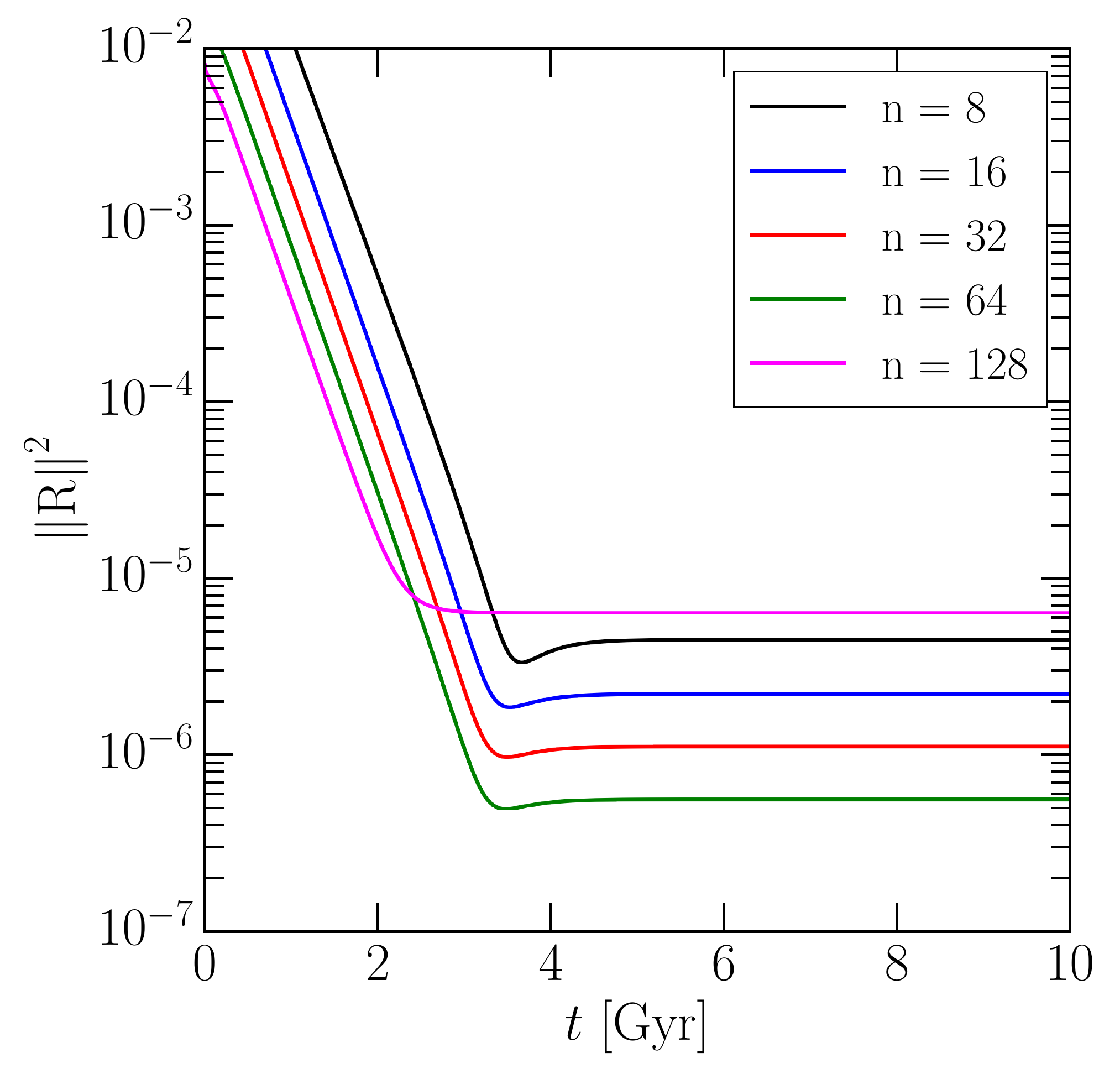}
\hspace{\stretch{1}}
\includegraphics[width=0.49\textwidth]{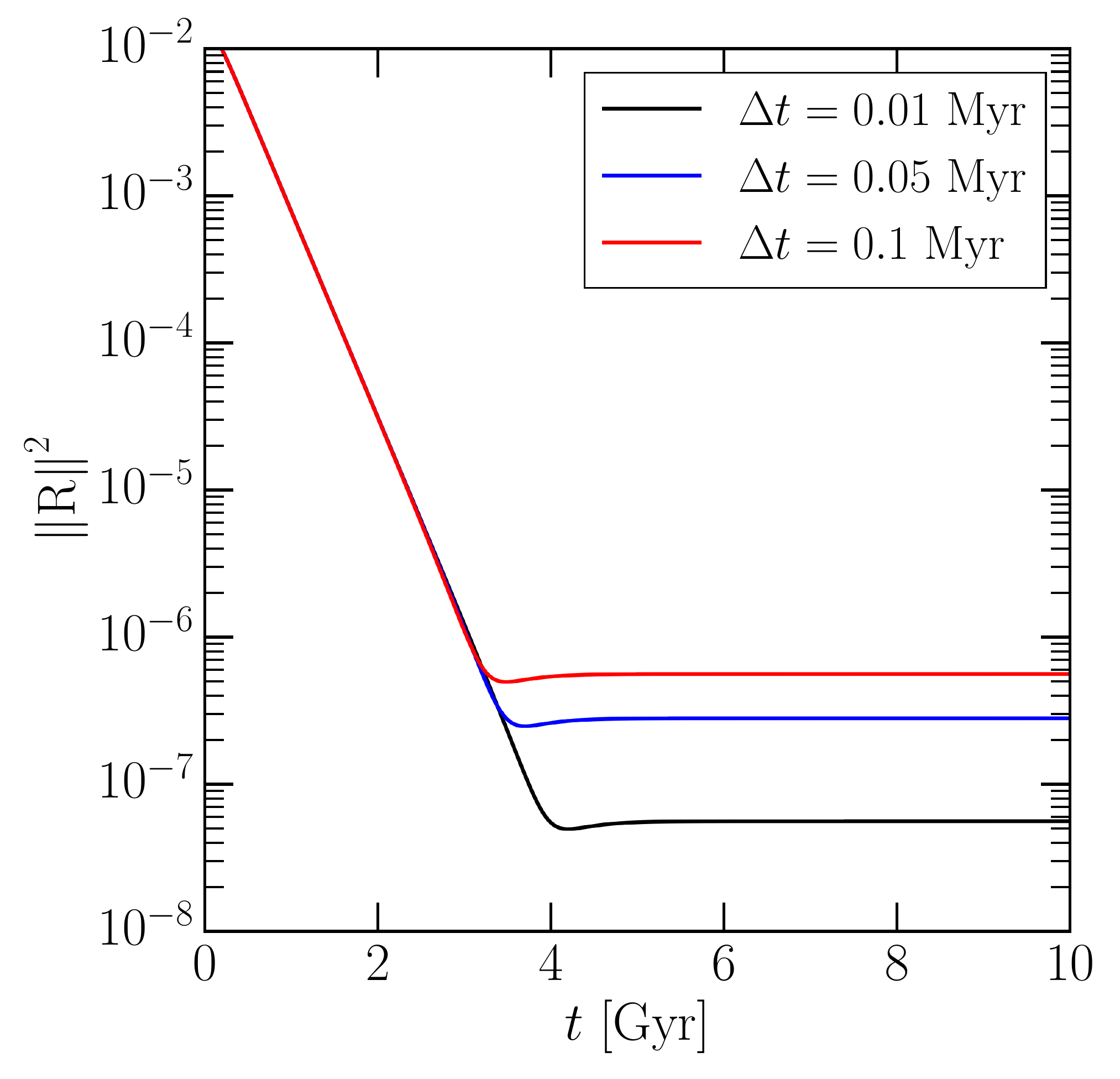}
\caption{{\it Left panel}: the squared norm of the residual $\|\mathcal{R}\|^2$ as a function of the simulation time for different spatial grid resolutions and for $\Delta t = 0.1$~Myr. {\it Right panel}: the same for a fixed value of the grid resolution $n = 64$ and for different time steps.}
\label{fig:diff_residual}
\end{figure}

\begin{figure}[!t]
\centering
\includegraphics[width=0.49 \textwidth]{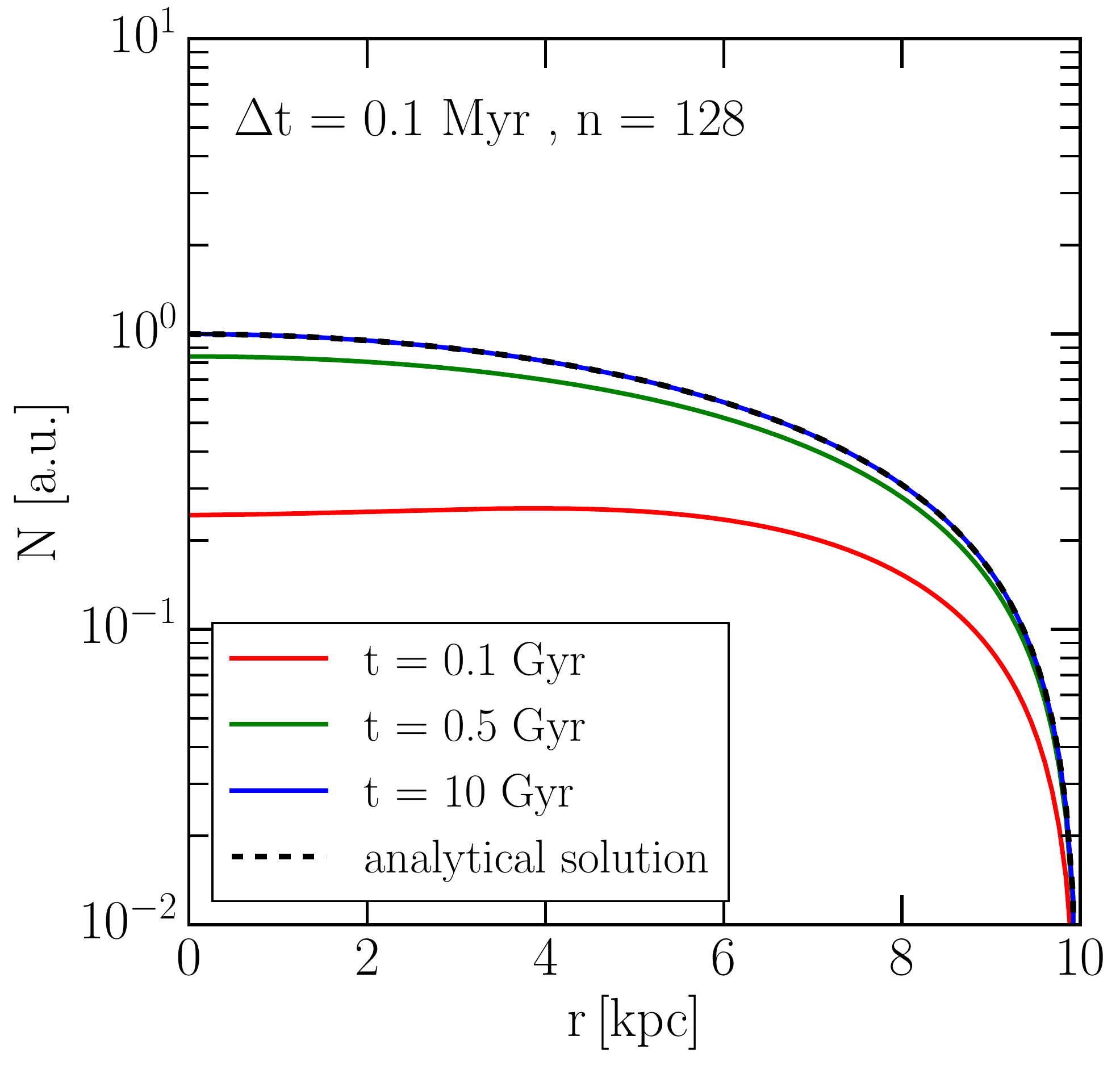}
\hspace{\stretch{1}}
\includegraphics[width=0.49 \textwidth]{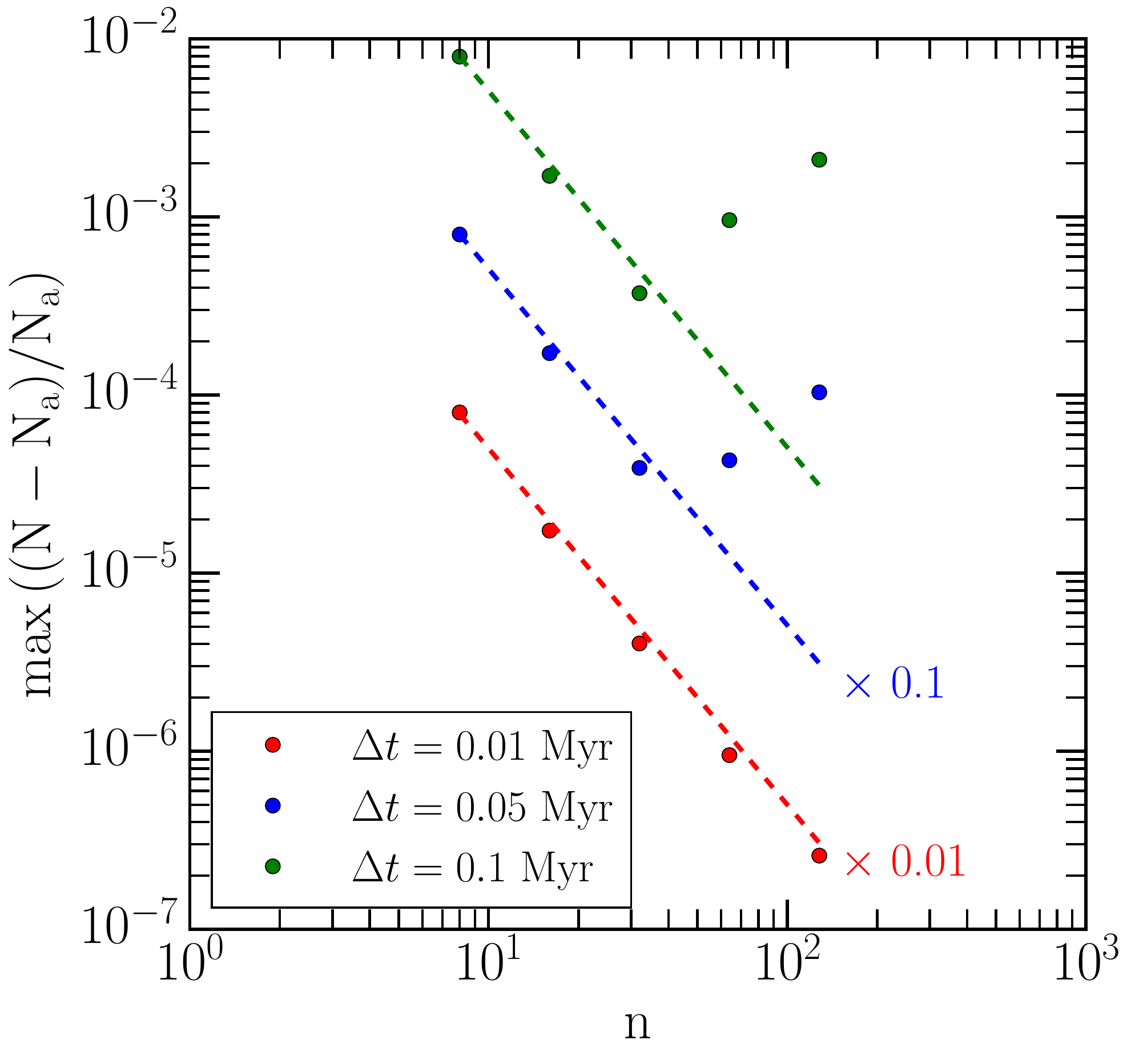}
\caption{{\it Left panel}: comparison between the numerical and the analytical solution at different times, for a fixed spatial resolution $n_r=n_z=128$ (radial profile). {\it Right panel}: relative error as a function of the spatial resolution for different timesteps.}
\label{fig:diff_solution}
\end{figure}

We show in the left panel of Fig.~\ref{fig:diff_solution} the comparison between the analytical and the numerical solutions at different times, for a fixed spatial and time resolution. 
In particular, the plot shows the profile along $r$ (and for $z=0$) of the two solutions. 
As it can be seen, the numerical solution reproduces remarkably well the analytical one.

The relative error as a function of the spatial resolution and for different timesteps $\Delta t$ is shown in right panel of Fig.~\ref{fig:diff_solution}.
The relative error decreases proportionally to the grid step squared, as a result of the discretisation of the operators $\mathcal{L}_z$ and $\mathcal{L}_r$ being accurate up to the second order in $r$ and $z$ for a regular grid.
We observe that the scaling remains valid down to a minimum resolution, $\Delta x \sim \sqrt{D_{xx} \Delta t}$, below which the round-off error dominates the truncation one.\footnote{It is a well-known fact (see, for example \cite{opac-b1077475}) that values of $\Delta t / D_{xx} \Delta x^2$ larger than 1 can introduce spurious oscillations in the numerical solution obtained with the Crank Nicolson scheme. In such condition, one does not expect the relative error to follow the scaling dictated by the discretization order.}

\subsection{Advection}
\label{sec:advection}

\begin{figure}[!t]
\centering
\includegraphics[width=0.49\textwidth,height=0.49\textwidth]{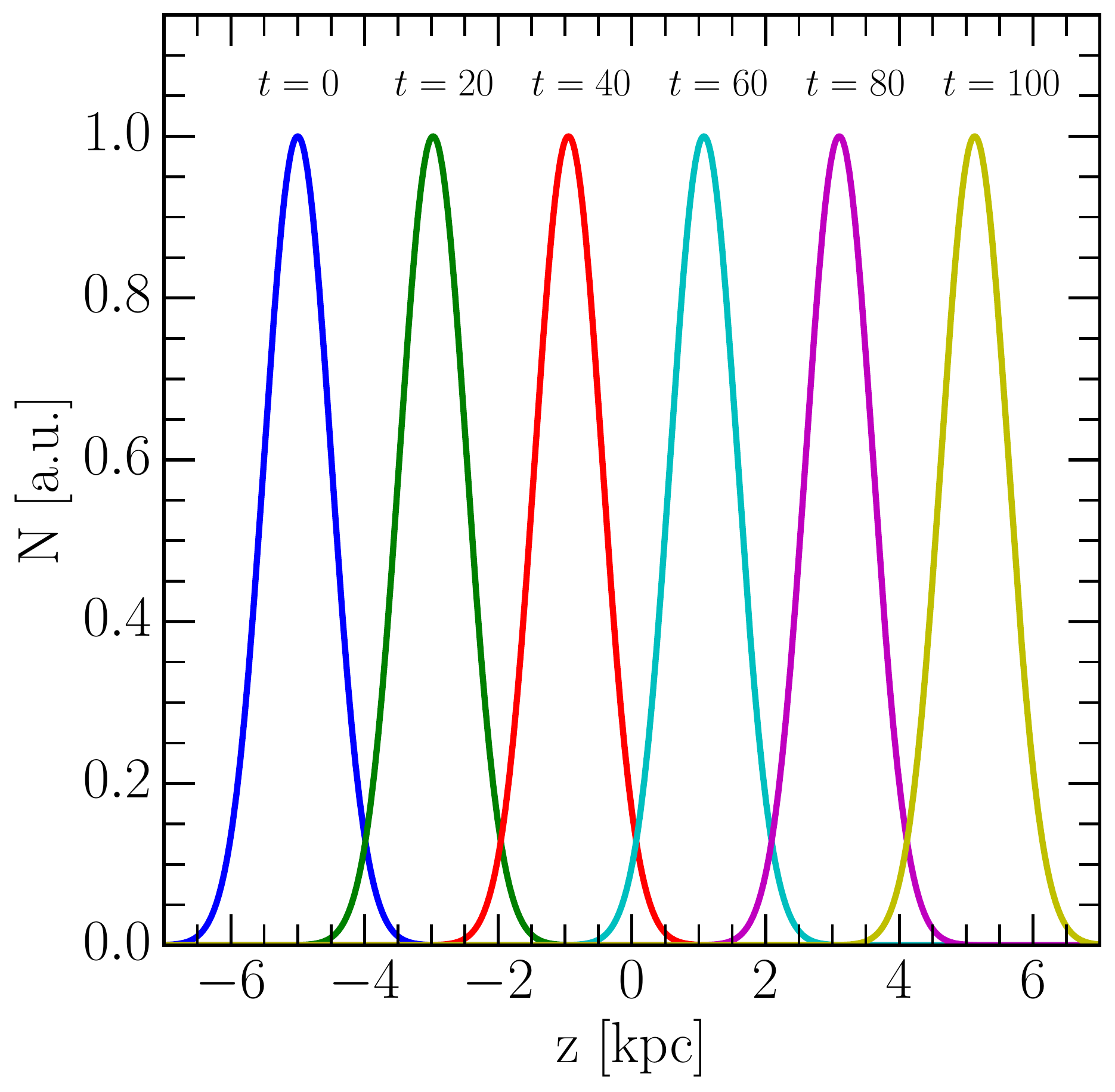}
\hspace{\stretch{1}}
\includegraphics[width=0.49\textwidth,height=0.49\textwidth]{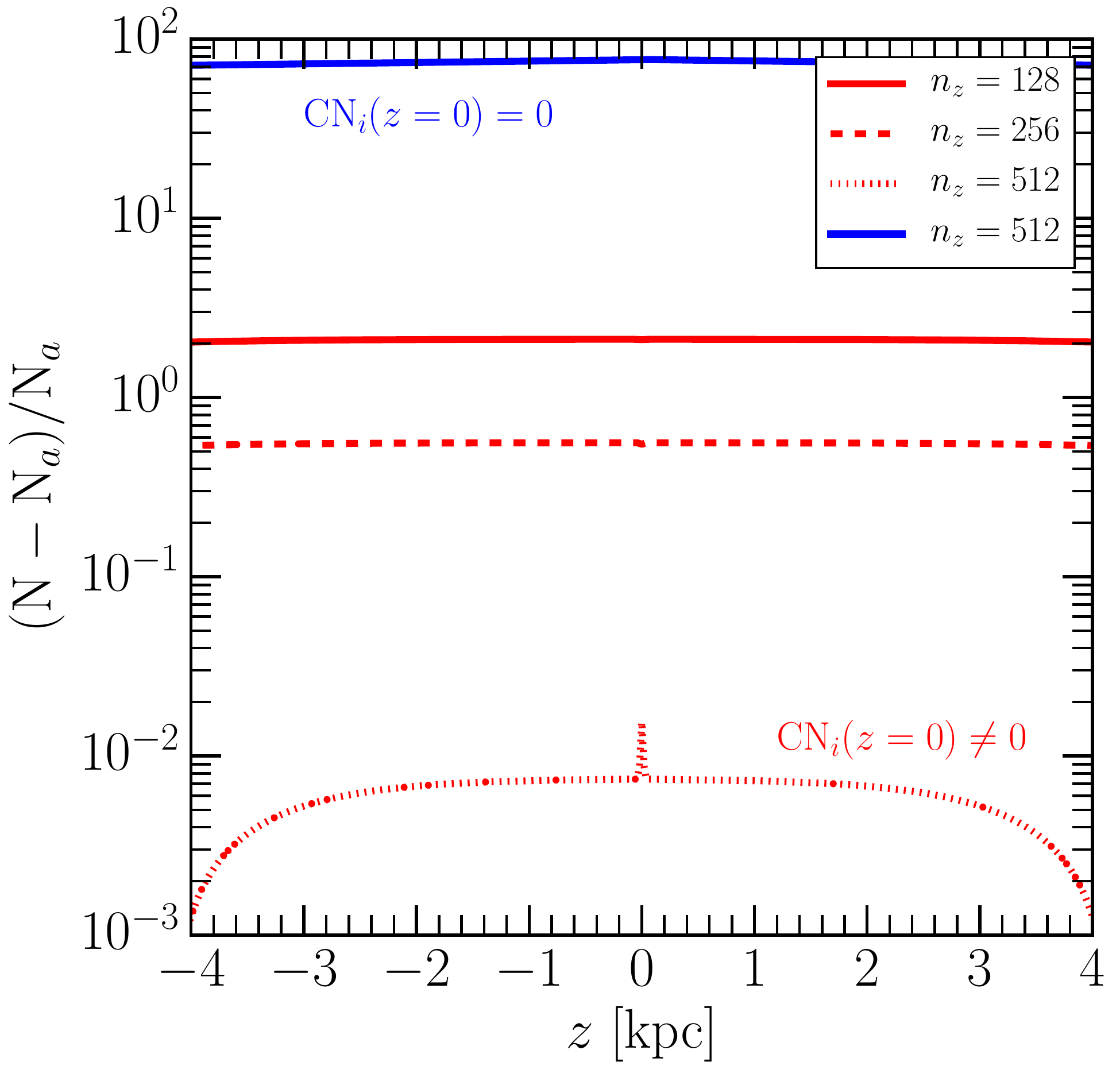}
\caption{{\it Left panel}: solution of the advection equation at different times for $v_w = 100$~km/s. {\it Right panel}: relative difference as a function of the spatial resolution (red lines). Relative difference corresponding to the case with null CN coefficients at $z = 0$ (blue solid line).}
\label{fig:advection_test}
\end{figure}

We test the advection operator by assuming a Gaussian IC with $\sigma = 500$~pc and a constant (both in intensity and direction) advective wind along $z$ with $v_w = 100$~km/s.
Differently than the other cases in which we compare with a steady-state solution, this is a typical Initial Value problem where a differential equation is given together with the unknown function in a given point of the solution domain (in this specific case $t=0$).
 
The left panel of Fig.~\ref{fig:advection_test} shows the density profile at different times. As expected from the analytical solution (see~\ref{App:analyticalsol_advec}), at each time the solution corresponds to the rigidly advected IC for $\Delta z = v_w t_s$, where $t_s$ is the simulation time.
The maximum relative difference between the IC and the solution obtained at $t_s = 100$~Myr is of $\sim 10^{-3}$ for $\Delta z = 20$~pc and it exhibits a scaling with the grid size as given by the truncation order. 

A second test is performed to test specifically the case with discontinuous advective velocity at $z = 0$ as it is the case of a constant Galactic wind.
The analytical solution is given by equation~\ref{Eq:discontinuos_advection}.

We discretize the advective operator with a backward scheme for $z > 0$ and a forward scheme for $z < 0$.
The CN coefficients at $z = 0$ are computed by summing the contribution of the $z > 0$ and $z < 0$ semi-intervals computed in the forward and backward schemes respectively (see table~\ref{sec:cntables}).

In the right panel of Fig.~\ref{fig:advection_test} we show the relative error of the numerical solution with respect to the analytical one for different grid sizes. The solution with $n_z = 512$ points reproduces the analytical function better than $1\%$.  
In the same plot we compare the numerical solution obtained by assuming vanishing CN coefficients at $z=0$ as implemented in the previous version of the code. 

\subsection{Momentum diffusion}
\label{sec:reacctest}

We consider the reacceleration equation in~\ref{Eq:reacc_ana} with normalization of $5.1 \cdot 10^{-16}$}~GeV$^2$/s ($v_A = 50$ km/s) and $Q_0 = 1$. The corresponding analytical solution is obtained in~\ref{App:analyticalsol_reacc} and the discretization scheme is reported in table~\ref{sec:cntables}.

\begin{figure}[!t]
\begin{center}
\includegraphics[width=0.49\textwidth]{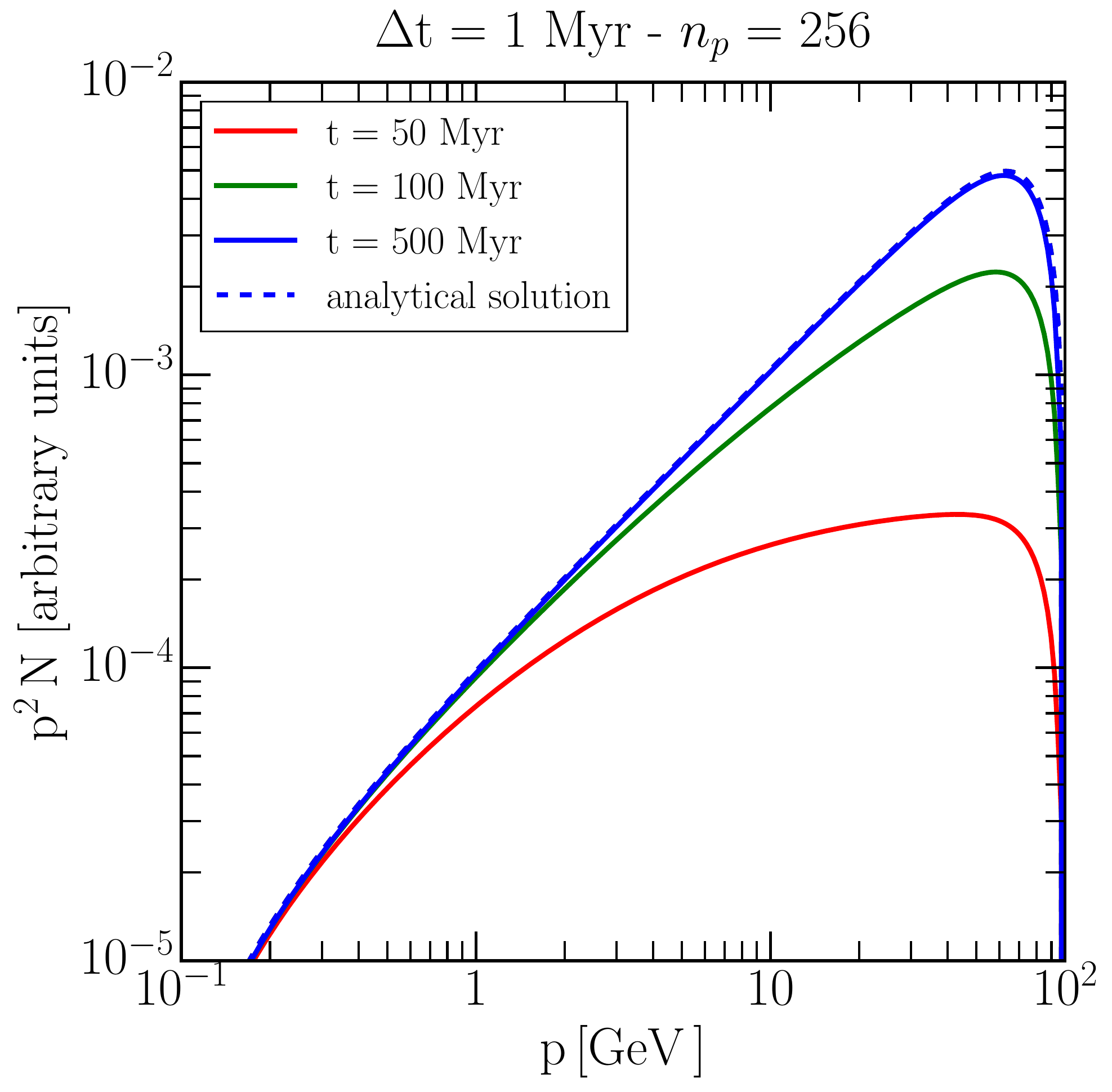}
\hspace{\stretch{1}}
\includegraphics[width=0.49\textwidth]{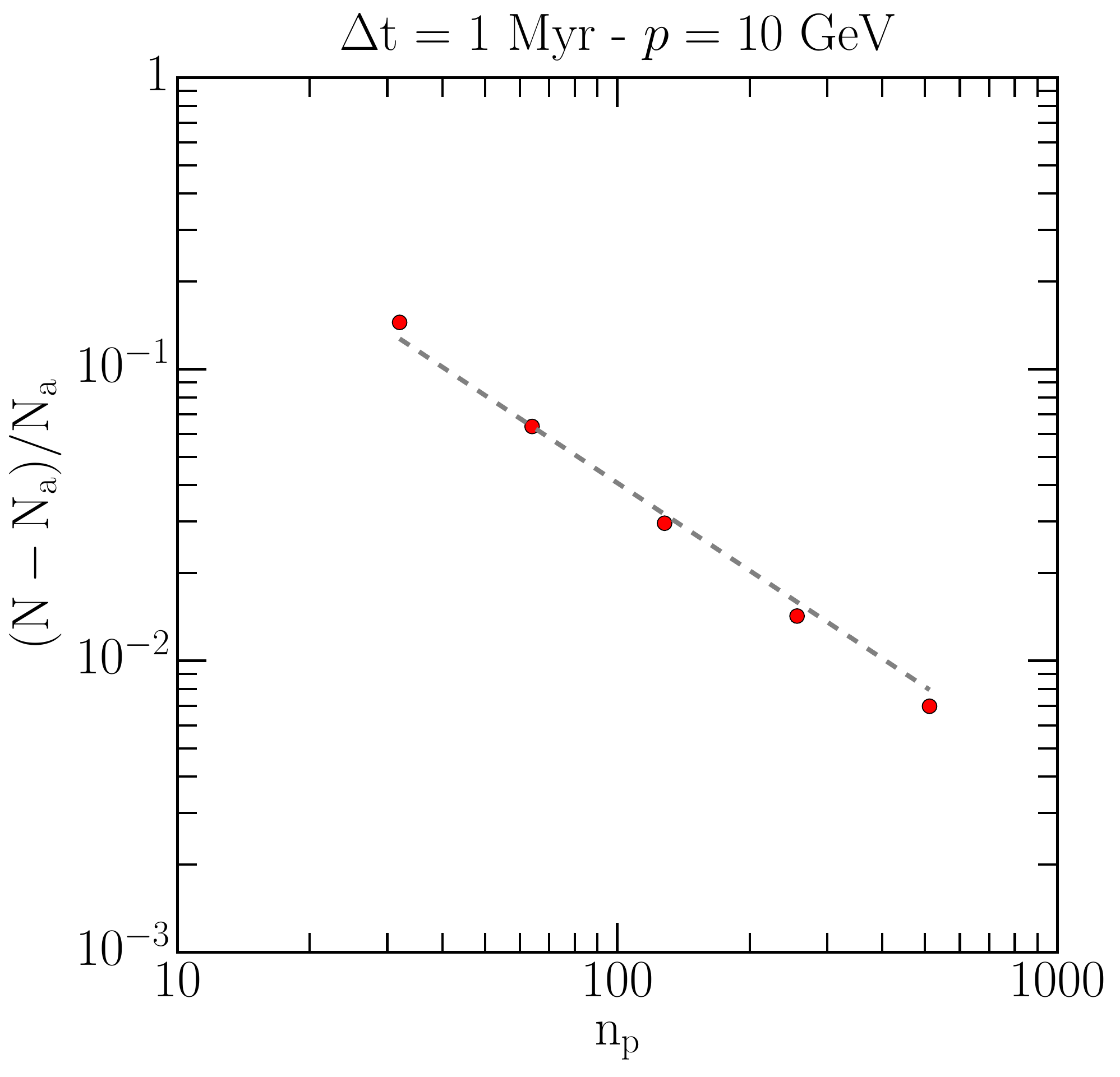}
\caption{{\it Left panel}: comparison between the numerical and the analytical solution of the reacceleration equation for $\Delta t = 1$ Myr. {\it Right panel}: the relative error as a function of the number of momentum grid points for $p = 10$~GeV.} 
\label{fig:solution_reacc}
\end{center}
\end{figure}

Following the same strategy detailed in \ref{sec:spatialdiff}, we consider different values of $\Delta t$ and different grid sizes; the momentum interval is $0.1 \div 10^2$ GeV, the number of intervals we consider is $32 \div 512$.
We show in Fig.~\ref{fig:solution_reacc} (left panel) the comparison between the numerical and the analytical solution for $\Delta t = 1$ Myr. 

We notice that the time needed to reach convergence is much larger at larger energies ($> 10^5$ iterations in the case considered in the plot for $p > 10$ GeV). 
However, at momenta larger than $\sim 10$~GeV the reacceleration operator is usually subdominant when spatial diffusion is also taken into account.

We show in Fig.~\ref{fig:solution_reacc} (right panel) the scaling of the relative error with the number of grid points. 
We obtain the $\propto n_p^{-1}$ scaling as expected from the truncation order of the first derivative term in $\mathcal{L}_p$.

\subsection{Energy losses}
\label{sec:elosstest}

The analytical solution of the energy-loss equation~\ref{Eq:loss_ana} is shown in Sec.~\ref{App:analyticalsol_eloss}.

We compare here the two methods presented in Sec.~\ref{Sec:CN}: a first-order Crank-Nicolson scheme as implemented in the previous version of the code, and a second-order more accurate scheme. Similarly as in the previous sections, we perform the numerical tests considering different values of $\Delta t$ and different grid sizes.

\begin{figure}[!t]
\begin{center}
\includegraphics[width=0.49\textwidth]{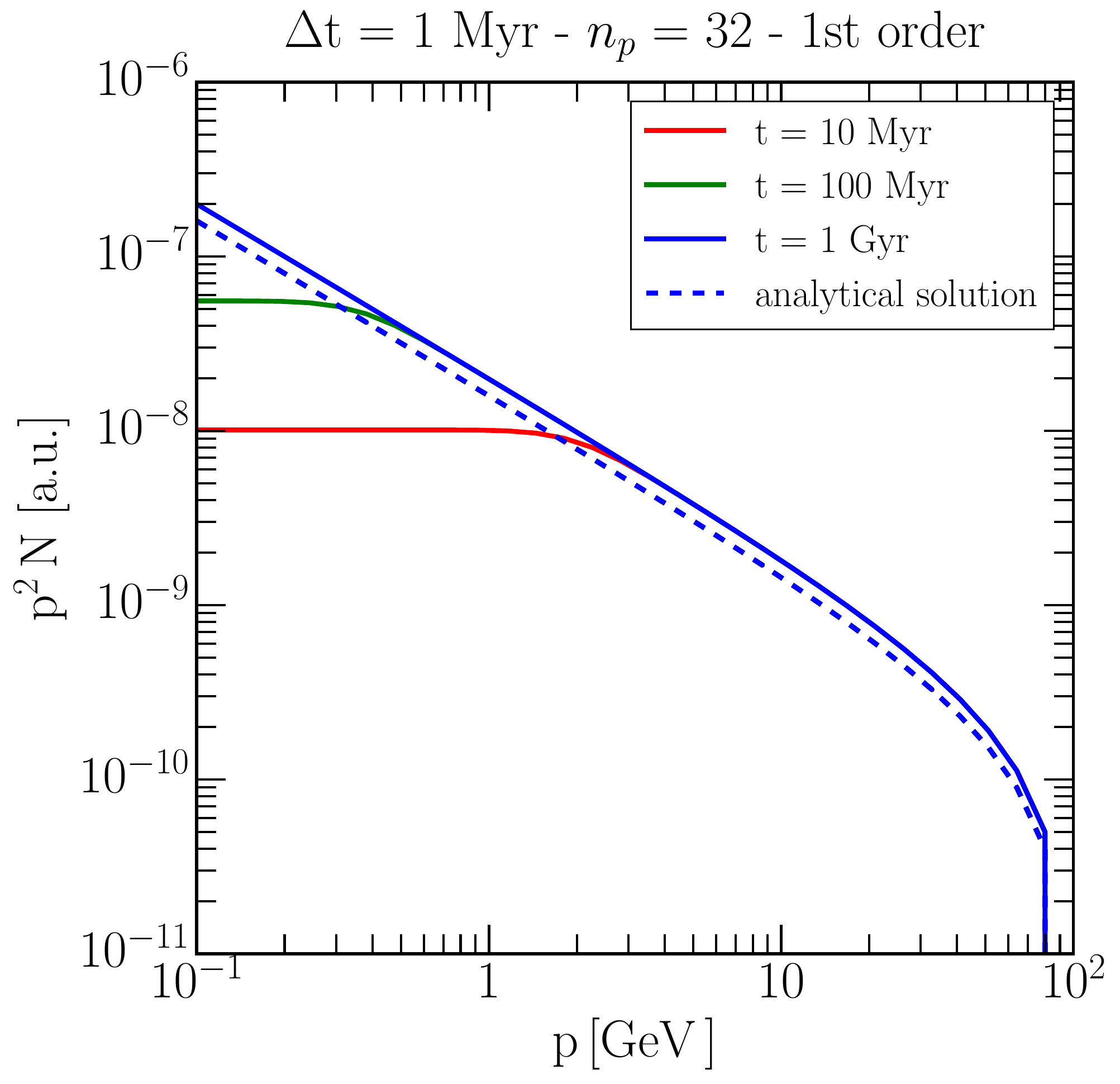}
\hspace{\stretch{1}}
\includegraphics[width=0.49\textwidth]{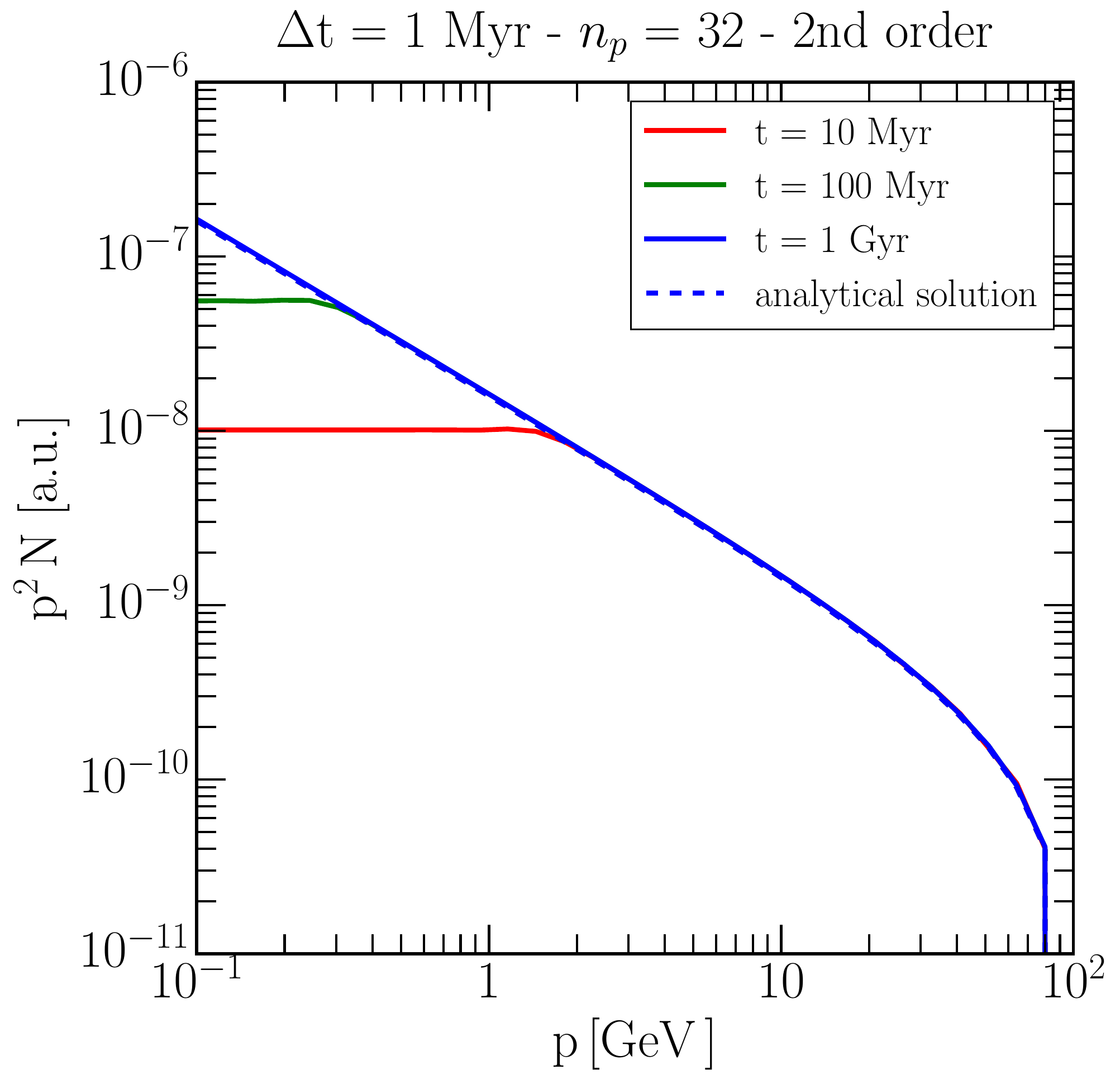}
	\caption{Comparison between the numerical and analytical solution for $\Delta t = 1$ Myr in both the first- (\textit{left panel}) and second-order (\textit{right panel}) cases, for energy loss term.} 
\label{fig:solution_eloss}
\end{center}
\end{figure}

\begin{figure}[!t]
\begin{center}
\includegraphics[width=0.49\textwidth]{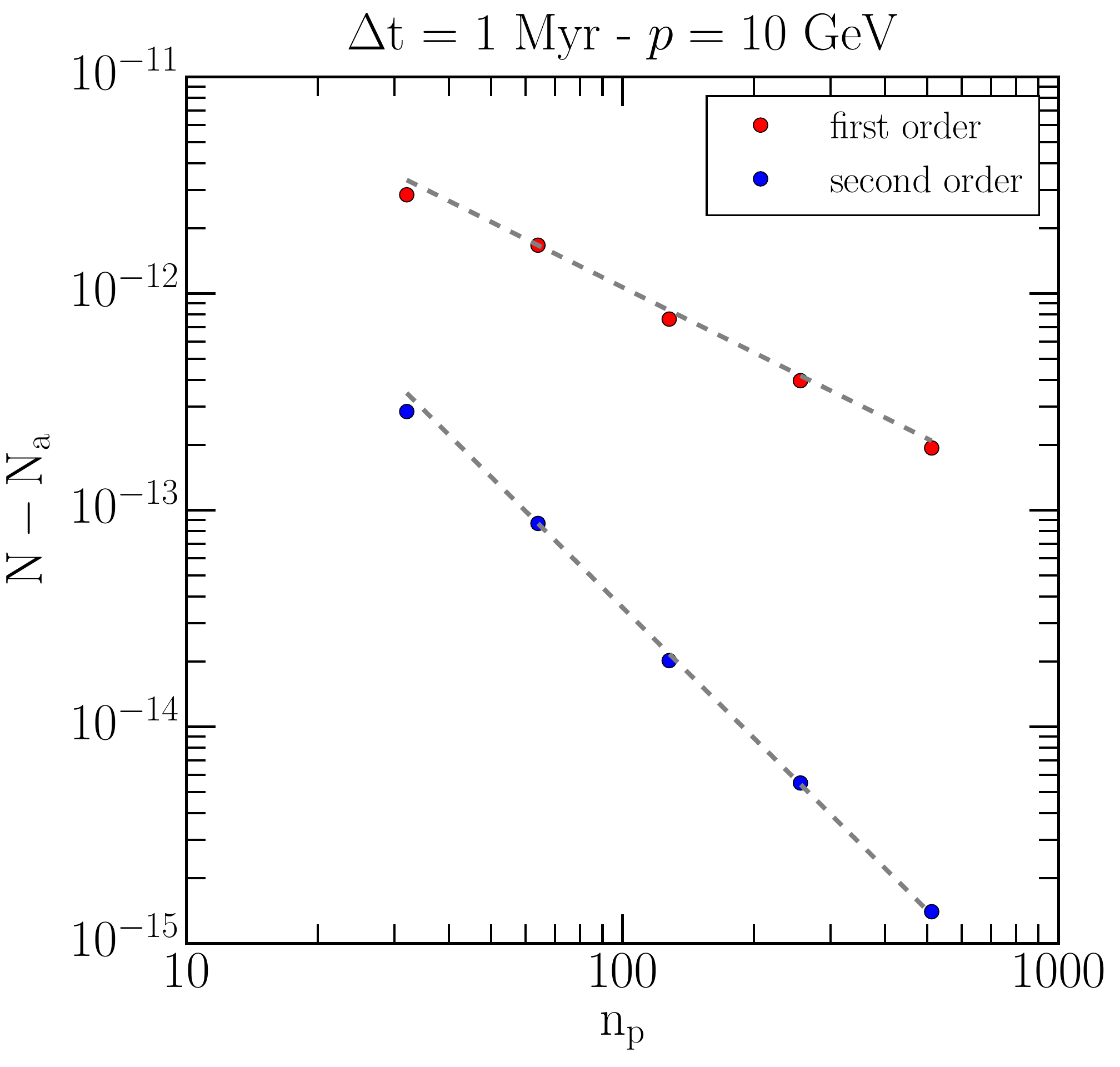}
\hspace{\stretch{1}}
\includegraphics[width=0.49\textwidth]{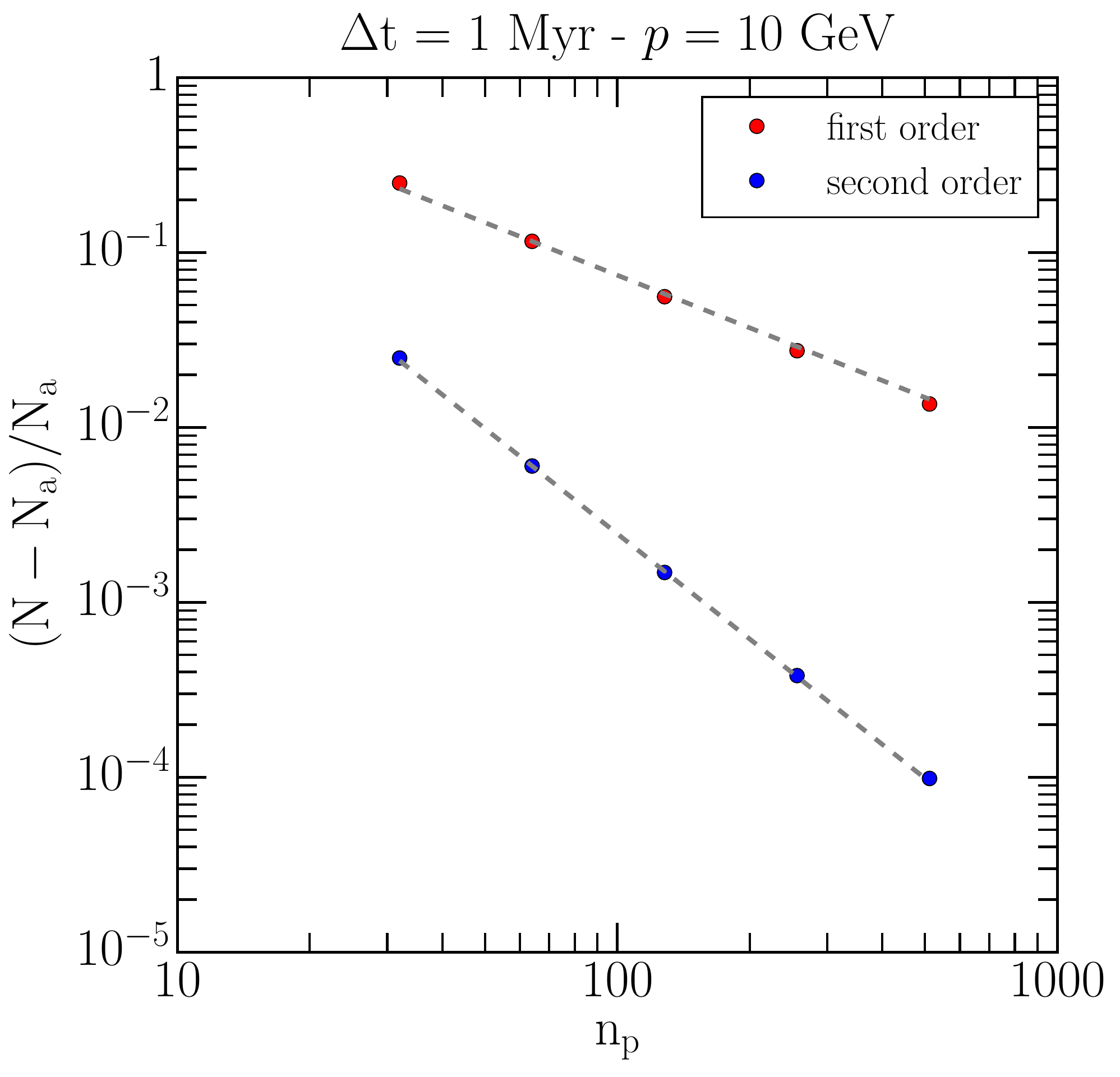}
\caption{Absolute (\textit{left panel}) and relative error (\textit{right panel}) scaling with the number of grid points, for both second and first-order schemes.} 
\label{fig:scaling_eloss}
\end{center}
\end{figure}

We compare in Fig.~\ref{fig:solution_eloss} the analytical solution with the numerical one as obtained for $\Delta t = 1$ Myr with the first- and the second-order discretisation schemes. 
The reader can appreciate the better accuracy of the second-order approach. With $n_p = 32$ grid points in the given momentum range, the first-order scheme produces a $\sim 25\%$ error with respect to the analytical solution, while the second-order scheme accounts for a $\sim 2.5\%$ error only. 
We remark that this improvement in accuracy is obtained without a significant increase in the number of iterations required to reach convergence.

Absolute and the relative errors at $p = 10$~GeV as a function of the grid size is shown in Fig.~\ref{fig:scaling_eloss}. 
Expected scalings are correctly reproduced by the numerical solution. 
Remarkably, the second-order scheme provides a $\lesssim 10^{-2}$ accuracy already with $64$ grid points, while the same accuracy is reached by the first order scheme with more than $512$ points.

\subsection{Convergence and variable time step}
\label{convergence}

In a realistic simulation, the code should be able to propagate particles in a huge range of energies and timescales.

As an example, spatial diffusion is determined by a diffusion coefficient whose dependence on the particle momentum is a power-law with slope $\delta \sim 0.3 \div 0.7$ (see Appendix~\ref{sec:spatialcoefficient}). 
That implies that high-energy particles can diffuse much faster than the low-energy ones. 
In particular, if the range of energies under scrutiny spans several orders of magnitude the difference in the diffusion timescales can be quite large. As an example, with $\delta=0.5$, $D_0 \sim 1.2 \times 10^{28}$~cm$^2$\,s$^{-1}$ and an halo $H = 4$~kpc, a 10 TeV particle have a diffusion timescale of $t_d = 2$~Myr, which is 100 times faster than a 1 GeV particle. 

In terms of the numerical solution of the diffusion equation, the code must evolve with a time-step $\Delta t$ that has to be sufficiently small to correctly follow the diffusion of high-energy particles and it requires a time to reach convergence which is of the order of the longest timescale at the smallest energy.
Because of this, the number of iterations that must be performed in order to find the correct numerical solution through all the energy spectrum could be extremely (and unnecessary) large. 

\begin{figure}
\centering
\includegraphics[width=0.49\textwidth, height=0.45\textwidth]{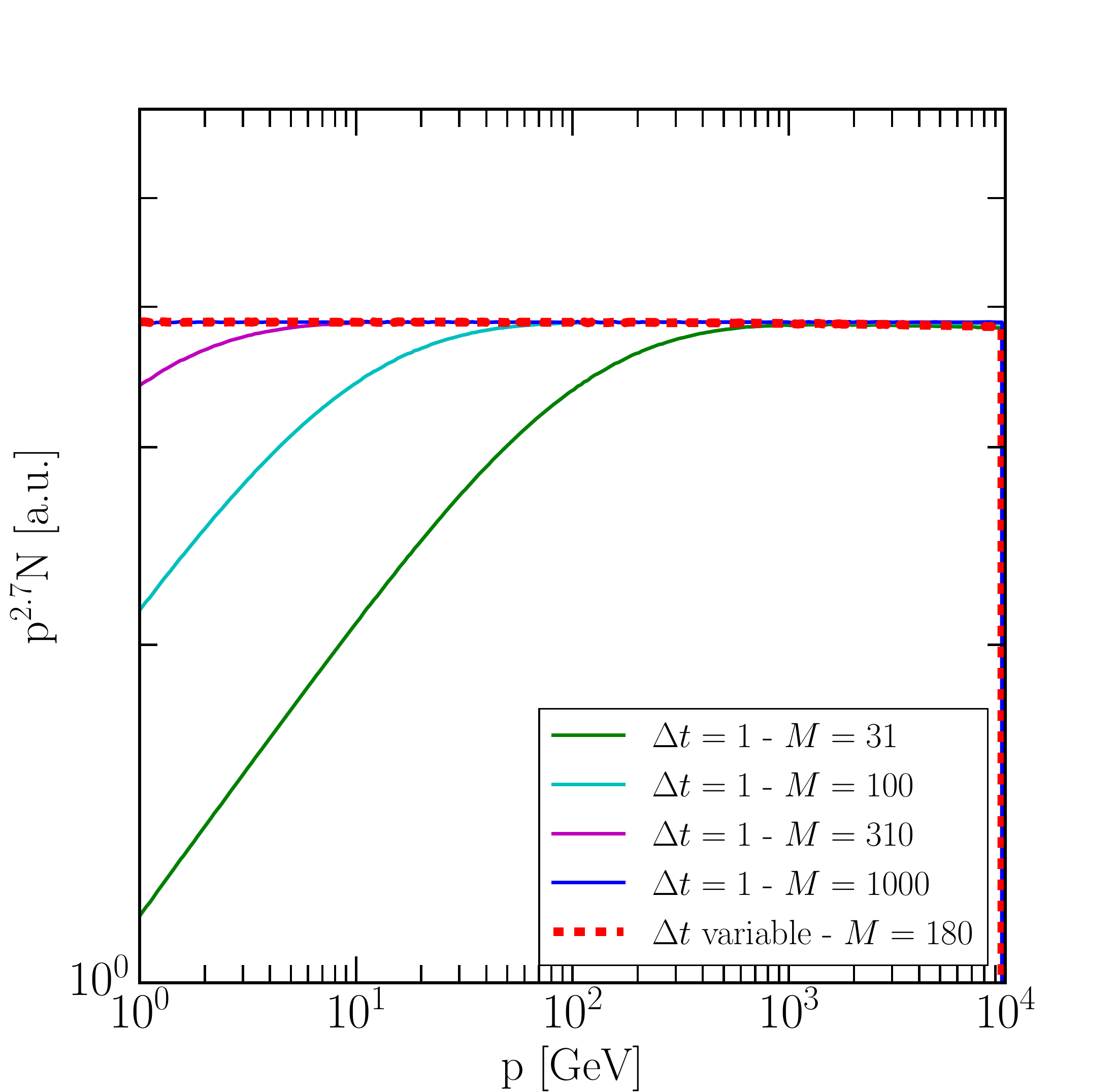}
\hspace{\stretch{1}}
\includegraphics[width=0.49 \textwidth, height=0.45\textwidth]{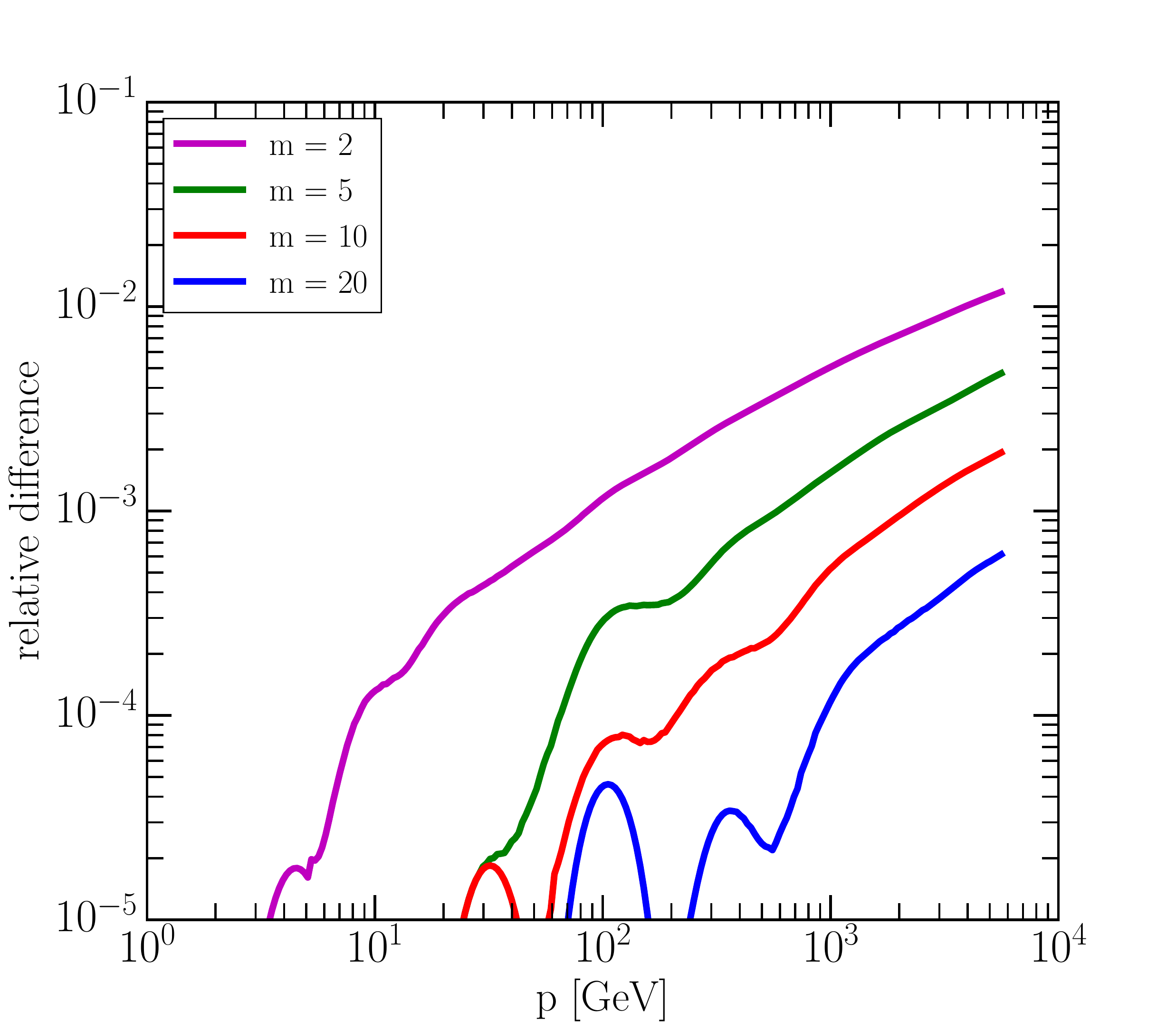}
\caption{{\it Left panel}: momentum spectra obtained by using a constant time-step $\Delta t$ (at different times) are shown in comparison with the solution found with a variable $\Delta t$. {\it Right panel}: the relative difference with respect to the analytical solution is shown as a function of the number of iterations.}
\label{Fig:diff_dtvar}
\end{figure}

An efficient algorithm to significantly reduce the computational effort has been proposed by~\cite{Strong1998} and a key feature of {\tt GALPROP}. 
%
The basic idea is to change the value of $\Delta t$ within the single run, starting from a large time step and reducing it after a certain number of iterations is performed. 

In the example we are considering here, we further assume source distribution having a Gaussian profile along $z$ with characteristic scale $100$~pc. 
We then start our simulation with $\Delta t_{\rm max} = 100$~Myr and we reduce the time step by a factor of $1.1$ each $m = 5$ iterations, until $\Delta t_{\rm min} = 1$~Myr is reached. 
We remind from Sec.~\ref{sec:spatialdiff} that the stability condition for diffusion reads: $\Delta t / t_d < 1$, and we choose $\Delta t$ as half of the corresponding timescale.  

We evaluate the performance of the code under this approach by comparing the local spectrum against the one we obtain by using a constant time-step with $\Delta t_{\rm const} = \Delta t_{\rm min}$.  
The results of the comparison are shown in Fig.~\ref{Fig:diff_dtvar}. 

As one can clearly see, by using a constant time-step, the numerical solution reproduces the analytical one only after $\sim 10^{3}$ iterations.
In fact, the use of a variable $\Delta t$ provides a very efficient way to reduce the number of iterations and an accurate numerical solution is found after $180$ iterations only. 

By showing the right panel of Fig.~\ref{Fig:diff_dtvar}, we point out how the overall accuracy found by using a variable $\Delta t$ is affected by changing the value of $m$, namely the number of iterations for each time step.
Since the total number of iterations is simply proportional to $m$, a trade-off between accuracy and computational time is required to choose this parameter.

\begin{figure}[!t]
\begin{center}
\includegraphics[width=0.5\textwidth]{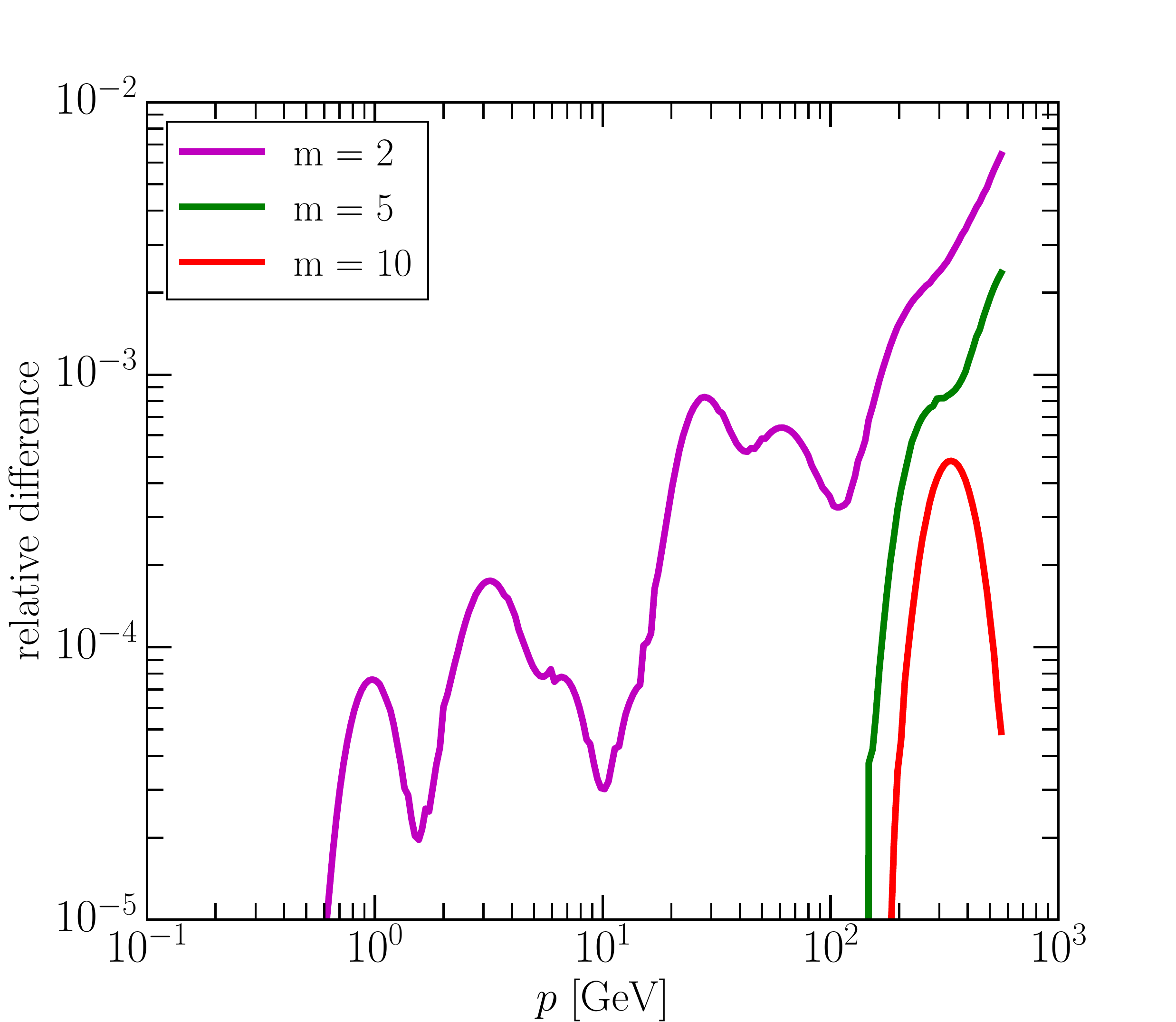}
\caption{The relative difference with respect to the analytical solution for the energy-loss equation is shown as a function of the number of iterations.} \label{fig:nrep_losses}
\end{center}
\end{figure}

As a second example, we consider the loss equation.  
We choose a typical value for the energy loss normalisation, $b_0 = 10^{-16} \, {\rm GeV/s}$, corresponding to the synchrotron energy-loss rate of an electron propagating in a $\sim 5\, {\rm \mu  G}$ magnetic field.
Under these conditions, the associated timescale, $t_l$, ranges from $\sim 3\, {\rm Gyr}$ at the lowest momentum to $\sim 0.3\, {\rm Myr}$ at $1$~TeV.

As in the previous example, we compare the case with a constant $\Delta t_{\rm cst}$ = $t_{l, \rm min} / 2$ and a variable $\Delta_t$, starting from the largest timescale and decreasing until $\Delta t_{\rm cst}$ is reached.
The figure~\ref{fig:nrep_losses}, where we show the accuracy with respect to the analytical solution for different choices of the parameter $m$, confirms a similar behaviour as for the diffusion case.

In conclusion, we found that the variable $\Delta t$ scheme provides a large improvement in terms of number of iterations.
As pointed out by~\cite{Kissmann2014}, this procedure could however have a shortcoming. As shown in the above two examples, the numerical scheme parameters, as the largest and the smallest timestep, strongly depends on the physical parameters (e.g.,~diffusion coefficient, Alfv\'en speed, etc.).
Moreover, the same author discussed how changing the physical parameters in a \galprop~run could generate a final output far from convergence (see also the appendix A.1.1 in \cite{2015APh....70...39K}). 
In fact, \galprop~solicits the user to tune the numerical parameters governing the time steps accordingly to the physical model chosen.

On the contrary, \dragon~code computes the minimum and maximum time steps according to the active operator timescales.
This strategy ensures numerical convergence for most of the physical parameter space.

\section{Physical applications based on new features}
\label{sec:newfeatures}

In this section we present the capabilities of the new features in \dragon~by showing a few example applications.
These features are motivated by several pieces of evidence (partially discussed in the following sections) 
calling for a more realistic description of the CR transport in our Galaxy than the simplified {\it one-zone model} implemented in most of the literature.

This chapter is organised as follows.

\begin{itemize}

\item 
In Section~\ref{sec:gradient}, \ref{sec:slope}, and \ref{sec:slope_aniso} we focus our attention on spatial-dependent and anisotropic diffusion, i.e., the most characterising novelties of \dragon.
Besides being theoretically motivated, the need for implementing these features comes from $\gamma$-ray observations, currently the most effective tracers of the CR distribution in the Milky Way.
In particular, we focus here on the so-called {\it gradient problem} and the {\it slope problem}.

The first anomaly -- already noticed in pre-Fermi data (e.g., with COS-B~\cite{Strong1988aa}) -- consists in a discrepancy between the radial profile of the $\gamma$-ray emissivity (inferred by the $\gamma$-ray longitudinal profile) and the one computed with conventional propagation models, using an injection term based on pulsar or SNR catalogues: The former appears to be flatter along the Galactic plane at large Galactic radii. This problem was recently confirmed by Fermi-LAT observations~\cite{collaboration:2009ag,Collaboration:2010cm}. 
Although several possible explanations have been proposed (e.g., a flatter CR source distribution in the outer Galaxy or a strong Galactic wind), this discrepancy may be the signature of a faster perpendicular escape through the halo in the inner region of the Galaxy~\cite{Evoli:2012ha}.

The second anomaly is about the $\gamma$-ray spectrum: The comprehensive collection of {\tt GALPROP}-based models provided by the Fermi-LAT collaboration in \cite{Ackermann2012} underpredicts the data above 10 GeV in the inner Galactic plane; this discrepancy is present in all the considered setups, with different prescriptions for the gas and source distributions. This high-energy excess is the signature of a harder CR spectrum in the inner Galactic plane, as shown also in~\cite{TheFermi-LAT:2015kwa,Acero:2016qlg,Yang:2016jda}; according to \cite{2015PhRvD..91h3012G} the hardening can be explained by a progressively harder scaling of the diffusion coefficient in the inner Galaxy (see Appendix~\ref{sec:spatialcoefficient}).

\item 
In Section~\ref{sec:pion}, we investigate the consequences on the proton spectrum of including the hadronic energy losses due to pion production and we quantify the impact on the propagated proton slope.  
 
\item 
In Section~\ref{sec:transient} we briefly discuss the possibility of propagating {\it transient} sources. This feature is relevant when CR injection cannot be assumed to be a continuous process in space and time.
The imprint of a nearby recent SN event on the locally observed CR spectrum has been recently revised in \cite{Kachelriess2015} and \cite{Tomassetti2015}. Under peculiar circumstances (in particular, highly anisotropic diffusion along regular magnetic field lines connecting the source to the Solar System), the bulk of low-energy hadronic CRs observed locally could be the result of a single recent event.
Motivated by these considerations, we include in our code the possibility to follow the propagation of CRs emitted by time-dependent sources, and present in detail a physical case with anisotropic diffusion.

\item 
As mentioned in Section~\ref{2Ddiscretization}, one of the key features of the \dragon~code is the possibility to perform CR transport over a non-uniform spatial grid. This property proves useful whenever one wants to study a certain region of the Galaxy (e.g., {\it the Galactic Centre}) with much higher resolution than the rest of the domain. In Section~\ref{sec:NEB} we present two illustrative examples in which the non-uniform grids can be convenient in terms of computational time.

\item 
The relevance of reacceleration in CR propagation is still unknown: Local CR observables cannot provide strong constraints on this process given the uncertainty on our understanding of $\sim$~GeV diffusion.
Since the new version of the \dragon~solver includes an alternative treatment of the boundary conditions in momentum, and a second-order scheme for the energy losses, it is interesting to compare the outcomes with what we would obtain by using the older solver. To this end, we describe the case of leptonic propagation, where the interplay of diffusion, energy losses (due to synchrotron emission and Inverse Compton), and reacceleration produces a peculiar feature in the spectrum at intermediate energies. We discuss this issue in Section~\ref{sec:hump}.

\end{itemize}

\begin{figure}[!t]
\begin{center}
\includegraphics[width=0.5\textwidth]{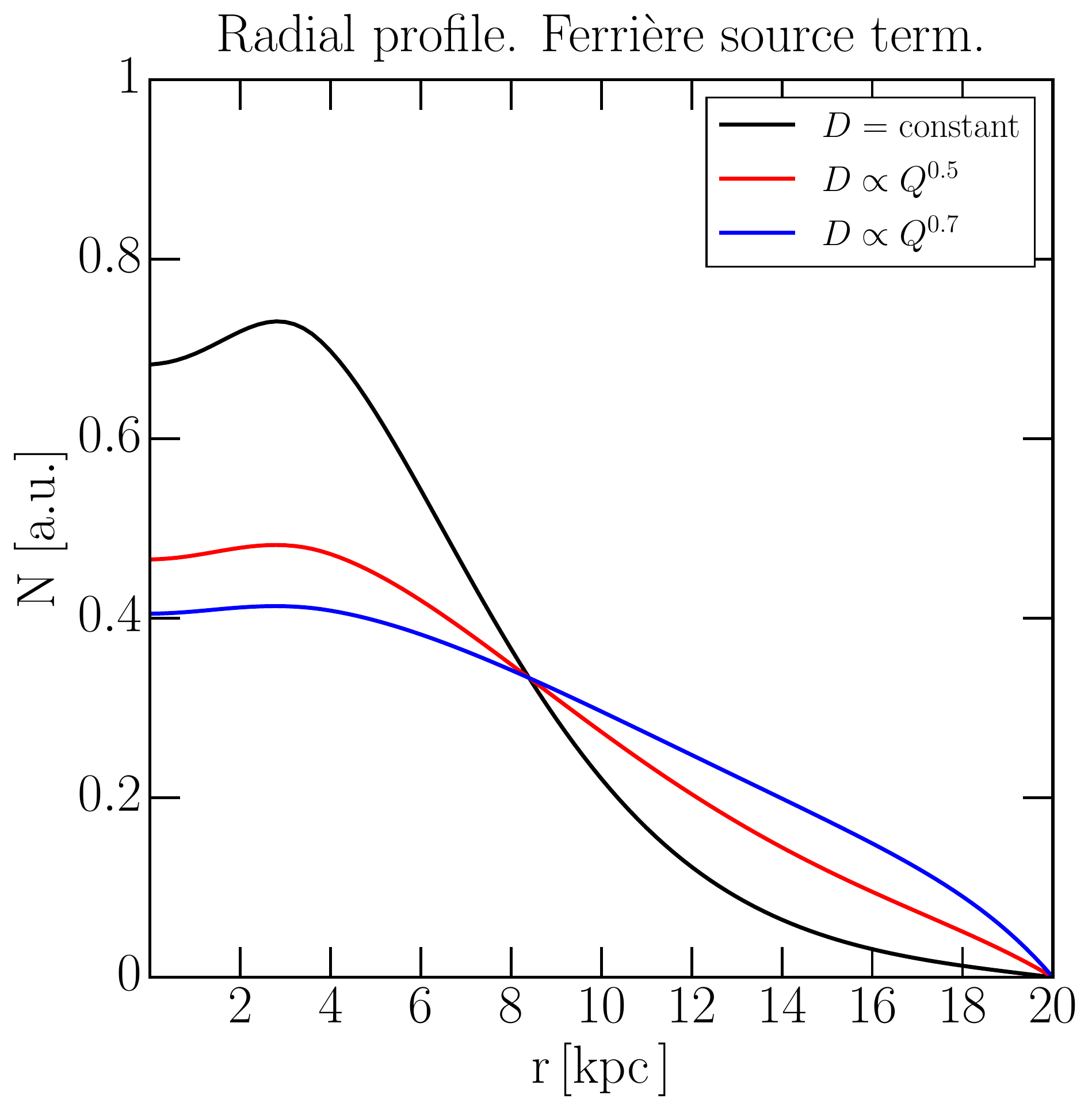}
\caption{Radial profile of CR proton density along the Galactic plane, for three different choices of the parameter $\tau$ setting the dependence of $D_\perp$ on the CR source density. }
\label{fig:gradient}
\end{center}
\end{figure}

\subsection{Inhomogeneous diffusion coefficient: The \textit{gradient problem}}
\label{sec:gradient}

Here we show how to solve the gradient problem in a position-dependent scenario with \dragon~working in 2D mode.
{\tt DRAGON}~featured already inhomogeneous diffusion, this numerical test is shown to confirm one of the main results obtained with the previous version of the code.

Assuming cylindrical symmetry and a purely azimuthal structure of the regular GMF, only the \textit{perpendicular} diffusion coefficient ($D_\perp$) needs to be considered.

We assume $D_\perp$ to be spatially correlated to the turbulence strength, hence to the CR source density $q(r)$. 
Being the exact relation between those quantities poorly known, we parametrise it as in the {\tt PropToSourceTerm} option described in Appendix \ref{sec:spatialcoefficient}, with $\delta = 0.5$ and $\eta = 1$.
For the source term we choose the {\tt Ferriere2001} model, with a slope for momentum of $2.3$.

We solve the 2D diffusion equation by adopting a variable time step $\Delta t$, starting from the largest time step $\Delta t = 16$ Myr and reducing it by a factor $0.5$ each 100 iterations, until $\Delta t = 10^{-4}$ Myr is reached.

\begin{figure}[!t]
\begin{center}
\includegraphics[width=0.49\textwidth]{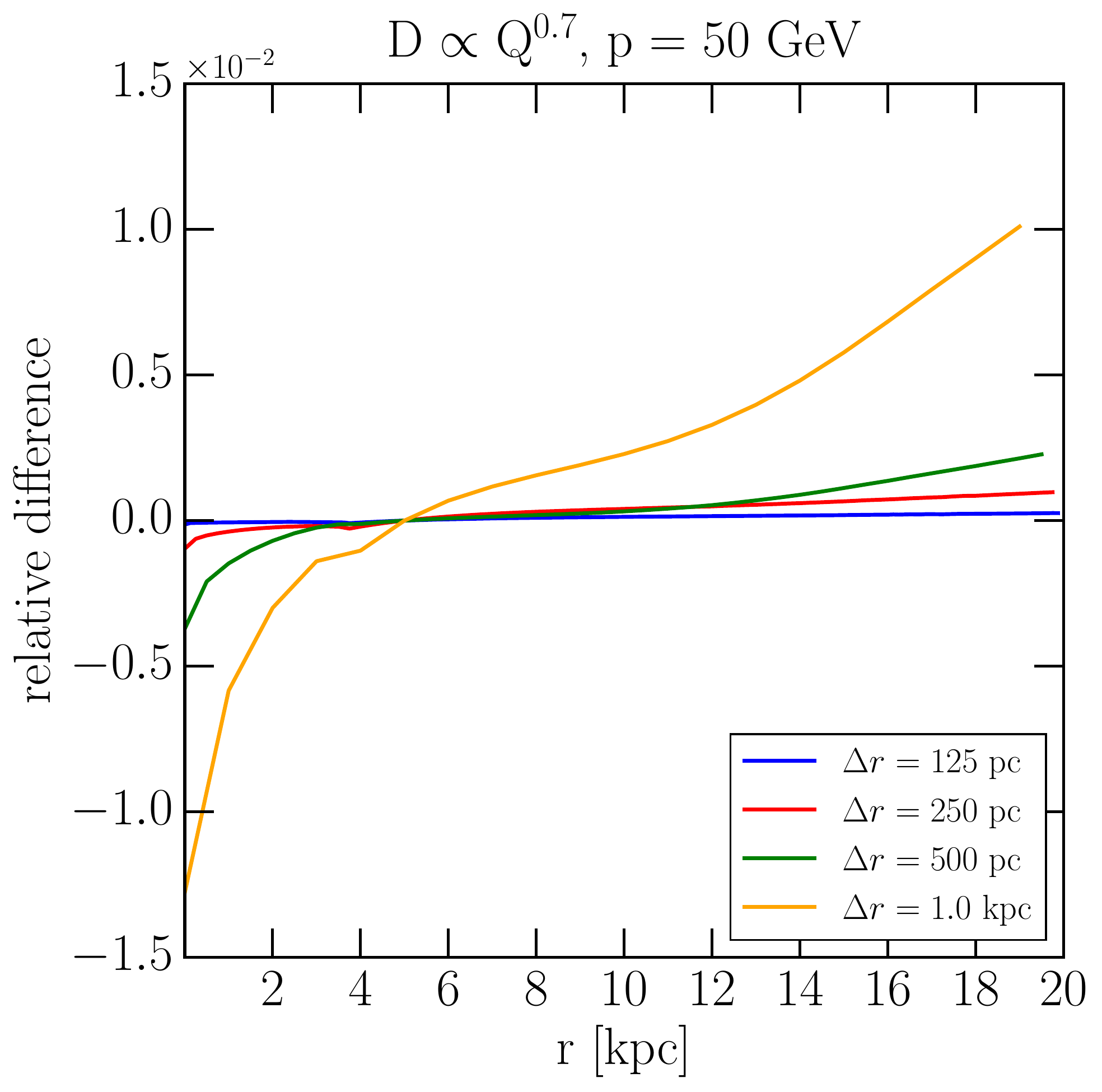}
\hspace{\stretch{1}}
\includegraphics[width=0.49\textwidth]{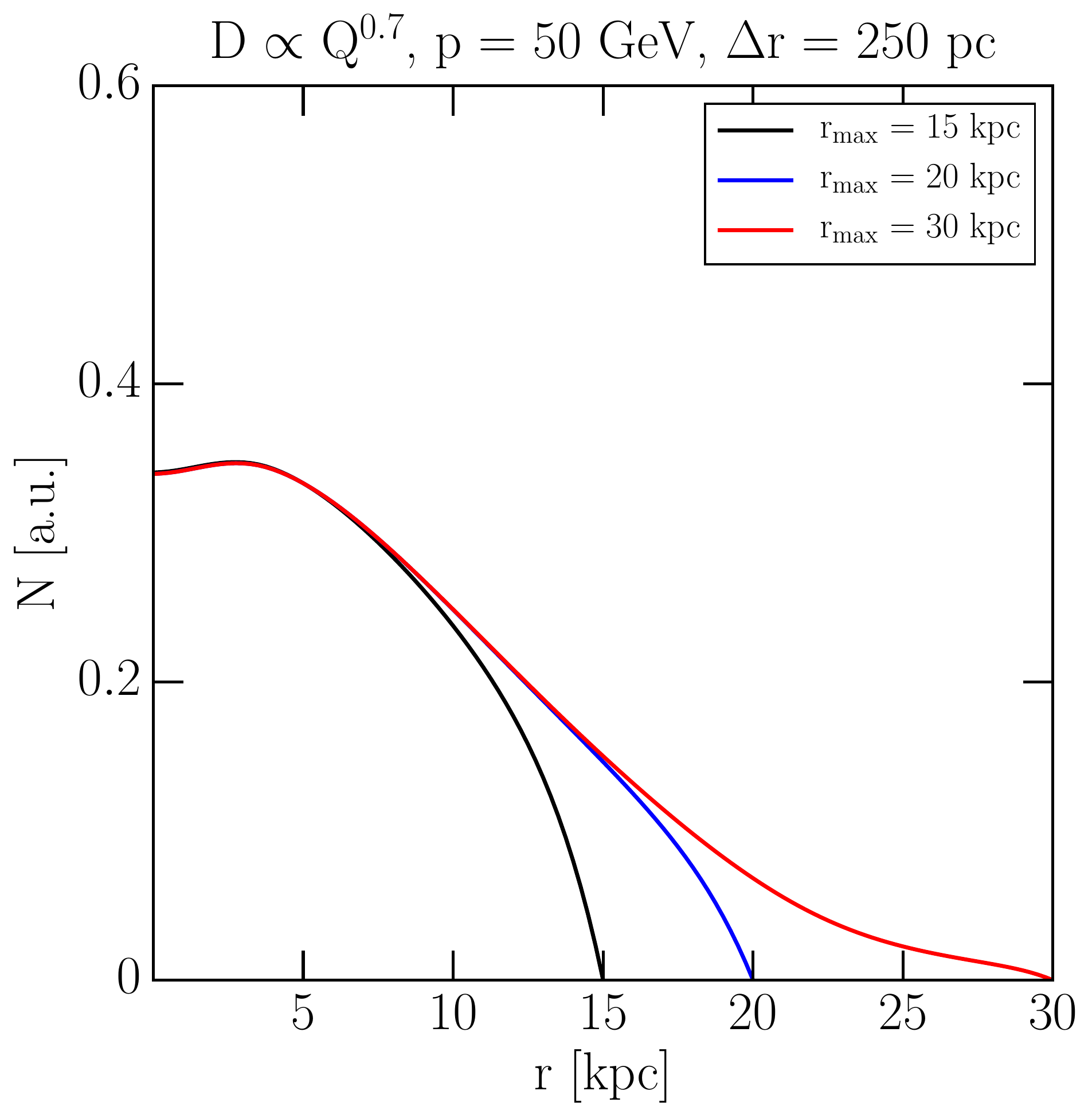}
\end{center}
\caption{{\it Left panel:} relative difference between the proton radial profiles computed with different spatial resolutions and with $\tau = 0.7$. {\it Right panel:} proton radial profile as given by changing the radial boundary position.}
\label{fig:Qtau_tests}
\end{figure}

The main result is reported in Fig.~\ref {fig:gradient}: proton radial profile flattens, thus ameliorating the gradient problem, when the value of $\tau$ is increased (see also \cite{Evoli:2012ha}).

We show further tests in Fig.~\ref{fig:Qtau_tests}, where we focus on the model characterised by $\tau = 0.7$.

In left panel we show the effect of different spatial resolutions on the accuracy of the solution. A reference run with a very fine grid ($\Delta r = 60$~pc) is used as a comparison. It is interesting to notice that, due to the smooth distributions of all the relevant quantities, a percent-level accuracy all through the Galactic plane can be reached with a grid with $\Delta r = 1$ kpc. 
In fact, we find that in most of the applications, a spacing of $\sim 200 \div 500$ pc is adequate to obtain an accurate solution, well below the level of $1\%$.

In right panel we consider the impact of changing the radial boundary condition. 
We find that the solution, in the region of interest ($r < 10$ kpc), is mildly dependent on the position of the boundary, and choosing a value for $r$ larger than $15$~kpc provide a negligible improvement in the accuracy of the solution.

\subsection{Variable scaling of the diffusion coefficient}
\label{sec:slope}

According to~\cite{2015PhRvD..91h3012G}, the $\gamma$-ray longitudinal profile can be  successfully reproduced by a progressively harder scaling of the diffusion coefficient in the inner Galaxy.

In \dragon, this scenario can be obtained by using the model {\tt VariableSlope} for the diffusion coefficient (see Appendix~\ref{sec:spatialcoefficient}). In~\cite{2015PhRvD..91h3012G}, a similar result was obtained with {\tt DRAGON}~by adding to the CN coefficients the additional terms containing the spatial derivatives of the rigidity power-law.

\begin{figure}[!t]
\begin{center}
\includegraphics[width=0.49\textwidth,height=0.49\textwidth]{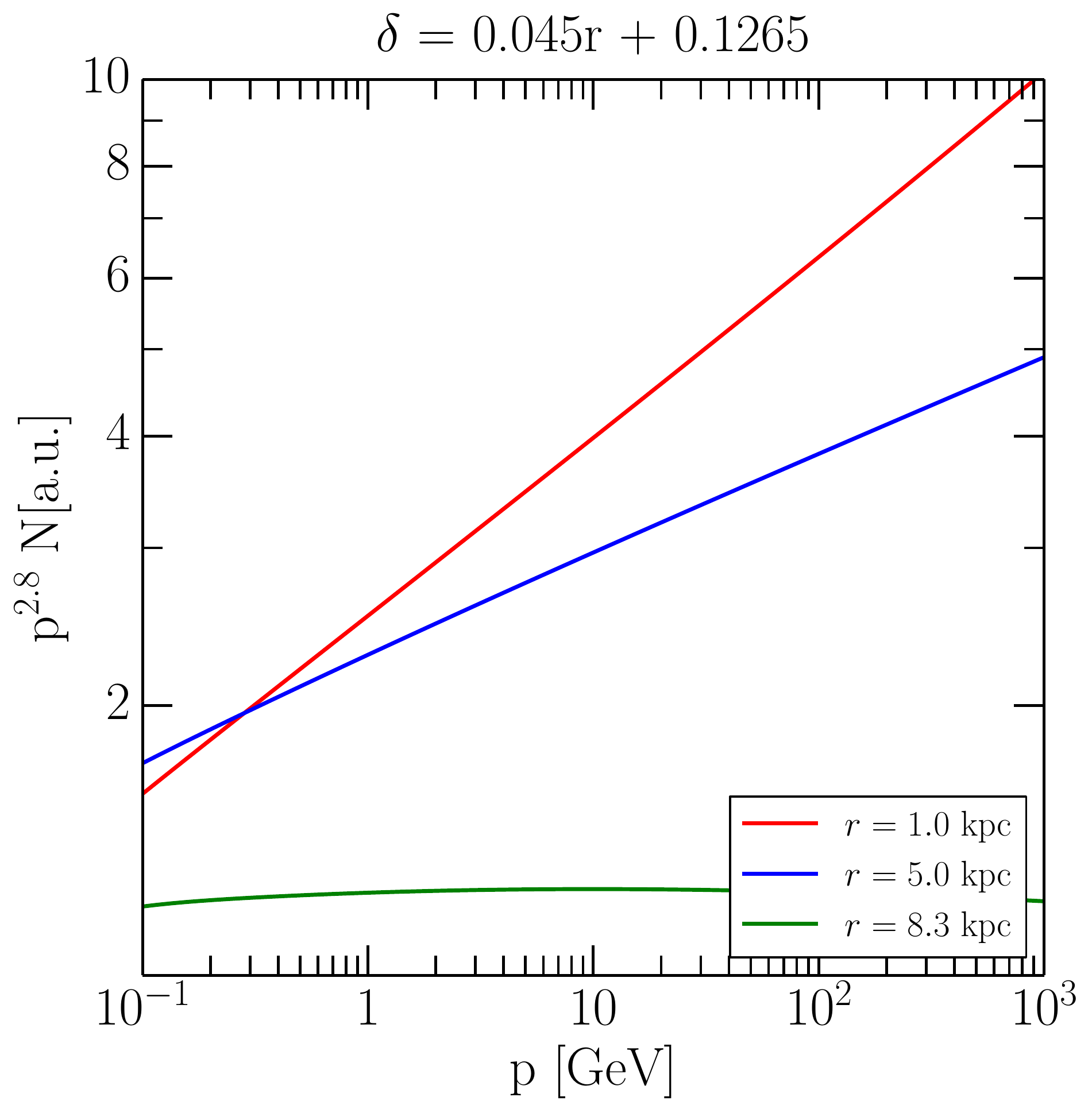}
\hspace{\stretch{1}}
\includegraphics[width=0.49\textwidth,height=0.49\textwidth]{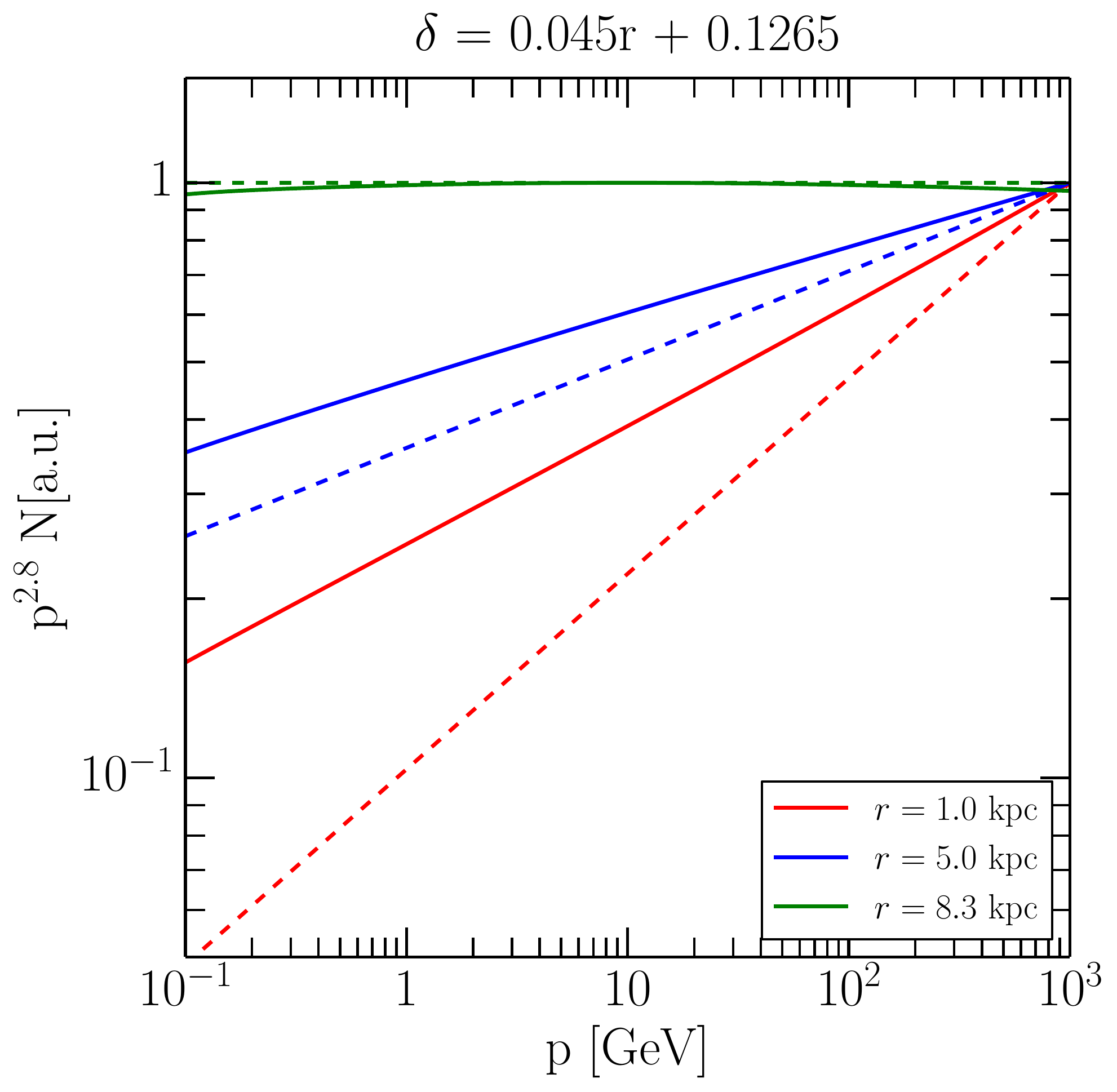}
\caption{{\it Left panel:}  CR proton spectrum on the Galactic plane at three different position as computed with a model with variable $\delta$. {\it Right panel:} comparison of the proton spectra obtained with \dragon~(solid lines) with the power-laws given by Eq.~\ref{eq:gamma_naive} (dashed lines). All curves are normalised to unit at 1 TeV.}
\label{fig:delta_variable_test}
\end{center}
\end{figure}

The spectra at different locations are shown in Fig.~\ref{fig:delta_variable_test}.
In left panel we reproduce the case as in~\cite{2015PhRvD..91h3012G} with $a = 0.045$~kpc$^{-1}$ and $b = 0.126$ (giving $\delta(r_\odot) = 0.5$). These values provide a significant hardening in the inner Galaxy, compatible with the $\gamma$-ray data (see ~\cite{2015PhRvD..91h3012G} for the details) .

The CR proton slope at different radii can be compared with the following prescription:
\begin{equation}\label{eq:gamma_naive}
\gamma(r) = \gamma_{\rm inj} + \delta(r),  
\end{equation}
which would be the generalisation of the leaky-box model, and valid in the thin disk approximation. 

In right panel of Fig.~\ref{fig:delta_variable_test} we compare propagated proton spectra with those expected from Eq.~\ref{eq:gamma_naive}. 
In the specific case we are considering, the relation between injected and propagated slopes remains valid at large radii (in particular, it is obtained at the Sun position).
As a consequence, a conventional propagation model tuned on local CR observables is still in agreement with local measurements (e.g., B/C) even if the spatially-dependent scaling is introduced.

\subsection{Anisotropic diffusion coefficient}
\label{sec:slope_aniso}

In this section we show how a radial dependence of the CR spectral index, similar to that discussed in the previous Section, may be obtained in a simplified framework of anisotropic diffusion.

Since CR diffusion is caused by their scattering off the magnetic field perturbations, and since such perturbations are usually expected to have components both in the perpendicular plane and along the parallel direction (with respect to the direction of the regular GMF), CRs can experience both a parallel and a perpendicular diffusion.  As discussed in \cite{DeMarco2007a}, numerical simulations of charged particle propagation in turbulent MFs found that these two components of the diffusion tensor are characterised by a different scalings with respect to the CR momentum. 

\begin{figure}[!t]
\begin{center}
\includegraphics[width = 0.49\textwidth]{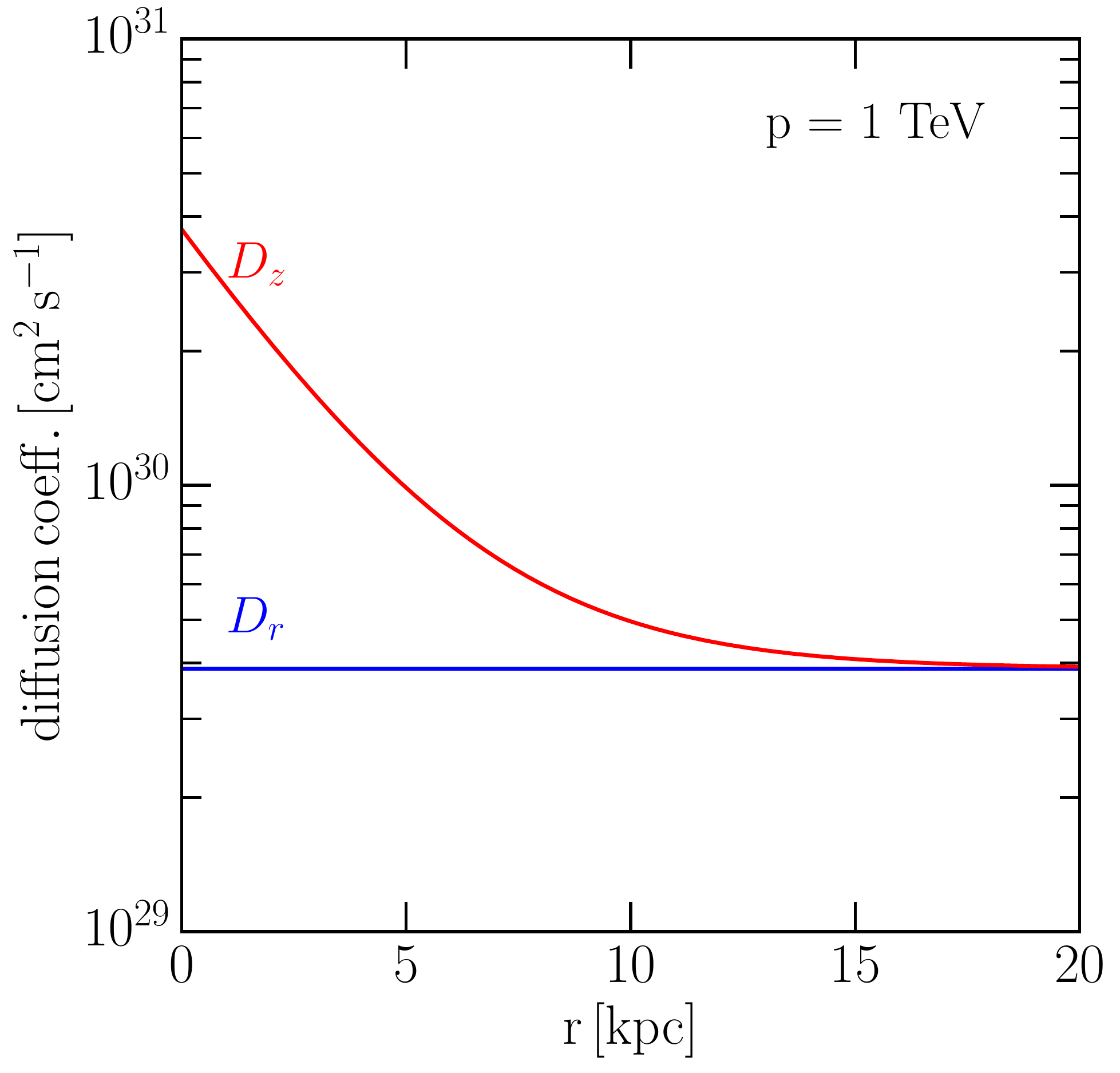}
\includegraphics[width = 0.49\textwidth]{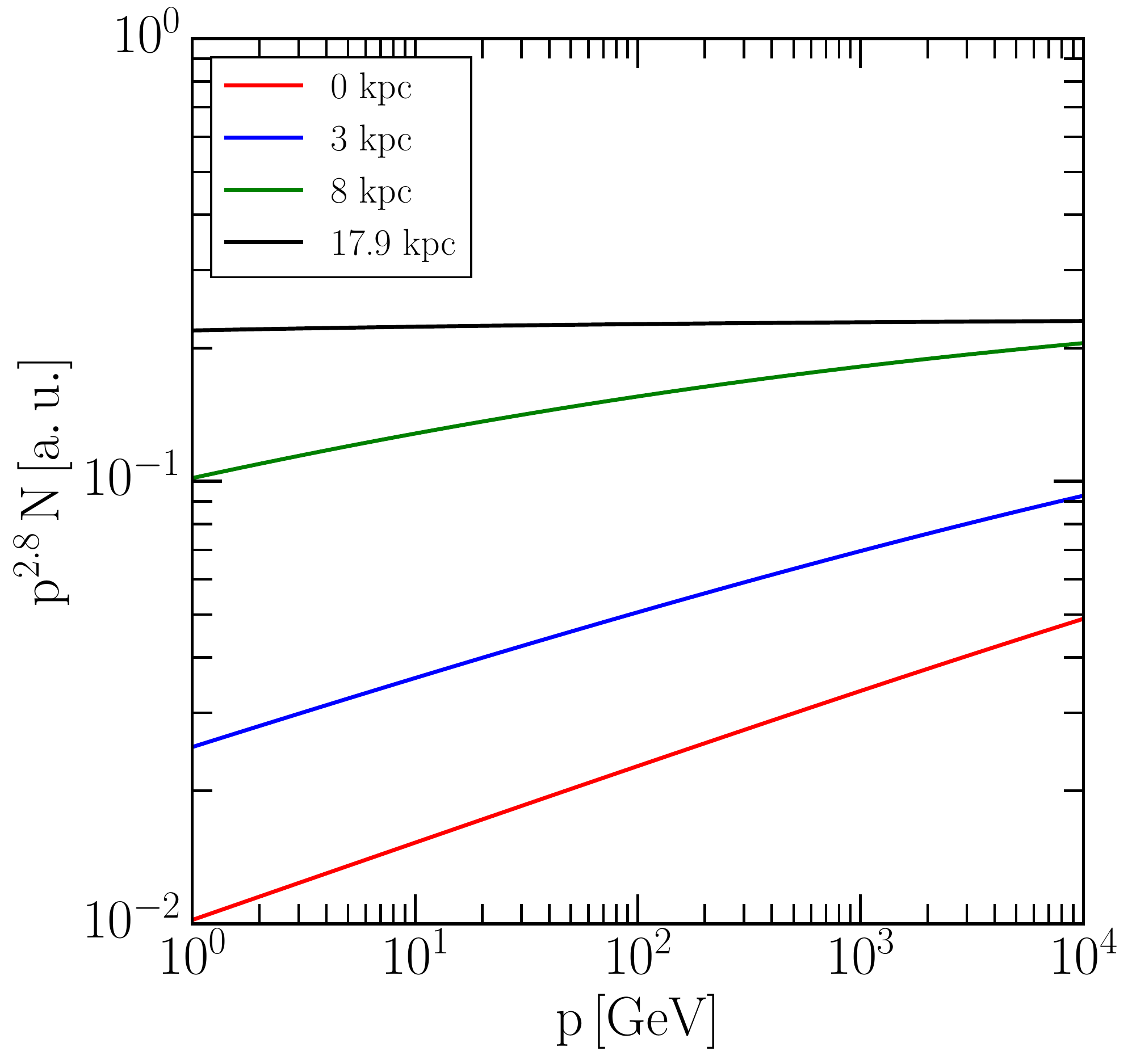}
\caption{{\it Left panel:} profiles of the diffusion coefficients along $r$ and $z$ for particles with $p = 1 TeV$ are shown. {\it Right panel:}  energy spectrum computed at different radial distances from the Galactic Centre.}
\label{fig:test_aniso}
\end{center}
\end{figure}

In order to test such scenario, we consider a GMF with two components:

{\begin{itemize}
\item{A purely azimuthal component, lying on the Galactic disk.}
\item{An out-of-plane component, directed along the $z$-axis and confined within the bulge ($R  < 2.9$ kpc).}
\end{itemize} 

This model is a simplified version of the {\tt Jansson2012} model actually implemented in \dragon ~and based on \cite{2012ApJ...757...14J} (more details in Appendix~\ref{regMF}). 

We expect therefore diffusion along the $r$-direction to be purely \emph{perpendicular}, since the GMF has no radial component, while the diffusion coefficient along the $z$-axis is given by the sum of a \emph{parallel} and a \emph{perpendicular} term: 

\begin{equation}
\begin{aligned}
D_r & = D_{0,\perp} \left(\frac{p}{p_0} \right)^{\delta_{\perp}}\\
D_z & = D_{0,\perp} \left(\frac{p}{p_0} \right)^{\delta_{\perp}} + D_{0,\parallel}\; \mathrm{exp}\left(-\frac{r}{R_0}\right) \left(\frac{p}{p_0} \right)^{\delta_{\parallel}}
\end{aligned}
\end{equation}

with $p_0 = 1$ GeV. 

For the normalization of the diffusion coefficients, we assume $D_{0,\parallel} = 10^{28}$ cm$^2$ s$^{-1}$ and, following~\cite{DeMarco2007a}, $D_{0,\parallel}/D_{0,\perp} = 30$. 
As already said, parallel and perpendicular diffusion are characterised by a different dependence on particles momentum: following again the results found in~\cite{DeMarco2007a}, we assume that $\delta_{\parallel}$ = 0.33 and $\delta_{\perp}$ = 0.5. The profile along $r$ of the diffusion coefficients $D_r$ and $D_z$ for particles with $p = 1$~TeV is shown in the left panel of Fig.~\ref{fig:test_aniso}. 

Concerning the geometry of the halo, we assume its radius to be 20 kpc, and its height to be 4 kpc.
We assume as a source term a Gaussian disk with momentum power-law injection:
\begin{equation}
Q(p,r,z) = \frac{1}{\sqrt{2\pi}z_s} \mathrm{exp} \left( -\frac{z^2}{z_s^2}\right)\;\left(\frac{p}{p_0}\right)^{-2.3}. 
\end{equation} 
with $z_s = 0.1$~kpc.

We show in right panel of Fig.~\ref{fig:test_aniso} the steady-state energy spectrum on the Galactic plane at different distances from the GC. For low values of $R$ (within the bulge), parallel diffusion dominates and, as a consequence, a significant hardening can be noticed in the propagated slope; on the other hand, for larger values of $R$, the slope is steeper, and tends to $2.8$, i.e. the value expected for perpendicular diffusion only. 

\subsection{Pion momentum losses}
\label{sec:pion}

A novelty introduced in \texttt{DRAGON2} with respect to earlier versions is a new implementation of pion production as a continuous loss term, in addition to ionisation and Coulomb losses.

In our numerical tests we consider a homogeneous source term confined in a disk (with scale height $\simeq 100$~pc); the energy-loss due to pion production relies on the analytical parametrisation reported in Section~\ref{sec:pion_pion}, with the scale height $z_{\rm losses} = 100$ pc. 

\begin{figure}[!t]
\begin{center}
\includegraphics[width=\textwidth]{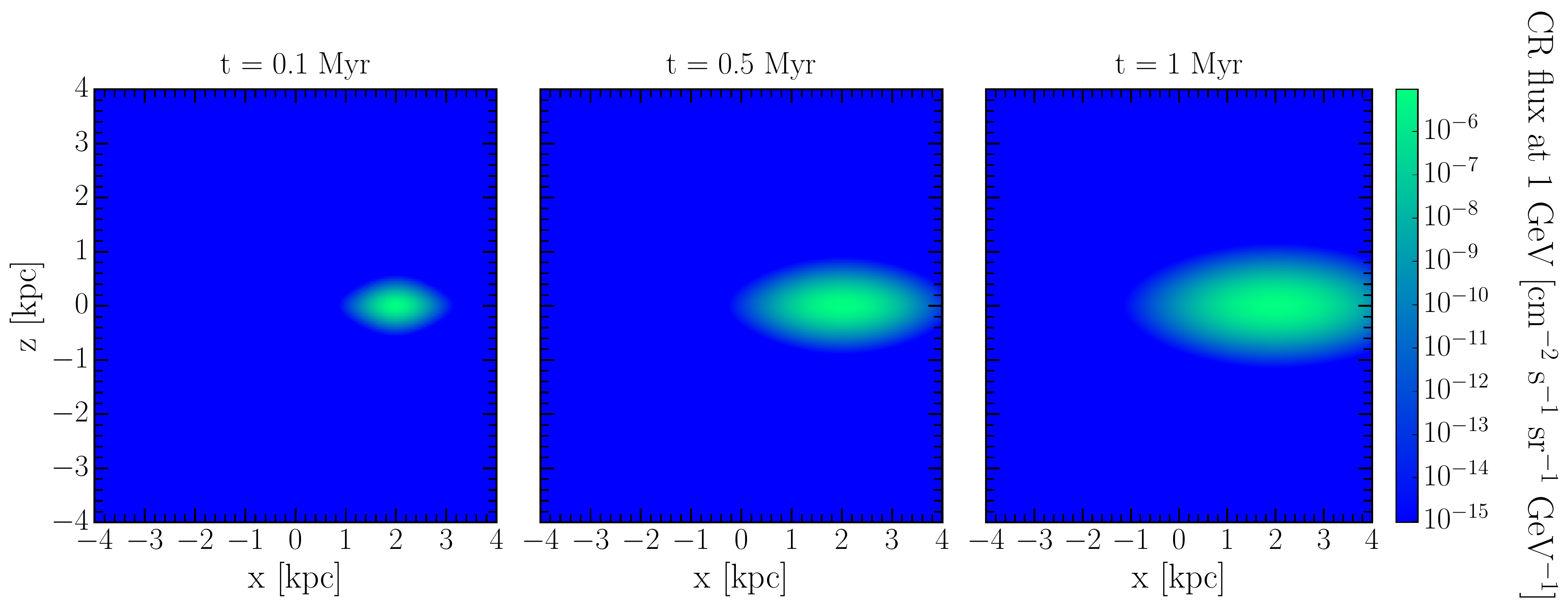}
\caption{CR density in the $x$-$z$ plane at different times; CRs are propagating anisotropically, with slower diffusion in the $z$ direction.} \label{fig:map}
\end{center}
\end{figure}

Concerning the diffusion coefficient, we refer to the Eq.~\ref{Eq:dcost}, and implement a standard diffusive regime corresponding to  $D_0 = 1.8 \cdot 10^{28}$ cm$^2$~s$^{-1}$ at $1$ GV, and $\delta = 0.5$; The diffusive halo height is set to $H = 4$ kpc.

In Fig.~\ref{fig:hadronic_loss_test} we show the results of our numerical tests: We notice that the impact of the pion-production energy loss term on the proton flux ranges from $\simeq 5\%$ at $\simeq 100$ MeV down to few percent at energies larger than $100$ GeV. The impact on the slope is $\lesssim 0.5\%$.

\begin{figure}[!t]
\begin{center}
\includegraphics[width=0.49\textwidth,height=0.49\textwidth]{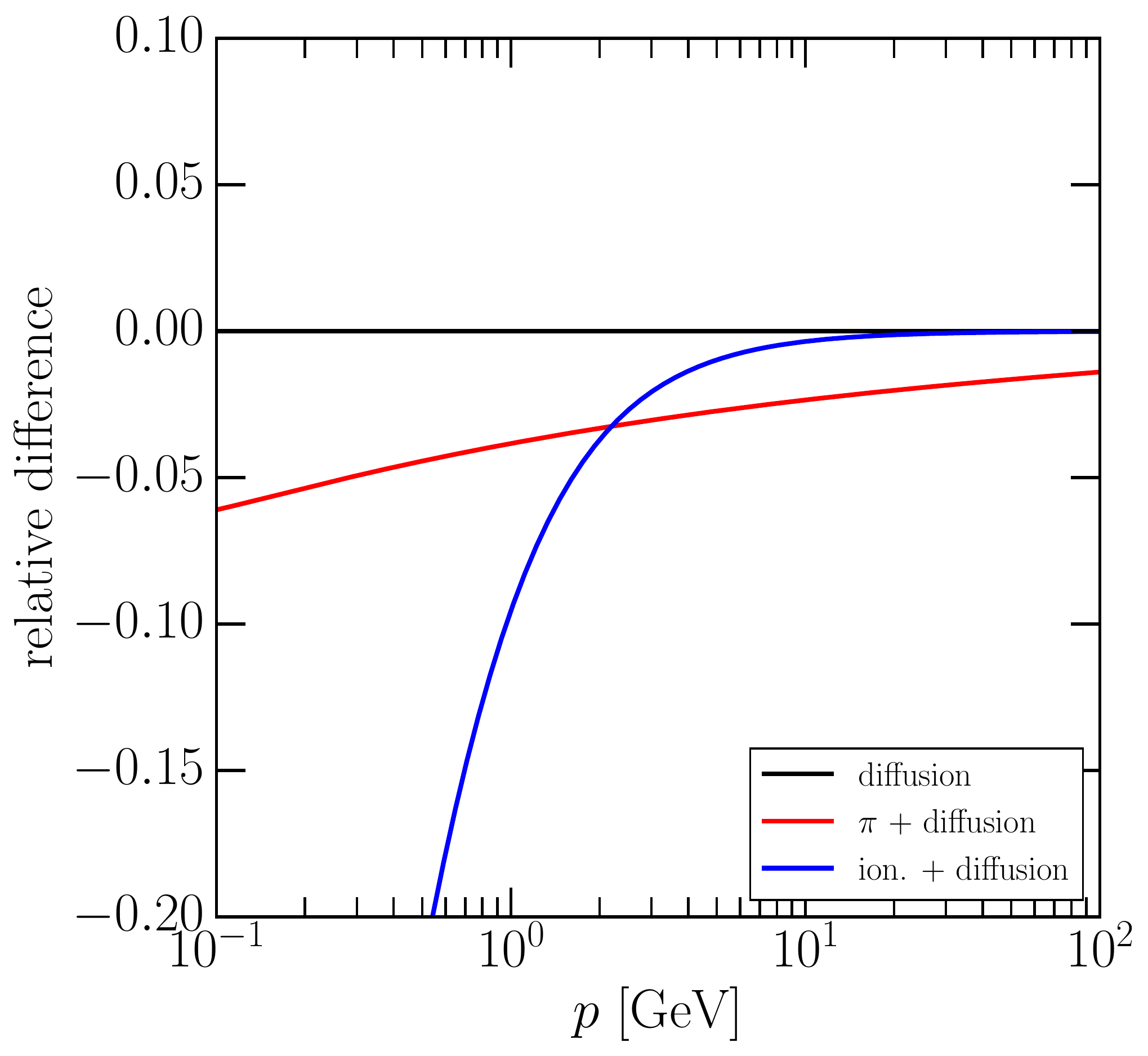}
\caption{Relative difference on the proton spectrum with and without the pion-production and ionization energy loss terms and assuming $D_{0} = 1.8 \cdot 10^{28}$ cm$^{2}$/s at $1$ GV.} 
\label{fig:hadronic_loss_test}
\end{center}
\end{figure}

\subsection{Anisotropic diffusion from a transient source}
\label{sec:transient}

The study of a transient source is relevant in many different context (e.g., to describe the Galactic centre activity). For this reason, we show here how \dragon~is able to follow the evolution of CRs emitted by an energetic source in a short event. Since the source is point-like and, in general, far from the centre of the coordinate system, we exploit the 3D mode; moreover, we consider anisotropic propagation, with $D_r > D_z$, inspired by a quasi-linear theory scenario dominated by parallel diffusion along the regular magnetic field line on the plane.

Our choice of the parameters is comparable with an average SNR event, as summarised here: 
\begin{itemize}
\item kinetic energy released: $10^{51}$~erg
\item efficiency of CR injection: $10\%$
\item CR injection spectrum: $\Phi = \Phi_0 (p/p_0)^{-2.3}$ in the range $0.1 \div 10^5$~GeV
\item diffusion coefficient slope: $\delta = 0.5$
\item diffusion coefficient normalization on the plane: $D_0^{xx} \,=\, D_0^{yy} \,=\, 10^{28}$ cm$^2$/s at 1 GeV
\item perpendicular diffusion coefficient normalization: $D_0^{zz} \,=\, 10^{27}$ cm$^2$/s at 1 GeV
\end{itemize}

The source is active from $t = 0$ to $t = 5 \cdot 10^4$ years. The evolution of the CR density in the x-z plane can be seen in Fig.~\ref{fig:map}: the signature of anisotropic diffusion is well clear in the central and right panel, corresponding to $t = 0.5$ and $t = 1$ Myr respectively.

In Fig.~\ref{fig:spectrum} we show the evolution of the spectrum on the $x$ axis, at a distance of $1$ kpc from the source. 
For each rigidity the flux peaks after a time $t$ compatible with the order-of-magnitude estimate given by $D(p) = L^2/(6t)$. 
The plot clearly shows the different timescales associated to CR diffusion at different rigidities, with the bulk of high-energy particles ($\simeq 100$ GeV) arriving at the observer's position after $\sim 0.5$ Myr, and low-energy ones ($\simeq 100$ MeV) after $\sim 10$ Myr.

\begin{figure}[!t]
\begin{center}
\includegraphics[width=0.5\textwidth]{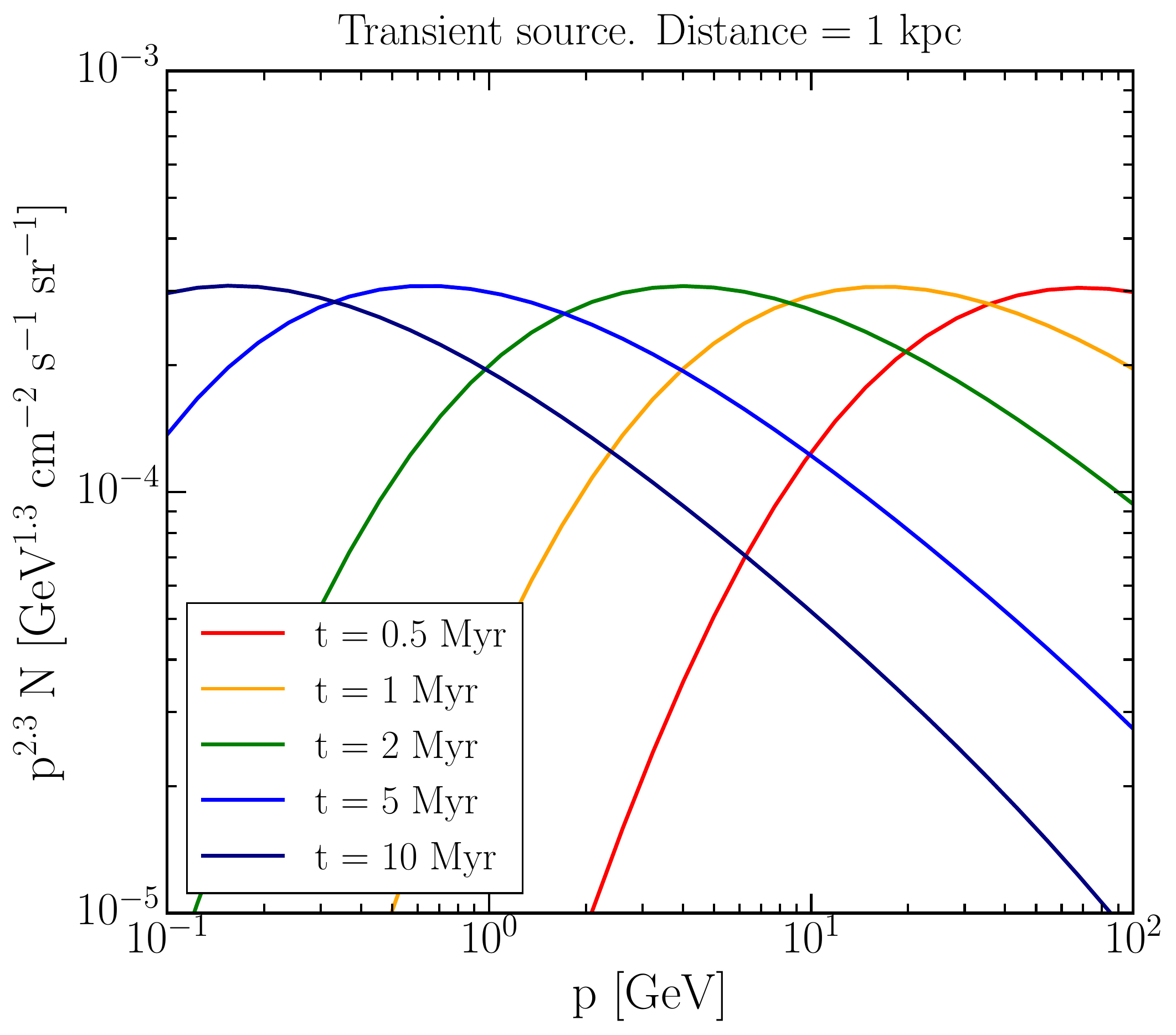}
\caption{CR spectra at fixed distance and different times. The source is transient and is located at $1$ kpc from the observer, on the Galactic plane. CRs diffuse anisotropically, with a larger coefficient along the plane and a lower one in the perpendicular direction.} \label{fig:spectrum}
\end{center}
\end{figure}

\subsection{Non-equidistant binning}
\label{sec:NEB}

\paragraph*{Sources at the GC (3D mode)} \hspace{0pt} \\

We consider a three-dimensional source term given by: 

\begin{equation}
Q(x,y,z) = \frac{1}{\sqrt{2 \pi}} \frac{1}{x_s y_s z_s}{\mathrm{exp}}  \left( -\frac{x^2}{2x_s^2} -\frac{y^2}{2y_s^2}-\frac{z^2}{2z_s^2}\right) 
\end{equation}   

where $x_s = y_s = 250$ pc and $z_s = 100 $ pc represent the size of the source along the $x$, $y$ and $z$ axes, respectively. We consider a purely diffusive and isotropic scenario characterized by a diffusion coefficient defined as: 

\begin{equation}
D_{xx,yy,zz}(p) = D_0 
\end{equation}  

with $D_0 = 10^{28}$ cm$^2$ s$^{-1}$.  

We study the diffusion of CRs in this setup by performing a run characterised by  a time-step $\Delta t $ = 0.01 Myr within a cube of edge 10 kpc. We find that, independently on the spatial grid used in the run, the numerical solution becomes constant after 2$\times 10^4$ iterations. We set the spatial resolution along the $x$ and $y$ axes to 156 pc (corresponding to 65 bins in each directions) and we test three different setups for the grid along the $z$ axis: 

\begin{itemize}
\item{An Equidistant Binning (EB) with $n_z =201$, corresponding to a constant resolution of 50 pc from $z=-5$ kpc to $z=5$ kpc.}
\item{A Not Equidistant Binning (NEB) with $n_z=27$, where the bins width grows from 50 pc for $|z| \le 100$ pc to 100 pc in the interval $100 \,\mathrm{pc} \, \le |z| \le \,500\, \mathrm{pc}$ and then up to 1.36 kpc in the range $500 \,\mathrm{pc} \, \le |z| \le \,5\, \mathrm{kpc}$ (going to larger values of $z$, the width of each bin is larger than the previous one by a 50\%).}
\item{An EB with the same number of bins as the NEB setup that we have just described (27), corresponding to a spatial resolution of 385 pc.}
\end{itemize}   

The three different grids in the $ |z| \le \,2\, \mathrm{kpc}$ region are depicted in left panel of Fig. \ref{fig:NEB_3D}. 

The results of the different runs are shown in the right panel of Fig. \ref{fig:NEB_3D}. In particular, the plot shows the CR density profile along the $z$ axis, computed at $x\,= \,y \,= \,0\,$pc. As it can be seen, the numerical solution found with  NEB is indistinguishable from the one given by  EB with $n_z=201$. A crucial factor that has to be taken into account, however, is that in the former case the solution is found in 9222 seconds, while in the latter the run lasts for 1196 seconds before a stable solution is found. As for  EB setup with $n_z = 27$, the runtime is approximately the same as in the NEB case, but the solution found with such a binning can be slightly wrong, in particular in the peak region, where the error can reach 5\%.
  
From these results, one could infer that in a scenario like the one that we considered in our test, which is characterised by a relatively spread CR distribution, using NEB  determines a small increase in the accuracy of the numerical solution. Still this improvement comes at no computational cost, making the feature serviceable also in this case.

\begin{figure}[t]
\begin{center}
 \includegraphics[width = 0.42 \textwidth]{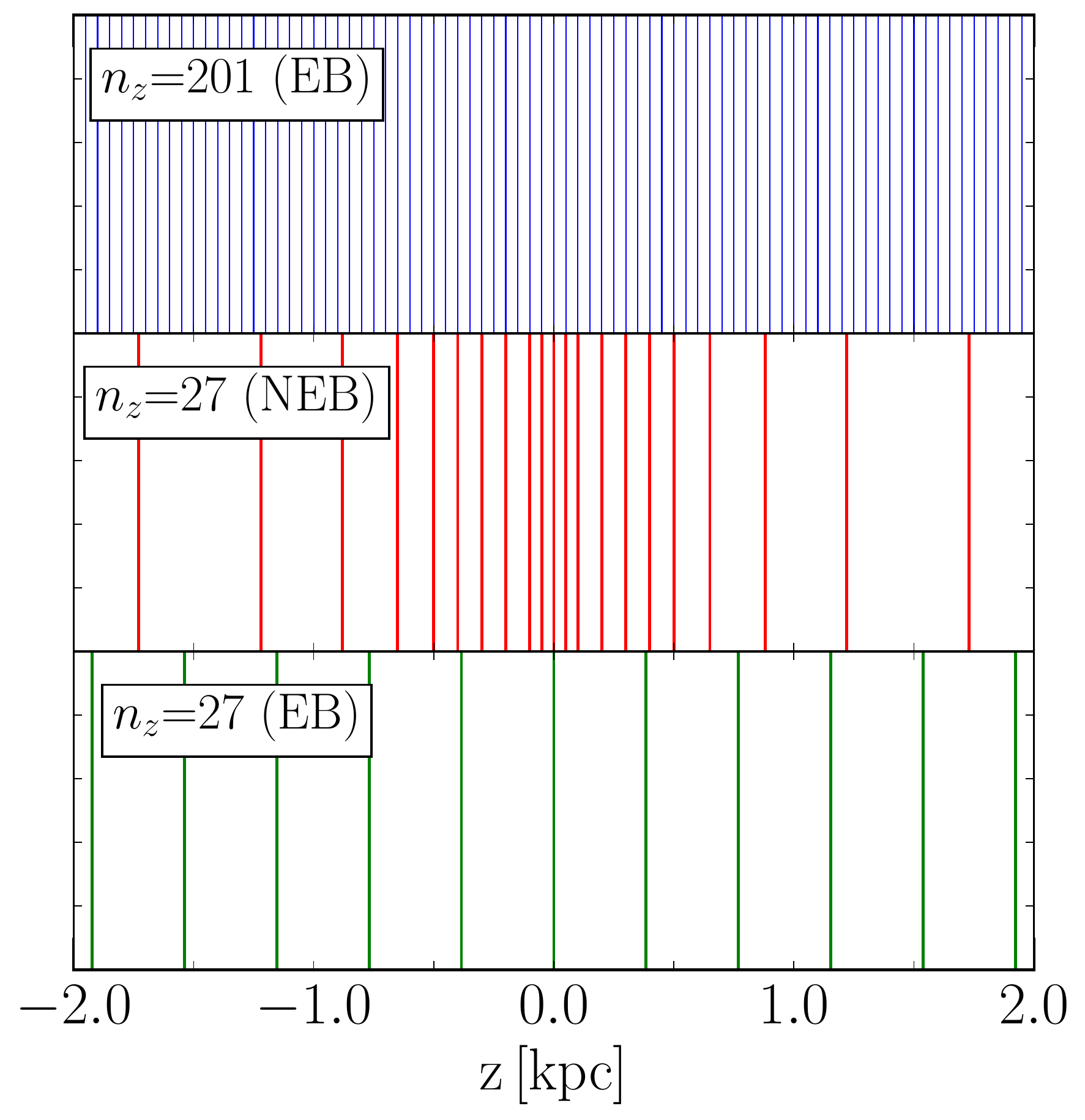}
 \includegraphics[width = 0.45 \textwidth]{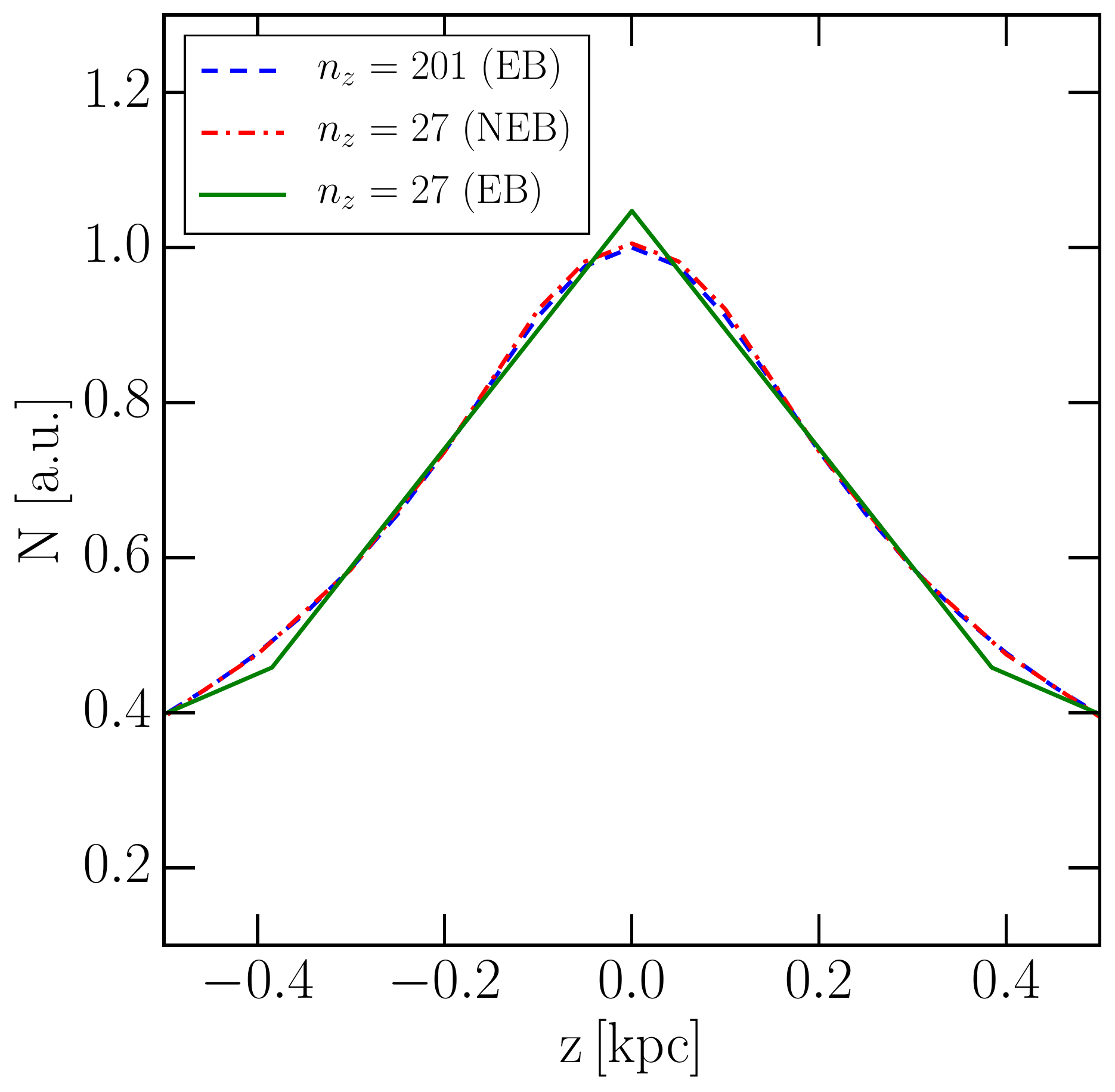}
\caption{ In {\it left panel} the different setups for the binning of the $z$ axis in  $ |z| \le \,2\, \mathrm{kpc}$ region are shown. {\it Right panel} shows the CR density profiles along the $z$ axis that are obtained for such setups. Both panels refer to the uniform and isotropic diffusion coefficient example.} \label{fig:NEB_3D}
\end{center}
\end{figure}

\paragraph*{Sources on a disk (2D mode)} \hspace{0pt} \\

\begin{figure}[t]
\begin{center}
 \includegraphics[width = 0.42 \textwidth]{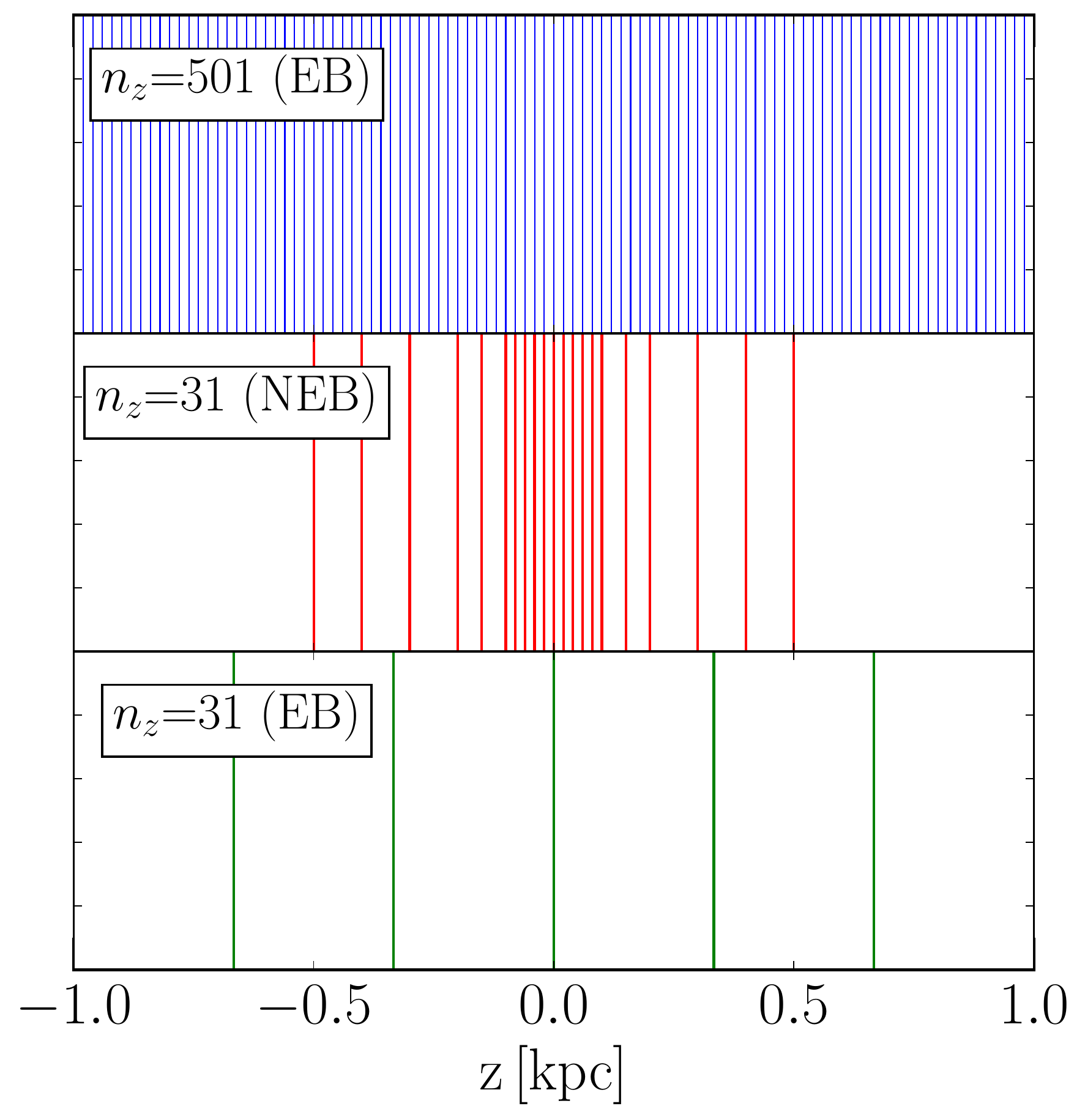}
 \includegraphics[width = 0.45 \textwidth]{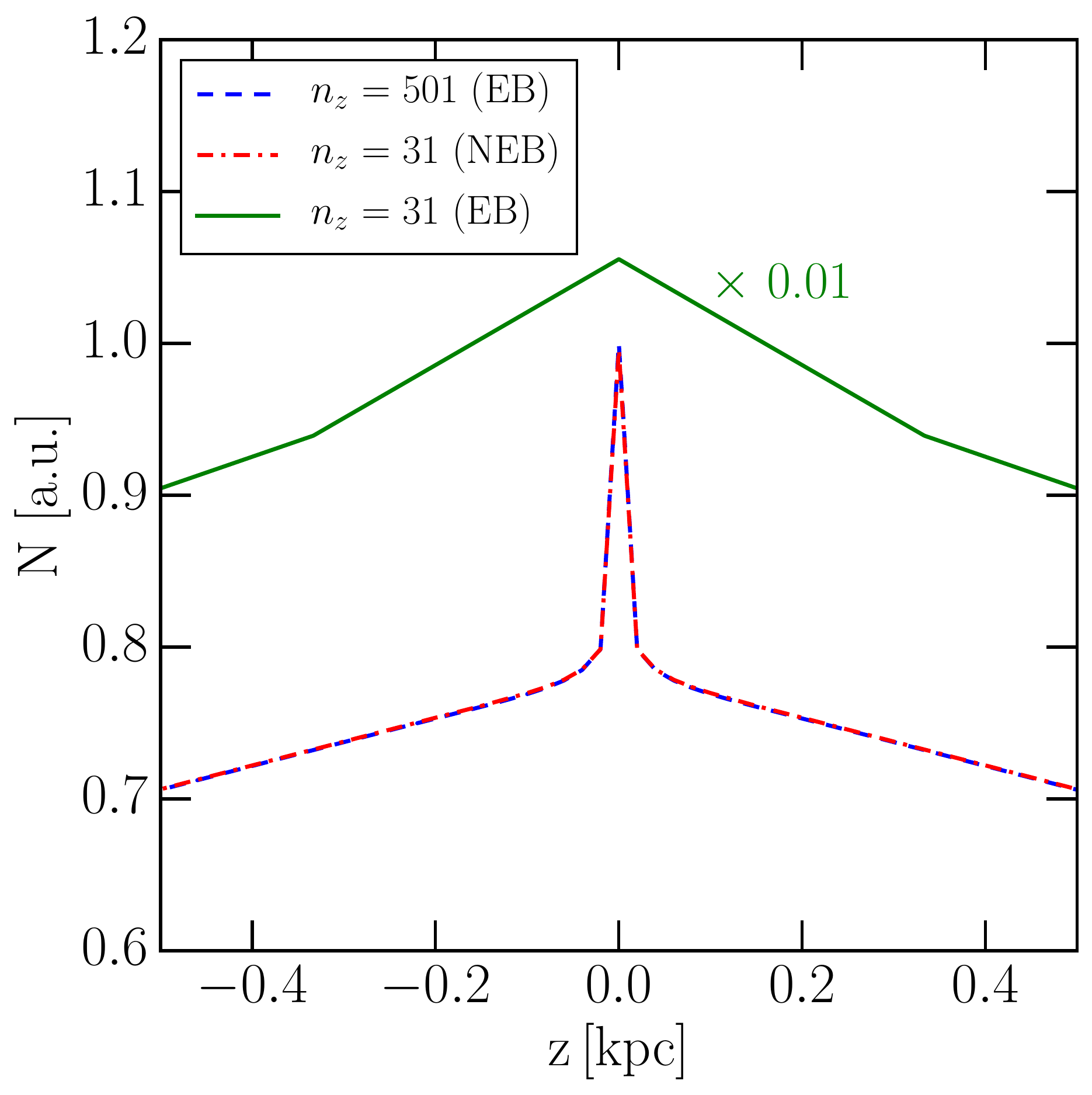}
\caption{ In {\it left panel} the different setups for the binning of the $z$ axis in the  $ |z| \le \,2\, \mathrm{kpc}$ region are shown. {\it Right panel} shows the CR density profiles along the $z$ axis that are obtained for such setups. Both panels refer to the non-uniform diffusion coefficient example.} \label{fig:NEB_1D}
\end{center}
\end{figure}

When particles are confined in a region that is much smaller than the size of the Galaxy,  NEB can represent a very useful and necessary instrument to obtain a precise solution in a relatively short time. To provide an illustrative example of such a case, we consider a one-dimensional scenario in which we have a Gaussian source term defined as: 

\begin{equation}
Q(z) = \frac{1}{\sqrt{2 \pi}} \frac{1}{z_s} {\mathrm{exp}} \left( -\frac{z^2}{z_s^2} \right)
\end{equation}  

where $z_s$ = 100 pc. As before, we consider a purely diffusive case, but this time we assume that the diffusion coefficient drops by up to three orders of magnitude in the source region: 

\begin{equation}
D(z) = D_0 \left[ 1 - 0.99 \left( \frac{z^2}{z_s^2}\right) \right]
\end{equation}  

where $D_0 = 10^{29}$ cm$^2$ s$^{-1}$. Under a physical point of view, one can motivate this decrease in the diffusion coefficient as a consequence of the stronger turbulence that characterises the region of the source. 

We study the propagation of CRs in the -5 kpc $\le $ z $\le $ 5 kpc region for three different setups of the binning along the z-axis:  

\begin{itemize}
\item{An EB with $n_z =501$, corresponding to a constant resolution of 20 pc.}
\item{A NEB with $n_z=31$, where the bins width is 20 pc for $|z| \le 100$ pc and then grows to 50 pc, 100 pc, 500 pc and 1 kpc as larger values of z are considered.}
\item{An EB with $n_z=31$, {\it i.e.} the same number of bins of the NEB setup described above. This number of bins corresponds to a spatial resolution of 333 pc.}
\end{itemize}

Left panel of Fig. \ref{fig:NEB_1D} illustrates the binnings corresponding to the three setups described above in the region $|z| \le 1$ kpc. 

The profile along the z-axis of the numerical solutions obtained for the three cases are shown in right panel of Fig.\ref{fig:NEB_1D}. As it can be clearly seen, the solution that can be obtained with NEB overlaps perfectly with the one that is found in the EB case with $n_z =501$, while the solution that characterises the EB with $n_z =501$ appears to be wrong by more than two orders of magnitude. The advantage of using NEB is here evident, since by going from $n_z =501$ to $n_z =31$ the runtime decreases considerably (it goes from 46 seconds to 4 seconds), without any loss in the accuracy of the solution.     

\subsection{Interplay of diffusion, reacceleration, and leptonic energy losses: the role of the boundary condition in momentum.}
\label{sec:hump}

The lepton spectrum is significantly shaped by the energy losses ($\propto E^2$) due to synchrotron emission and inverse Compton scattering (see~\ref{Sec:elosses} for the relevant formulas). 
We consider here a realistic setup in which the  energy loss term is coupled with other operators: reacceleration and diffusion. 

A significant reacceleration, in combination with ICS and synchrotron losses, produces a hump-like feature in the spectrum: In this section we aim at characterising this feature in a realistic setup.

As for the hadronic tests, we consider a homogeneous source term confined in a disk (with scale height  $\simeq 200$ pc); the energy-loss term is taken as follows:

\begin{equation}
\dot{p} \,=\, \ \exp{\left(-\frac{z^2}{2 z_{\rm l}}\right)} \,\cdot\, \left[b_0 + b_2 (p/p_0)^2\right]
\end{equation}

with $b_0 = 10^{-17}$ GeV/s, $b_2 = 10^{-15}$ GeV/s, $p_0 = 1$ GeV; the losses scale height is $z_{\rm l} = 1$ kpc.
The diffusion coefficient is taken as in Eq.~\ref{Eq:dcost}, with $D_0 = 10^{28}$ cm$^2$ /s and $\delta = 0.5$; the Alfv\'en velocity is $50$ km/s. 
The diffusive halo height is $H = 4$ kpc.

We are interested in the [$10$ MeV -- $100$ GeV] momentum range; within this interval the energy loss timescale is [$\simeq 30$ -- $\simeq 0.3$ Myr], the reacceleration timescale is [$\simeq 6$ -- $\simeq 600$ Myr], the diffusion timescale is  [$\simeq 6000$ -- $\simeq 50$ Myr]. The solution is obtained with the variable $\Delta t$ described in~\ref{convergence}.

\begin{figure}[!t]
\begin{center}
\includegraphics[width=0.49\textwidth]{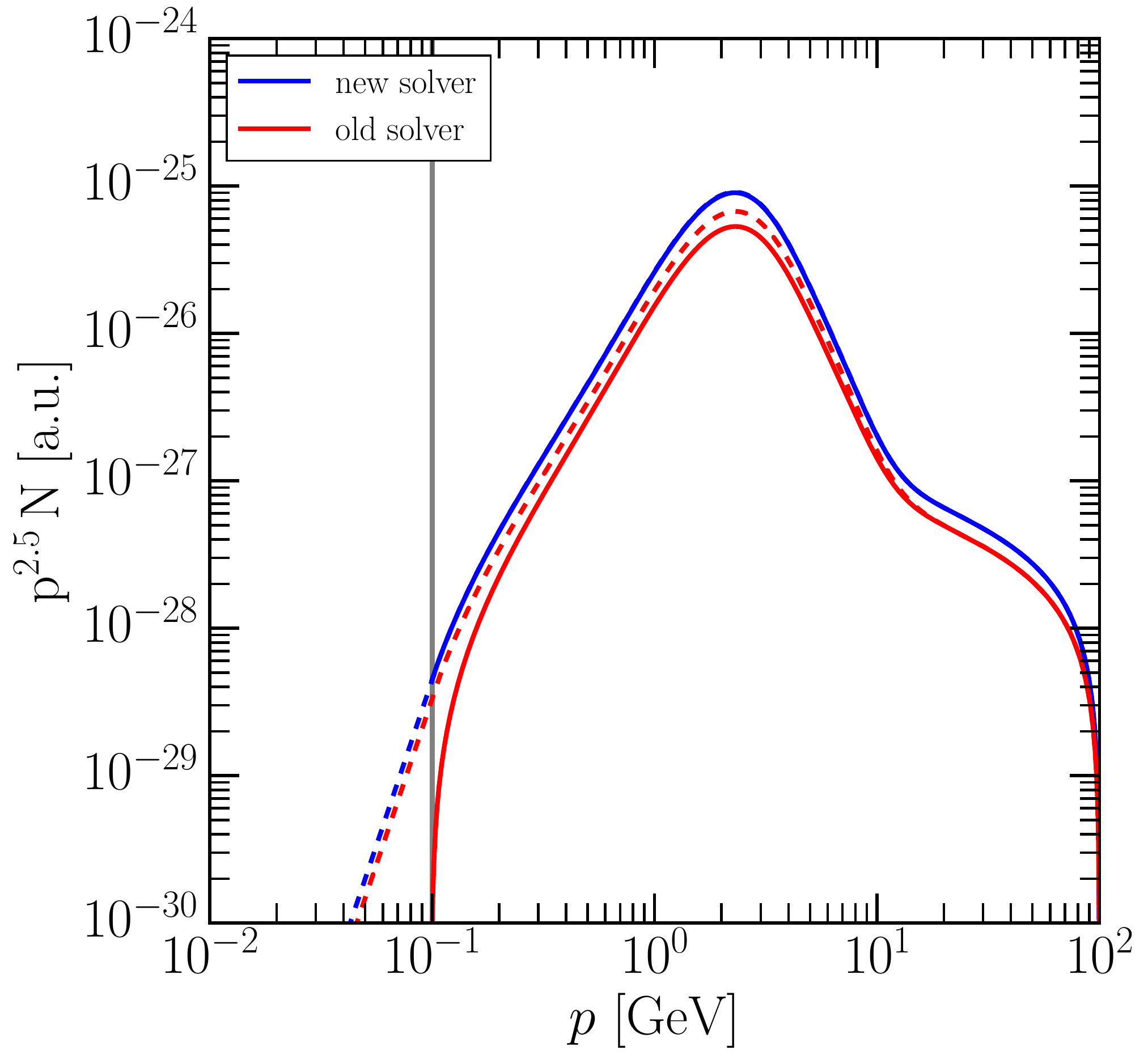}
\includegraphics[width=0.49\textwidth]{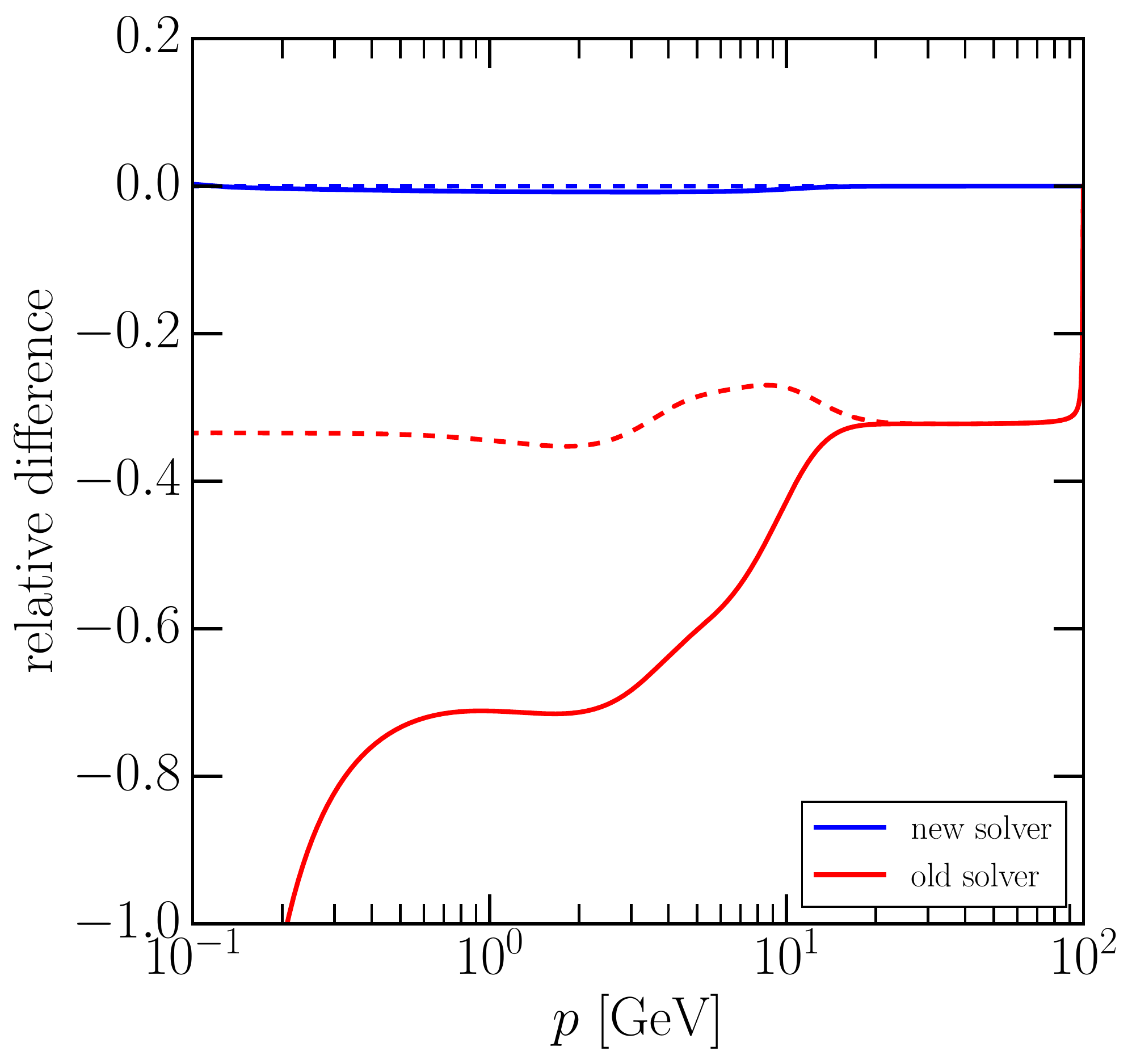}
\caption{{\it Left panel:} CR spectrum at the Sun position in arbitrary units. Red solid line: old solver (first-order energy loss, old prescription for the boundary condition); blue solid line: new solver (second-order energy loss, new prescription for the boundary condition). In both cases the CR injection is modelled as a power-law starting from $p_{\rm min} = 100$ MeV (grey vertical line). Dashed lines refer to the momentum boundary {\it of the simulation} set at $10$ MeV. {\it Right panel:} relative difference with respect to the new solver, line legend as in the left panel.} \label{fig:dragon_galprop}
\end{center}
\end{figure}

In Fig.~\ref{fig:dragon_galprop} we show the CR spectrum at the Sun position, and we compare the new prescriptions for the discretisation of reacceleration and energy loss operators (described in detail in section~\ref{sec:discretization}) with the ones implemented in the first version of {\tt DRAGON}. 

The hump is clearly visible in the spectrum in both cases, and give rise to a peak at $\simeq 2.5$ GeV: slightly higher than the energy where the energy-loss and reacceleration timescales are the same ($\sim 1$ GeV). This feature is caused by the interplay between the two competing effects of energy loss (which is responsible for a downwards flux in momentum space) and reacceleration (which is a diffusive term in momentum space, and is responsible for an upwards flux due to the monotonically decreasing injection spectrum).

The {\tt DRAGON} solver is characterised by a first-order discretisation for the energy losses, and the boundary conditions in momentum are obtained by imposing null derivatives for the Crank-Nicolson coefficients.
The \dragon~solver features a second-order discretisation for the energy losses, and as a boundary condition at the lowest momentum (see also section~\ref{App:analyticalsol_reacc}):
\begin{equation}
\frac{\partial}{\partial p} \left( \frac{N}{p^2} \right)_{p_{\rm min}} = 0
\end{equation}

The reference runs are shown in Fig.~\ref{fig:dragon_galprop}: We set the minimum momentum of CR injection at $p_{\rm min}^{\rm inj} = 0.1$ GeV and the boundary of the simulation at $p_{\rm min} = 0.1 * p_{\rm min}^{\rm inj}$.
In this setup, the two solvers exhibit a $30$\% difference at all energies due to the better accuracy of the second-order scheme.

We then compare each solver with the case in which $p_{\rm min} = p_{\rm min}^{\rm inj}$. We notice that the solution obtained with the new solver is practically unchanged by this prescription, while the solution of the {\tt DRAGON}~solver, with the old boundary condition, is significantly different at lower momenta. 
We conclude that the momentum boundary condition of \dragon~enables us to impose~$p_{\rm min} = p_{\rm min}^{\rm inj}$ with negligible effect on the solution accuracy.

In Fig.~\ref{fig:resolution}, we consider a reference run with $n_p$ = $512$ points/dex and consider the impact of lowering the resolution. From this plot, we came to the conclusion that a resolution of $n_p$ = $16$ points/dex (i.e.,~$64$ points in total in the momentum range where the source is injecting particles) yields an error as large as $15\%$ in some portions of the spectrum. Even with a second order discretisation method, a $\sim 1$\% accuracy in the whole momentum range requires a much larger resolution ($n_p \sim 128$ points/dex).

\begin{figure}[!t]
\begin{center}
\includegraphics[width=0.5\textwidth]{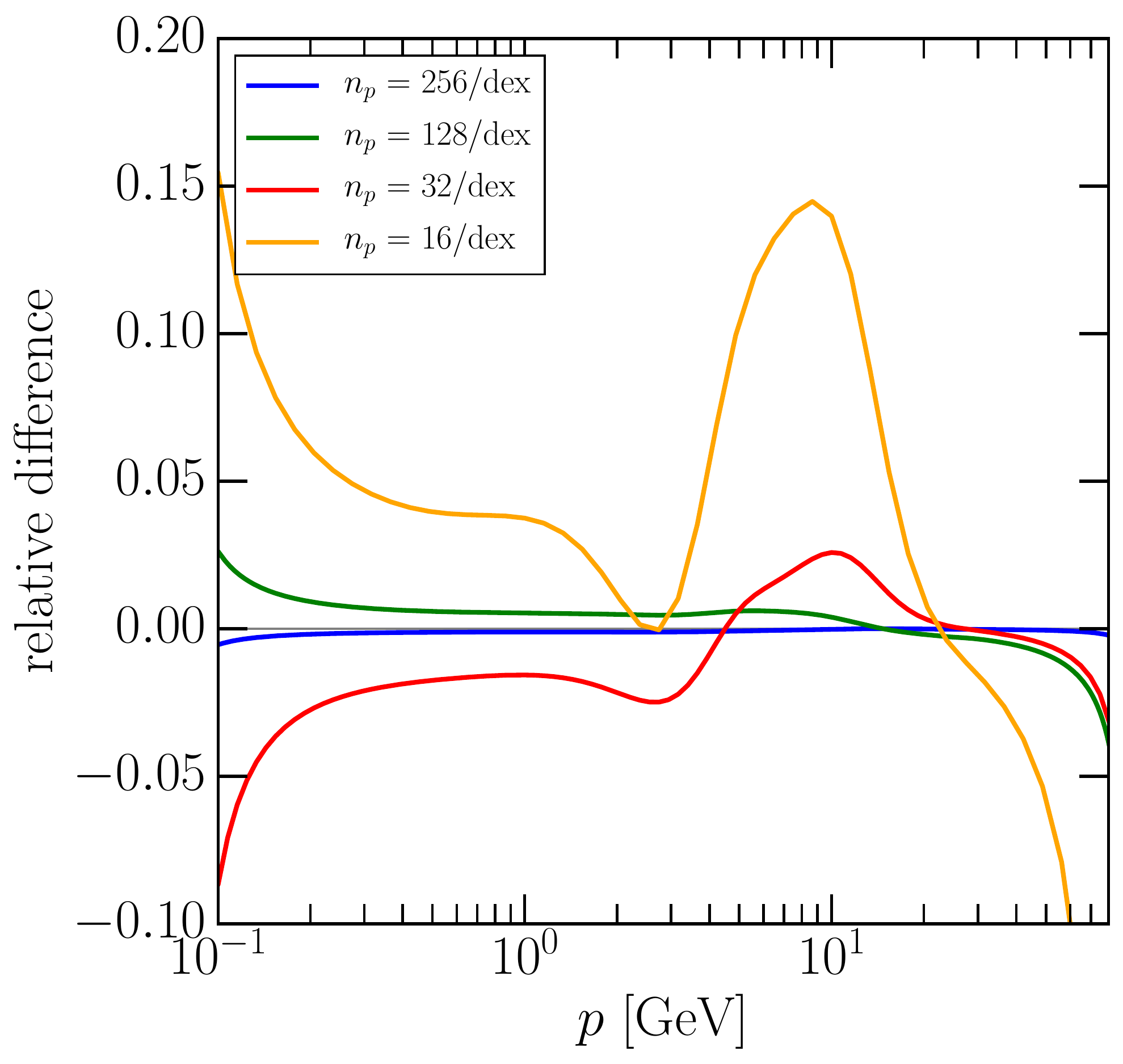}
\caption{The relative difference of the lepton spectrum for different energy grid sizes and the case $n_p = 512$/dex.} 
\label{fig:resolution}
\end{center}
\end{figure}

\section{Conclusions}

The simulation of galactic CR propagation plays an essential role in understanding the properties of the galactic sources and of the interstellar medium.
In this paper we introduced the new version of the publicly available galactic CR propagation code~\dragon. 

To interpret the recent measurements of primary CR fluxes and their secondary products, the 
code was rewritten to solve the complete CR transport equation in 2D and 3D, for both leptonic and hadronic species, under very general assumptions. 

The equation solver has been updated for all the operators: anisotropic and inhomogeneous diffusion, advection, reacceleration, and energy losses.
As discussed in the text, some of these features have been already exploited with modified version of {\tt DRAGON} to study diffusion emissions and local spectra (see, e.g.,~\cite{2015PhRvD..91h3012G,2016arXiv161006182F}), that are now fully tested and included in the public code release.
For each operator, we discussed the numerical scheme adopted in \dragon~and we tested against the corresponding analytical solution. 

In doing so, we provide a complete set of analytical solutions that can be used as a benchmark test for any CR transport framework.

\dragon~features a wide collection of up-to-date models for all the relevant astrophysical ingredients involved in the computation: source term, magnetic field, interstellar radiation field, gas; all the  distributions implemented in the code are presented in the paper. 

We emphasised the novel features of \dragon~with respect to its previous version by simulating relevant physical cases where these features play a crucial role.
In particular, we showed the implications of a different normalisation and rigidity scaling of the diffusion coefficient in different regions of the Galaxy, we described the case of CR anisotropic propagation from a transient source, we considered the possibility of a refinement of the grid in a peculiar interesting region, we discussed the interplay of diffusion, reacceleration and energy losses and the impact of these processes on the propagated leptonic spectrum.

\dragon~is an open source code under GNU general public license~\footnote{\url{https://www.gnu.org/licenses/gpl.html}} distributed as a git repository (see \url{www.dragonproject.org}), and it can be easily upgraded or re-used to describe a wider range of physical conditions with respect to those treated in this paper.

Forthcoming publications will focus on additional aspects not covered in this paper, e.g., the general (non-diagonal) treatment of anisotropic diffusion, and a comprehensive description of the cross-section network.

\section*{Acknowledgments}

We are grateful to Luca Maccione who was one of the main authors and driving force of the {\tt DRAGON} code and keeps providing us valuable advice and encouragement.
We are also indebted to Andy Strong for his support and many helpful discussions. 
We thank Iris Gebauer and her team (Matthias Weinreuter, Simon Kunz, Florian Keller) for their help with the debugging and testing of the solver, and for the implementation of the non-equidistant binning. 
We warmly thank 
Richard Bartels, 
Francesca Calore,
Elena Canovi, 
Silvio Sergio Cerri, 
Massimo Gaspari, 
Philipp Girichidis,
Ralf Kissmann, 
Paolo Lipari,
Giovanni Morlino,
Martin Pohl,
Marco Taoso, 
Alfredo Urbano,
Mauro Valli,
Silvia Vernetto, 
for useful discussions, bug reporting, and feedback on the code. 
A.V. acknowledges the Gottfried Wilhelm Leibniz programme of the Deutsche Forschungsgemeinschaft (DFG) and a Grant from the GIF, the German-Israeli Foundation for Scientific Research. 

We refer to our website {\tt www.dragonproject.org}, where a short history of the {\tt DRAGON} project is also reported, for a complete list of acknowledgements regarding routines and data extracted from external public codes.

\appendix

\newlength{\wcoll}
\newlength{\wcolc}
\newlength{\wcolu}
\newlength{\wcolbc}

\setlength{\wcoll}{0.4\textwidth}
\setlength{\wcolc}{0.22\textwidth}
\setlength{\wcolu}{0.4\textwidth}
\setlength{\wcolbc}{0.12\textwidth}

\renewcommand{\arraystretch}{4}

\vspace{-5cm}

\begin{landscape}

\section{CN Tables}
\label{sec:cntables}

\begin{table}[h]
\small 
\centering
\begin{tabular}{| c | c | c | c | c |}
\hline
\hline
Operator & $L_i$ & $C_i$ & $U_i$ & b.c. \\
\hline
\hline
$\mathcal{L}_r$ &
\begin{minipage}{\wcoll}
\begin{equation*}
\frac{D_{rr, i}}{\Delta r_c \Delta r_d} - \frac{D_{rr, i}}{2 r_i \Delta r_c} - \frac{D_{rr, i+1} - D_{rr, i-1}}{4 \Delta r_c^2}
\end{equation*}
\end{minipage} 
&
\begin{minipage}{\wcolc}
\begin{equation*}
\frac{D_{rr, i}}{\Delta r_c} \left[ \frac{1}{\Delta r_u} + \frac{1}{\Delta r_d} \right]
\end{equation*}
\end{minipage} 
&
\begin{minipage}{\wcolu}
\begin{equation*}
\frac{D_{rr, i}}{\Delta r_c \Delta r_d} - \frac{D_{rr, i}}{2 r_i \Delta r_c} - \frac{D_{rr, i+1} - D_{rr, i-1}}{4 \Delta r_c^2}
\end{equation*}
\end{minipage} 
&
\begin{minipage}{\wcolbc}
$N_{-1} = N_{1}$ \\ $N_{n - 1} = 0$
\end{minipage} 
\\

$\mathcal{L}_z$ & 
\begin{minipage}{\wcoll}
\begin{equation*}
\frac{D_{zz, i}}{\Delta z_c \Delta z_d} - \frac{D_{zz, i+1} - D_{zz, i-1}}{4 \Delta z_c^2}
\end{equation*}
\end{minipage} 
&
\begin{minipage}{\wcolc}
\begin{equation*}
\frac{D_{zz, i}}{\Delta z_c} \left[ \frac{1}{\Delta z_u} + \frac{1}{\Delta z_d} \right]
\end{equation*}
\end{minipage} 
&
\begin{minipage}{\wcolu}
\begin{equation*}
\frac{D_{zz, i}}{\Delta z_c \Delta z_d} - \frac{D_{zz, i+1} - D_{zz, i-1}}{4 \Delta z_c^2}
\end{equation*}
\end{minipage} 
&
\begin{minipage}{\wcolbc}
$N_0 = 0$ \\ $N_{n-1} = 0$
\end{minipage} 
\\

$\mathcal{L}_a$ &
\begin{minipage}{\wcoll}
\begin{equation*}
\begin{dcases}
\frac{v_{i-1}}{\Delta z_d}\,\, (z > 0) \\
\frac{-v_{i-1}}{\Delta z_c}\,\, (z = 0) \\
0\,\, (z < 0)
\end{dcases}
\end{equation*}
\end{minipage} 
& 
\begin{minipage}{\wcolc}
\vspace{0.6cm}
\begin{equation*}
\begin{dcases}
\frac{v_{i}}{\Delta z_d}\,\, (z > 0) \\
0\,\, (z = 0) \\
\frac{v_{i}}{\Delta z_u}\,\, (z < 0)
\end{dcases}
\end{equation*}
\end{minipage} 
& 
\begin{minipage}{\wcolu}
\begin{equation*}
\begin{dcases}   
0\,\, (z > 0) \\
\frac{-v_{i+1}}{\Delta z_c} \,\, (z < 0) \\
\frac{v_{i+1}}{\Delta z_u} \,\, (z < 0)
\end{dcases}
\end{equation*}
\end{minipage} 
& 
\begin{minipage}{\wcolbc}
$N_0 = 0$ \\ $N_{n-1} = 0$
\end{minipage} 
\\
$\mathcal{L}_p$ &
\begin{minipage}{\wcoll}
\begin{equation*}
- \frac{D_{pp,i+1}-D_{pp,i-1}}{4 \Delta p_c^2} + \frac{D_{pp,i}}{\Delta p_c \Delta p_d} + \frac{D_{pp,i-1}}{\Delta p_c p_{i-1}}
\end{equation*}
\end{minipage} 
&
\begin{minipage}{\wcolc}
\begin{equation*}
-\frac{D_{pp,i}}{\Delta p_c} \left[ \frac{1}{\Delta p_u} + \frac{1}{\Delta p_d} \right]
\end{equation*}
\end{minipage}  
& 
\begin{minipage}{\wcolu}
\begin{equation*}
\frac{D_{pp,i+1}-D_{pp,i-1}}{4 \Delta p_c^2} + \frac{D_{pp,i}}{\Delta p_c \Delta p_u} - \frac{D_{pp,i+1}}{\Delta p_c p_{i+1}}
\end{equation*}
\end{minipage}  
& 
\begin{minipage}{\wcolbc}
$N_0 = \frac{p_0^2}{p_1^2} N_1$ \\ $N_{n-1} = 0$
\end{minipage} 
\\
$\mathcal{L}_l$ & 
\begin{minipage}{\wcoll}
\begin{equation*}
0
\end{equation*}
\end{minipage} 
& 
\begin{minipage}{\wcolc}
\begin{equation*}
-\frac{\dot{p}_i}{p_{i+1} - p_i}
\end{equation*}
\end{minipage}  
& 
\begin{minipage}{\wcolu}
\begin{equation*}
-\frac{\dot{p}_{i+1}}{p_{i+1} - p_i} 
\end{equation*}
\end{minipage}  
& 
\begin{minipage}{\wcolbc}
$N_{n-1} = 0$
\end{minipage} 
\\
\hline
\hline
\end{tabular}
\caption{Crank-Nicolson coefficients and boundary conditions for the 2D case (
$\Delta x_c \equiv \frac{x_{i+1} - x_{i-1}}{2}$, 
$\Delta x_{\rm u} \equiv sx_{i+1} - x_{i}$,
$\Delta x_{\rm d} \equiv x_{i} - x_{i-1}$).}\label{Tab:CN2D}
\end{table}
\end{landscape}

\begin{landscape}
\begin{table}[!t]
\small 
\centering
\begin{tabular}{| c | c | c | c | c |}
\hline
\hline
Operator & $L_i$ & $C_i$ & $U_i$ & b.c. \\
\hline
\hline
$\mathcal{L}_x$ & 
\begin{minipage}{\wcoll}
\begin{equation*}
\frac{D_{xx, i}}{\Delta z_c \Delta x_d} - \frac{D_{xx, i+1} - D_{xx, i-1}}{4 \Delta x_c^2}
\end{equation*}
\end{minipage} 
&
\begin{minipage}{\wcolc}
\begin{equation*}
\frac{D_{xx, i}}{\Delta x_c} \left[ \frac{1}{\Delta x_u} + \frac{1}{\Delta x_d} \right]
\end{equation*}
\end{minipage} 
&
\begin{minipage}{\wcolu}
\begin{equation*}
\frac{D_{xx, i}}{\Delta x_c \Delta x_d} - \frac{D_{xx, i+1} - D_{xx, i-1}}{4 \Delta x_c^2}
\end{equation*}
\end{minipage} 
&
\begin{minipage}{\wcolbc}
$N_0 = 0$ \\ $N_{n-1} = 0$
\end{minipage} 
\\
$\mathcal{L}_y$ & 
\begin{minipage}{\wcoll}
\begin{equation*}
\frac{D_{yy, i}}{\Delta y_c \Delta y_d} - \frac{D_{yy, i+1} - D_{yy, i-1}}{4 \Delta y_c^2}
\end{equation*}
\end{minipage} 
&
\begin{minipage}{\wcolc}
\begin{equation*}
\frac{D_{yy, i}}{\Delta y_c} \left[ \frac{1}{\Delta y_u} + \frac{1}{\Delta y_d} \right]
\end{equation*}
\end{minipage} 
&
\begin{minipage}{\wcolu}
\begin{equation*}
\frac{D_{yy, i}}{\Delta y_c \Delta y_d} - \frac{D_{yy, i+1} - D_{yy, i-1}}{4 \Delta y_c^2}
\end{equation*}
\end{minipage} 
&
\begin{minipage}{\wcolbc}
$N_0 = 0$ \\ $N_{n-1} = 0$
\end{minipage} 
\\
$\mathcal{L}_z$ & 
\begin{minipage}{\wcoll}
\begin{equation*}
\frac{D_{zz, i}}{\Delta z_c \Delta z_d} - \frac{D_{zz, i+1} - D_{zz, i-1}}{4 \Delta z_c^2}
\end{equation*}
\end{minipage} 
&
\begin{minipage}{\wcolc}
\begin{equation*}
\frac{D_{zz, i}}{\Delta z_c} \left[ \frac{1}{\Delta z_u} + \frac{1}{\Delta z_d} \right]
\end{equation*}
\end{minipage} 
&
\begin{minipage}{\wcolu}
\begin{equation*}
\frac{D_{zz, i}}{\Delta z_c \Delta z_d} - \frac{D_{zz, i+1} - D_{zz, i-1}}{4 \Delta z_c^2}
\end{equation*}
\end{minipage} 
&
\begin{minipage}{\wcolbc}
$N_0 = 0$ \\ $N_{n-1} = 0$
\end{minipage} 
\\
\hline
\hline
\end{tabular}
\caption{Crank-Nicolson coefficients and boundary conditions for the spatial coordinates in the 3D case (
$\Delta x_c \equiv \frac{x_{i+1} - x_{i-1}}{2}$, 
$\Delta x_{\rm u} \equiv sx_{i+1} - x_{i}$,
$\Delta x_{\rm d} \equiv x_{i} - x_{i-1}$).}\label{Tab:CN3D}
s\end{table}
\end{landscape}

\renewcommand{\arraystretch}{1}

\section{Analytical solutions to the transport equations}
\label{App:A}

In this section, we provide different analytical solutions to the transport equation defined in \ref{eq:prop}. In order to derive these solutions we will consider a series of simplified scenarios in which only one operator of the transport equation plays a role, the others being set to zero.  
Analytical solutions derived in this way are valuable to test the performances of the numerical solver for each transport operator. As mentioned in Section~\ref{sec:solver}, we can compare the numerical scheme of each operator against its analytical solution, allowing to evaluate the quality of the convergence reached by the corresponding discretisation scheme.

\subsection{Two-dimensional inhomogeneous and anisotropic diffusion equation}
\label{App:analyticalsol_diff}
We provide here an analytical solution for the two-dimensional diffusion equation:
\begin{equation}\label{Eq:2D_ana}
- D_{zz} \frac{\partial^2 N}{\partial z^2} - \left[ \frac{D_{rr}}{r} + \frac{\partial D_{rr}}{\partial r} \right] \frac{\partial N}{\partial r} - D_{rr} \frac{\partial^2 N}{\partial r^2} = Q(r,z)
\end{equation}
in the domain extending from $0$ to $R$ in $r$ direction and from $-L$ to $L$ in $z$ direction, and boundary condition~$N(z = \pm L, r = R) = 0$ and $\partial N/\partial r(r = 0) = 0$.

We assume inhomogeneous and anisotropic diffusion coefficients, namely:
\begin{eqnarray*}
D_{xx} (r)  \,&=&\, D_0 \left( 1 + \frac{r^2}{R^2} \right)\\
D_{zz} \,&=&\, f D_0
\end{eqnarray*}
where $f$ is the parameter defining the anisotropy between diffusion in the $z$ and in the $r$ direction.

Following~\cite{Kissmann2014}, we prescribe an analytical solution as:
\begin{equation}
N_a (r,z) = \cos \left( \frac{\pi r}{2R} \right) \cos \left( \frac{\pi z}{2L} \right)
\end{equation}
and we find the corresponding source function that fulfils Eq.~\ref{Eq:2D_ana} as:
\begin{equation}
Q(r,z) = \left[ \frac{D_0}{R^2} \left( 1 + \frac{r^2}{R^2} \right) + \frac{D_{zz}}{L^2} \right] \frac{\pi^2 N_a}{4} + \frac{\pi D_0}{2 R^2} \left( \frac{R}{r} + \frac{3r}{R} \right) \sin \left( \frac{\pi r}{2R} \right) \cos \left( \frac{\pi z}{2L} \right)
\end{equation}

\subsection{Three-dimensional anisotropic diffusion equation}

We consider the equation: 

\begin{equation}\label{Eq:3D_ana}
- D_{xx} \frac{\partial^2 N}{\partial x^2} - D_{yy} \frac{\partial^2 N}{\partial y^2} - D_{zz} \frac{\partial^2 N}{\partial z^2} = Q(x,y,z), 
\end{equation}

which describes a homogeneous and anisotropic three-dimensional diffusion. The diffusion is assumed to be confined in the domain which extends from $-L_x$ to $L_x$ in $x$ direction, from $-L_y$ to $L_y$ in $y$ direction and from $-L_z$ to $L_z$ in $z$ direction. The diffusion coefficients can be written as: 

\begin{eqnarray*}
D_{xx} \,&=&\, D_0 \\
D_{yy} \,&=&\, f_y D_0 \\
D_{zz} \,&=&\, f_z D_0 
\end{eqnarray*}

where $f_y$ and $f_z$ are parameters that quantify the anisotropy of the diffusion along the $y$ and $z$ directions. In analogy with the two-dimensional case described above, we prescribe an analytical solution given by: 

\begin{equation}
N_a(x,y,z) = \cos\left( \frac{\pi x}{2L_x} \right) \cos\left( \frac{\pi y}{2L_y} \right) \cos\left( \frac{\pi z}{2L_z} \right).
\end{equation}

In order to satisfy Eq.~\ref{Eq:3D_ana}, the source term must take the following form:

\begin{equation}
Q(x,y,z) = \frac{\pi^2}{4} \left( \frac{D_{xx}}{L_x^2} + \frac{D_{yy}}{L_y^2} + \frac{D_{zz}}{L_z^2} \right) \cos\left( \frac{\pi x}{2L_x} \right) \cos\left( \frac{\pi y}{2L_y} \right) \cos\left( \frac{\pi z}{2L_z} \right)
\end{equation}

\subsection{Energy-loss equation}
\label{App:analyticalsol_eloss}

Taking only losses and source terms into account, the steady-state equation becomes:
\begin{equation}\label{Eq:loss_ana}
\frac{\partial}{\partial p} \left( \dot{p} N \right) \, = \, Q(p) 
\end{equation}

We assume different power-laws for both the loss term rate
\begin{equation}
\dot{p} \,=\, -b_0 \left( \frac{p}{p_0} \right)^2 \qquad (\text{with } b_0 > 0)
\end{equation}
and for the source term:
\begin{equation}\label{Eq:source_ana}
Q(p) \,=\, Q_0 \, \left( \frac{p}{p_0} \right)^{-\alpha} \qquad (\text{with } \alpha > 2) 
\end{equation}
where $p_0$ is a reference momentum.

The analytic solution of Eq.~\ref{Eq:loss_ana} can be found by integrating both sides in momentum:
\begin{equation}
\int \frac{dp}{p_0} \frac{\partial}{\partial p/p_0} \left(\frac{p}{p_0}\right)^2 N(p) = -\frac{p_0 Q_0}{b_0} \int \frac{dp}{p_0} \left(\frac{p}{p_0}\right)^{-\alpha}
\end{equation}
which gives:
\begin{equation}
N(p) = \frac{p_0 Q_0}{(\alpha-1) b_0} \left(\frac{p}{p_0}\right)^{-\alpha-1} - C \left(\frac{p}{p_0}\right)^{-2}
\end{equation}

The integration constant $C$ is found by imposing the boundary condition~$N(p = p_{\rm max}) = 0$:
\begin{equation}
C = \frac{p_0 Q_0}{(\alpha-1) b_0} \left(\frac{p_{\rm max}}{p_0}\right)^{-\alpha+1}
\end{equation}

\subsection{Reacceleration equation}
\label{App:analyticalsol_reacc}

In order to find an analytical solution for the reacceleration equation in the momentum range $[p_{\rm min},p_{\rm max}]$:
\begin{equation}\label{Eq:reacc_ana}
-\frac{\partial}{\partial p} \left[ p^2 D_{\rm pp} \, \frac{\partial}{\partial p} \left( \frac{N}{p^2} \right) \right] = Q(p)
\end{equation}
we assume a simplified expression for the momentum diffusion coefficient:
\begin{equation}
D_{\rm pp} = D_0 \left( \frac{p}{p_0} \right)^2
\end{equation}
while the source term is assumed to be the same as in~\ref{Eq:source_ana}.

The transport equation~\ref{Eq:reacc_ana} becomes:
\begin{equation}
\frac{\partial}{\partial x} \left( x^4 \, \frac{\partial}{\partial x} f \right) = -\frac{Q_0}{D_0} x^{-\alpha}
\end{equation}
where $x \equiv p/p_0$ and $f \equiv N/p^2$.

A first integration of both sides is performed :
\begin{equation}\label{Eq:firstint}
\frac{\partial}{\partial x} f = -\frac{Q_0}{D_0} \frac{x^{-\alpha-3}}{1-\alpha} - C x^{-4}
\end{equation}

The integration constant $C$ is given by the boundary condition~$\frac{\partial}{\partial x} f = 0$ for $x = x_{\rm min}$:
\begin{equation}
C = -\frac{Q_0}{D_0} \frac{x_{\rm min}^{-\alpha+1}}{1-\alpha}
\end{equation}

A further integration of Eq.~\ref{Eq:firstint} gives:
\begin{equation}\label{Eq:secondint}
f = \frac{Q_0}{D_0} \frac{x^{-\alpha-2}}{(1-\alpha)(\alpha+2)} + \frac{C}{3} x^{-3} - B.
\end{equation}
The second integration constant $B$ is derived under the assumption that $f = 0$ for $x = x_{\rm max}$:
\begin{equation}
B = \frac{Q_0}{D_0} \frac{x_{\rm max}^{-\alpha-2}}{(1-\alpha)(\alpha+2)} + \frac{C}{3} x_{\rm max}^{-3}
\end{equation}

Finally, by substituting the integration constants in Eq.~\ref{Eq:secondint}, we obtain:
\begin{multline}
\frac{N(p)}{p^2} = \frac{Q_0}{D_0(1-\alpha)} \left[ \frac{1}{\alpha+2} \left( \left( \frac{p}{p_0} \right)^{-\alpha-2} - \left( \frac{p_{\rm max}}{p_0} \right)^{-\alpha-2} \right) \right. \\ 
\left. + \frac{1}{3} \left( \frac{p_{\rm min}}{p_0} \right)^{-\alpha+1} \left( \left( \frac{p_{\rm max}}{p_0} \right)^{-3} - \left( \frac{p}{p_0} \right)^{-3} \right) \right]
\end{multline}

\subsection{Advection equation}
\label{App:analyticalsol_advec}

The advective transport of $N(z,t)$ by the velocity field $v_w(z)$ is described by the equation:
\begin{equation}\label{Eq:conv_ana}
\frac{\partial N}{\partial t} = - \frac{\partial (v_w N)}{\partial z} 
\end{equation}

The solution of~\ref{Eq:conv_ana}, assuming $v_w$ constant along $z$, is determined by an initial condition $N_0(z)$:
\begin{equation}
N(z,t) = N_0 (z-v_w t)
\end{equation}
which is just the initial function $N_0$ shifted by $v_w t$ to higher $z$ (for $v_w > 0$) or to smaller $z$ (for $v_w < 0$).

In Section~\ref{sec:advection} we assume a Gaussian initial condition centred at $z=0$:
\begin{equation}
N_0 (z) = (2\pi \sigma^2_z)^{-1/2} \, \exp \left( -\frac{z^2}{2\sigma_z^2} \right)
\end{equation}
and we assume positive $v_w$.

In addition, to test the advective transport by the velocity field $v(z) = {\rm sgn}(z) v_z$, as it is the case of a Galactic wind, we assume a simplified one dimensional equation for $N(z)$:
\begin{equation}\label{Eq:simple_advective}
- D_{\rm zz} \frac{\partial^2 N}{\partial z^2} + \frac{\partial (v N)}{\partial z} = Q_0 \delta(z) 
\end{equation}
with boundary condition~$N(|z| = L) = 0$.

The solution to this equation can be written as:
\begin{equation}\label{Eq:discontinuos_advection}
N(z) = N_0 \frac{1 - {\rm e}^{w \left( 1-|z|/L \right)}}{1 - {\rm e}^w}
\end{equation}
where $w$ can be found by solving Eq.~\ref{Eq:simple_advective} for $z > 0$ and it turns out to be:
\begin{equation}
w = - \frac{v_z L}{D}
\end{equation}

By integrating the Eq.~\ref{Eq:simple_advective} in the neighbourhood of the galactic disk (i.e., by computing $\lim_{\epsilon \rightarrow 0} \int_{\epsilon^-}^\epsilon \, dz \dots$), we end up with:
\begin{equation}
N_0 = \frac{Q_0}{2 v_z}.
\end{equation}

\section{Astrophysical ingredients}
\label{sec:astroingredients}

In this section we will present a review of all those {\it astrophysical ingredients} (gas and sources distributions, galactic magnetic fields, ISRF, relevant velocities and energy losses) that play a role in our numerical simulations. 

These ingredients are not always known with enough accuracy, and different models describing them can be found in literature. 
It is important to underline that using different models for the same ingredient can lead to systematic uncertainties in the outcome of our numerical simulations and for this reason we implement in the code at least two different models for each ingredient. 
The modular structure that we introduce in \dragon~allow the users to easily extend each module with a new model.  

Please note that the names that we associate to every model are the same as those with which these models have been called in the code.

\subsection{Gas distributions}
\label{sec:gasdistro}

Interstellar gas plays an important role in the process of CR energy loss and in secondary production. The radio emission is used to trace its distribution (for a review see~\cite{Ferriere2001}).
Here we shortly report the different distributions of the interstellar Hydrogen: ionised (HII), atomic (HI), and molecular (H$_2$) phase as implemented in the code. 
Radial distributions of the two most abundant components (HI and H$_2$) for the models implemented in \dragon~are shown in Fig.~\ref{fig:gas_radial}.  
As customary, we assume that Helium distribution follows the Hydrogen one with a ratio of $0.11$ between the number density of the two species.

\subsubsection{HI gas distributions}

\begin{figure}[!t]
\centering
\includegraphics[width=0.49\textwidth]{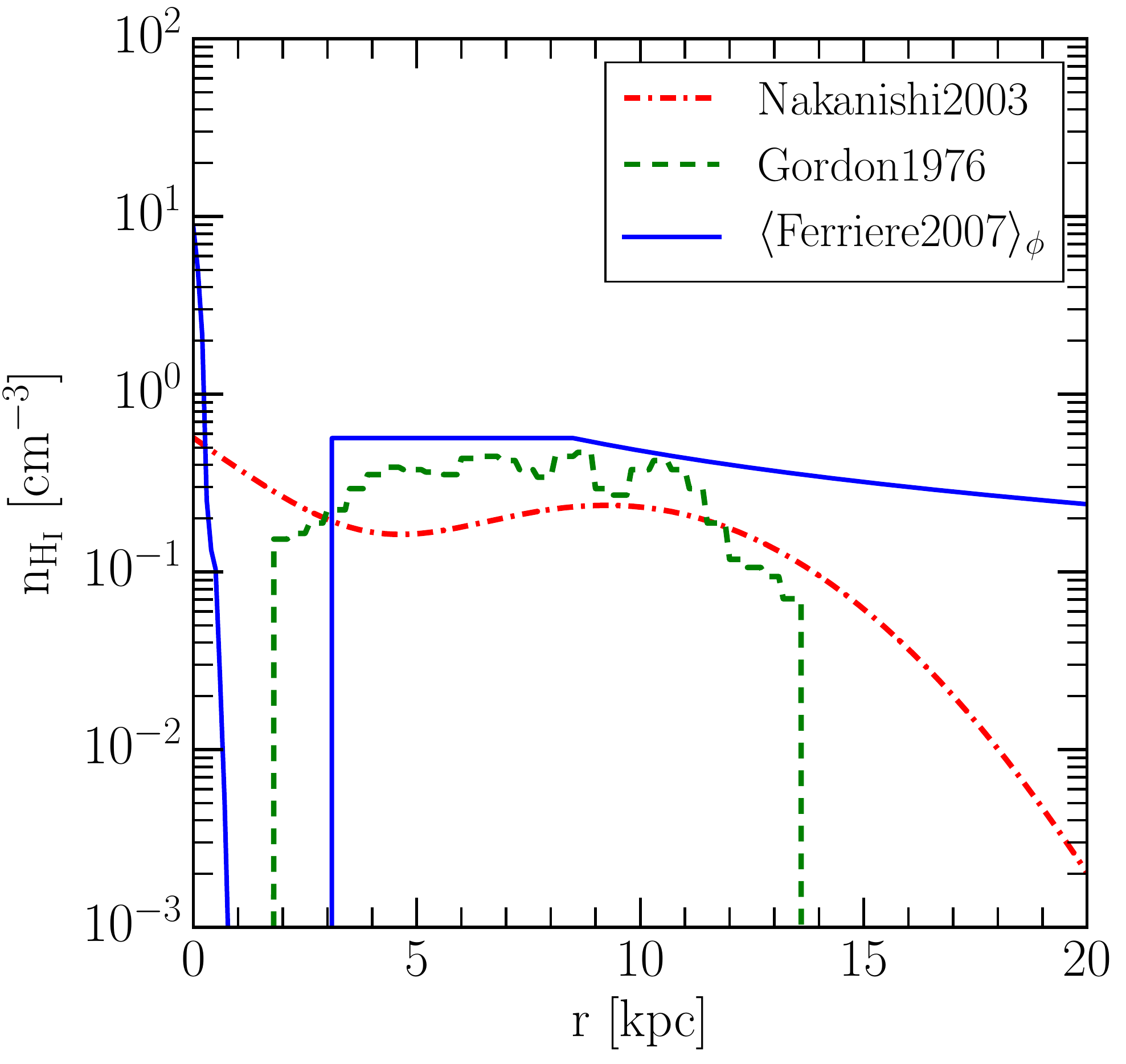}
\hspace{\stretch{1}}
\includegraphics[width=0.49\textwidth]{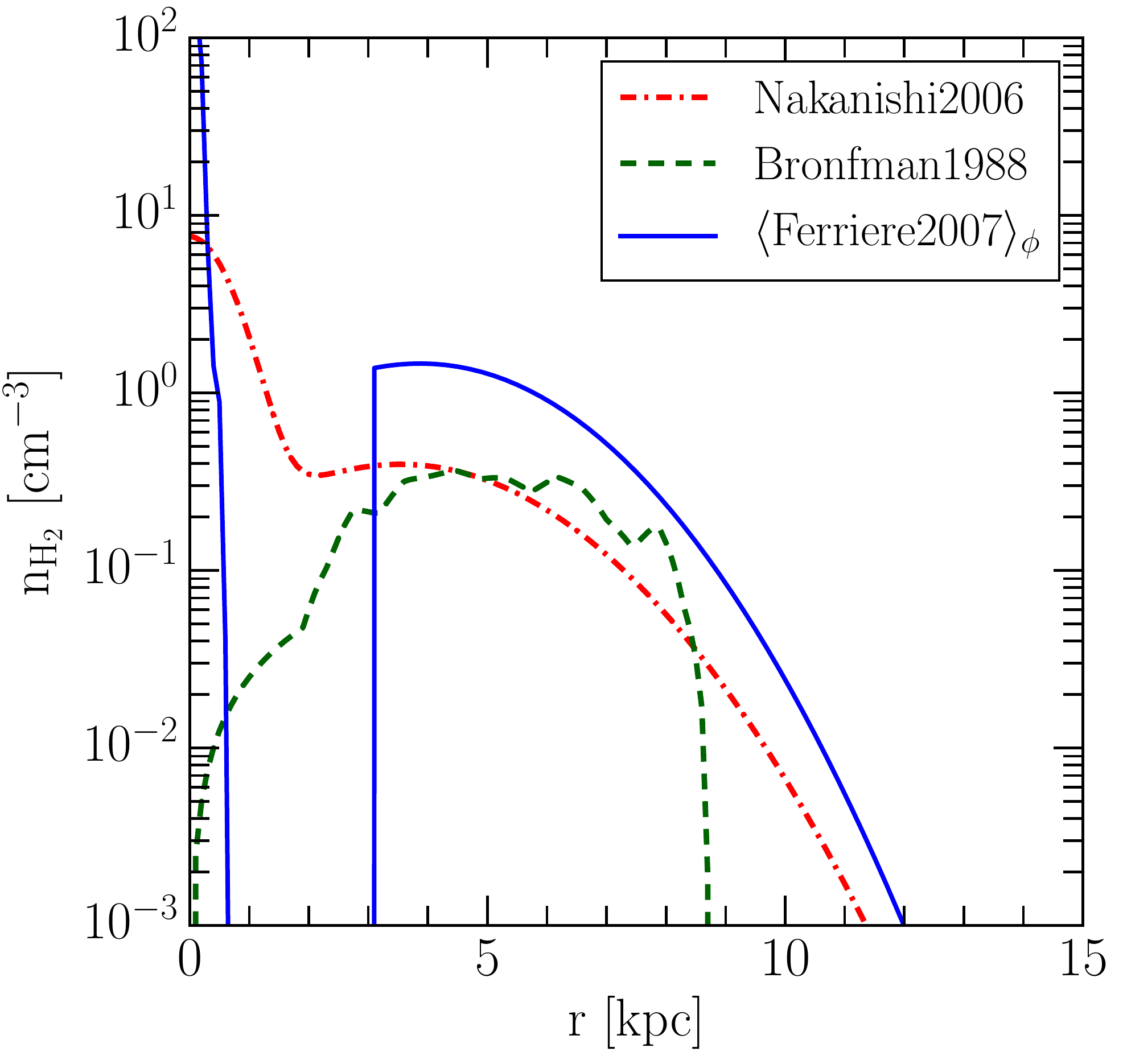}
\caption{Comparison of the gas radial distributions for the HI ({\it left panel}) and H2 ({\it right panel}) for the models implemented in the code.}
\label{fig:gas_radial}
\end{figure}

For the neutral Hydrogen we adopt the following models:

\begin{description}

\item[Gordon1976] This model is based on HI relative distributions as reported by~\cite{Gordon1976} (see their Table 1). 
The gas density is assumed to be a Gaussian along $z$ with radially binned plane density ($n_{0,i}$), with bin positions $r_i$ and half-width at half-maximum ($z_{1/2,i}$).
As pointed out by~\cite{Bronfman1988}, distances inside the solar circle (as $r_i$, $z_{1/2,i}$) scale proportionally to $R_\odot$, while density (as $n_{0,i}$) scales inversely to it. 
We adjust the model quantities according to the value of $R_\odot$ adopted in the simulation, i.e 8.5 kpc.

\item [Nakanishi2003] In~\cite{Nakanishi2003}, a three-dimensional distribution of HI gas in the Galaxy is derived from available HI survey data and the azimuthally averaged surface density and scale height as function of galactocentric radius are shown in their Figures 4 and 10.
We fit these quantities with a third-order polynomial (after re-scaling them to $R_\odot = 8.5$~kpc as described in the Appendix D of~\cite{Bronfman1988}) and we compute the gas density as:
\begin{equation}
n_{\rm HI} (r,z) = n_{\rm HI, 0}(r) \exp \left[ - \ln 2 \left( \frac{z}{z_{1/2}(r)} \right)^2 \right]
\end{equation} 

\item[Ferriere2007] We use the analytical fit provided by~\cite{Ferriere1998} for $r > 3$~kpc and based on HI distributions reported in~\cite{Dickey1990} for within the solar circle and in~\cite{Kulkarni1987} for the galactic region outside it. 
For the inner Galaxy ($r < 3$~kpc) we use here an azimuthally averaged version of the inner Galaxy HI three-dimensional distribution given in~\cite{Ferriere2007}.

\item[Pohl2008] The reconstructed spatial distribution of neutral gas in the Galaxy is obtained by~\cite{Pohl2008}, who have kinematically deconvolved the LAB HI survey by~\cite{2005A&A...440..775K} with appropriate modifications reflecting the larger single-cloud linewidth and the galactic warp and flaring in the outer Galaxy\footnote{The FITS file containing the gas density on a 3D grid is provided by the authors of~\cite{Pohl2008}.}. 
In Fig.~\ref{fig:gas_contour}, the gas density pattern along the Galactic Plane (averaged for $z$) is shown when three-dimensional propagation mode is selected.   

\end{description}

\subsubsection{H$_2$ gas distributions}

Molecular Hydrogen (H$_2$) is not directly observable and is traced by the $2.6$ mm CO emission line.
The CO-to-H2 conversion factor is called $X_{CO}$; both its normalisation and spatial dependence are affected by large uncertainties. 

For modelling the H$_2$ component we adopt the following distributions (all these models are rescaled for a constant $X_{\rm CO} = 10^{20}$~cm$^{-2}$ K$^{-1}$ / km s$^{-1}$ and they are multiplied by the $X_{\rm CO}$ profile chosen among the models described in~\ref{sec:XCO}):

\begin{description} 
\item[Bronfman1988] An axisymmetric model for the H$_2$ is provided by \cite{Bronfman1988}, as a fit of the combined Northern and Southern CO surveys performed with the Columbia Millimeter-Wave Telescope.  
The best-fit model parameters are given in their Tab.~3.

\item[Nakanishi2006] This model is described in~\cite{Nakanishi2006} and is based on the~\cite{Dame2001} CO survey. As for the corresponding Nakanishi2003 HI model, we take the radial profiles of the H$_2$ density at the midplane and of the FWHM from their Tab.~1 and we rescale for $R_\odot = 8.5$~kpc.  

\item[Ferriere2007] As for the Ferriere2007 HI model, we implement the distribution as given in~\cite{Ferriere1998} for $r > 3$~kpc and in~\cite{Ferriere2007} for $r < 3$~kpc.

\item[Pohl2008] The same technique utilized for the neutral gas was applied to derive the molecular gas distribution in the Galaxy~\cite{Pohl2008}. The analysis relies on the CO data by~\cite{Dame2001}, while for the inner Galaxy is based on a gas-flow simulation using smoothed-particle hydrodynamics (SPH), introducing a realistic barred gravitational potential derived from the observed COBE/DIRBE near-IR light distribution~\cite{2002MNRAS.330..591B}\footnote{The FITS file is available for download from: \url{http://www.app.physik.uni-potsdam.de/gas.html}}.
In Fig.~\ref{fig:gas_contour}, the gas density pattern along the Galactic Plane (averaged for $z$) is shown when three-dimensional propagation mode is selected.   

\end{description}

\begin{figure}[!t]
\centering
\includegraphics[width=0.491\textwidth]{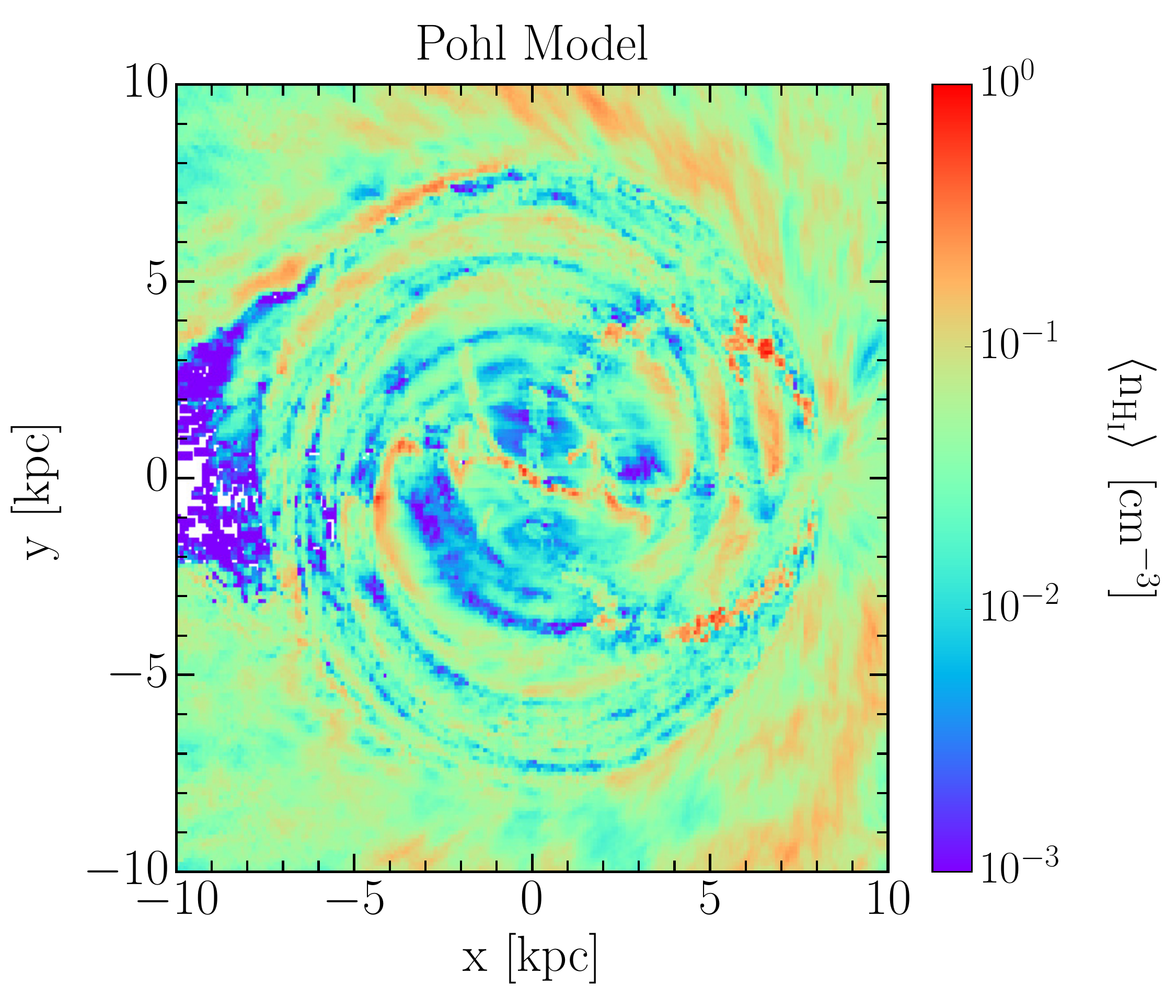}
\hspace{\stretch{1}}
\includegraphics[width=0.491\textwidth]{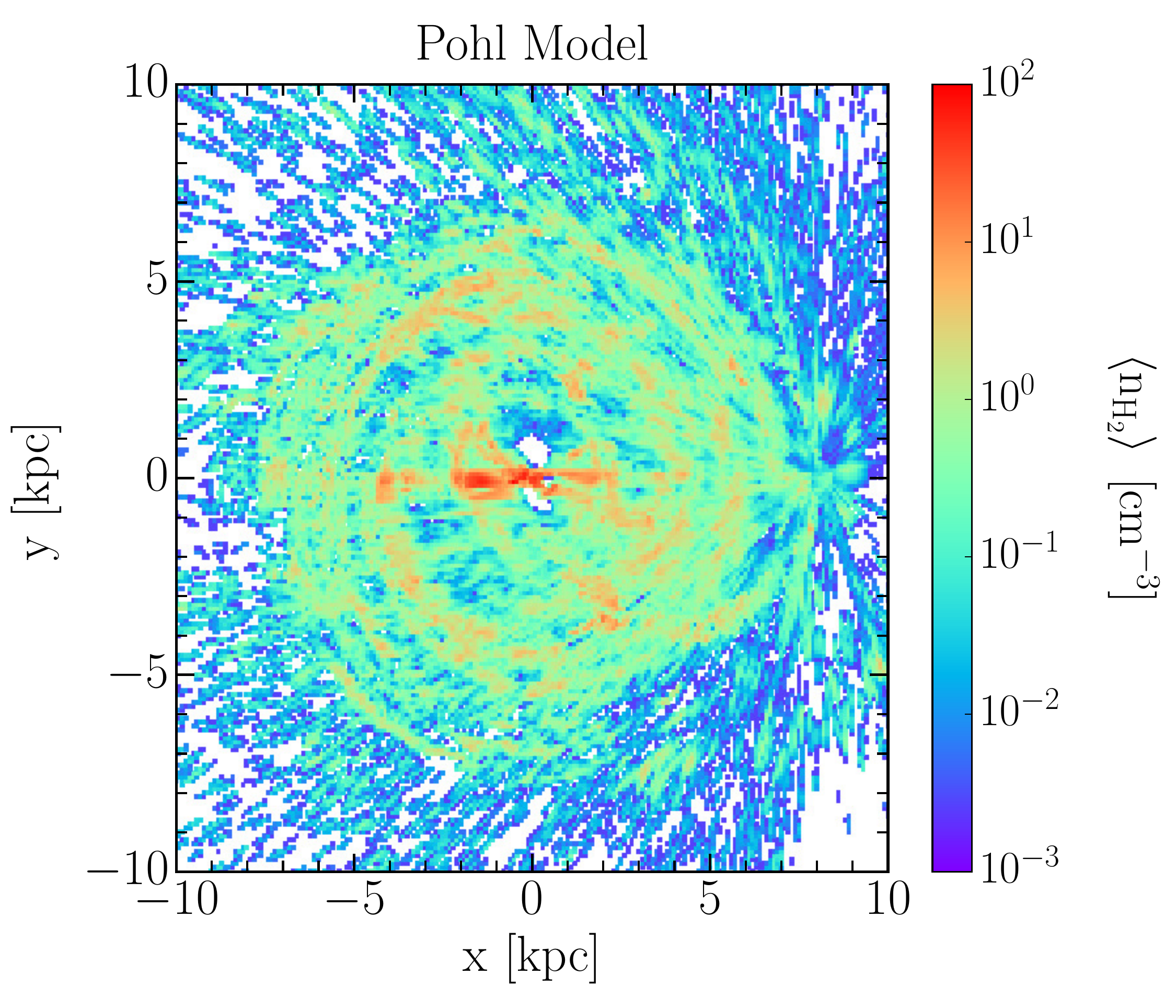}
\caption{The $z$-averaged gas density along the Galactic Plane for the {\tt Pohl2008} model (HI {\it right panel}, H$_2$ {\it left panel} ) in 3D propagation mode.}
\label{fig:gas_contour}
\end{figure}

\subsubsection{X$_{\rm CO}$ models}
\label{sec:XCO}

We then multiply H$_2$ distribution by the $X_{\rm CO, 20}$ (normalized to $10^{20}$~cm$^{-2}$ K$^{-1}$ / km s$^{-1}$) distribution; for the latter we adopt the following radial dependences:
\begin{description}
\item[Arimoto1996] The authors of \cite{Arimoto1996} studied the radial dependence of the conversion factor by comparing CO luminosities to the virial masses of a large number of giant molecular clouds situated at various galactocentric distances. They obtained the relation (rescaled to $r_\odot = 8.5$~kpc):
\begin{equation}
X_{\rm CO, 20} (r) = 0.9 \exp \left( \frac{r}{7.1\, \rm kpc} \right) 
\end{equation}

\item[Strong2004] We use the outward radial gradient in $X_{\rm CO}$ adopted in~\cite{Strong2004} to reconcile the predictions of the {\tt GALPROP} code, based on the assumption that the distribution of CR sources follows the observed distribution of pulsars, with the $\gamma$-ray profiles measured by EGRET/COMPTON.

\item[Ackermann2012] A similar analysis as in~\cite{Strong2004} was performed by making use of the FERMI data in~\cite{Ackermann2012}. The radial-dependent $X_{\rm CO}$ is shown in their Figure 25.  

\item[Evoli2012] In~\cite{Evoli2012}, a two-zone model for the $X_{\rm CO}$ is used. In this model the $X_{\rm CO}$ inner value, the outer value and the border position can be chosen by the user.
\end{description}

\subsubsection{HII gas models}

For the ionized Hydrogen component we feature:
\begin{description}
\item[Cordes1991] In~\cite{Cordes1991} the authors make use of different astronomical measurements ( pulsars emission, dispersion measurements and radio wave scattering measurement) to derive a cylindrically symmetric model for the Galactic distribution of free electrons. From this distribution is possible to infer the properties of the ISM ionized component.

\item[NE2001] The NE2001 model is described in~\cite{Cordes2002,Cordes2003}\footnote{The original source code has been downloaded from \url{http://www.nrl.navy.mil/rsd/RORF/ne2001/}}. 
The density of the Galactic free electrons has been derived from the combined fit of pulsar dispersion measures, temporal and angular broadening of radio pulses, and emission measures. 
The NE2001 model contains several components, and we consider the most relevant on a Galactic scale: (a) a thick disk with large Galactocentric scale height; (b) a thin ($\sim 140$~pc), annular disk in the inner Galaxy; (c) a Galactic centre component.

We additionally consider a correction proposed in~\cite{Gaensler2008}. By excluding sight lines at low Galactic latitude that are contaminated by HII regions and spiral arms, they find that the height of the thick disk roughly doubles to $\sim 2$~kpc.

\item[Ferriere2007] As for the Ferriere2007 HI model, we implement the distribution as given in~\cite{Ferriere1998} for $r > 3$~kpc and in~\cite{Ferriere2007} for $r < 3$~kpc.
\end{description}

\subsection{Magnetic fields models}
\label{sec:MF}

Galactic magnetic field plays a crucial role in lepton energy losses (see Section \ref{synchrotron_losses}), especially by synchrotron emission.
Moreover, as discussed in the Introduction, it enters to determine the diffusive properties of charged CR. 
 
There are several methods to measure and constrain the intensity and the orientation of this component: Zeeman splitting observations \cite{1999ApJ...520..706C}, infrared, synchrotron and starlight polarisation studies \cite{2010ApJ...722L..23N,2010MNRAS.401.1013J,1996ApJ...462..316H}, and measurements of the Faraday rotation measures of Galactic and extragalactic sources \cite{2006ApJ...642..868H,2011ApJ...738..192P}.

The Galactic magnetic field $\vec{B}$ is usually described as a sum of two components: a large-scale regular $\vec{B}_{\rm{reg}}$ and a small-scale random part $\vec{B}_{\rm{ran}}$ both having a strength of the order of $\mu$G in the Solar system neighbourhood (e.g.,~\cite{Beck2009}).
Away from this local region, however, the properties of those components are poorly known and one has to invoke phenomenological models. 

We consider models for the regular and random component separately.
  
\subsubsection{Regular component}
\label{regMF}

Regular magnetic field is divided in a disk, $\vec{B}_{\rm{disk}}$ extending few hundred parsec away from the GP, a thicker halo, $\vec{B}_{\rm{halo}}$, and a less intense vertical -- poloidal or X-shaped -- components. 
We implement in \dragon~five different models for the regular Galactic magnetic field. All those models are parametrised in cylindrical coordinates $(r,\theta,z)$:
There are two classes of models, depending on whether the direction of the field in two adjacent arms is the same (axisymmetric, or ASS model) or opposite (bisymmetric, or BSS model).

\begin{description}
\item[\textsf{Sun2007 ASS/BSS}] 

In this model the disk field takes the form of a coplanar and constant pitch angle spiral field with components~\cite{Sun2008} 
\begin{eqnarray}
\label{eq:regBr}
B^{r}_{\rm{disk}} &=& D_1(r,\theta,z) D_2(r,\theta,z) \sin{p} \\ 
\label{eq:regBtheta}
B^{\theta}_{\rm{disk}} &=& D_1(r,\theta,z) D_2(r,\theta,z) \cos{p} \\
\label{eq:regBzeta}
B^{z}_{\rm{disk}} &=& B^{z}_0,  
\end{eqnarray} 
where $p$ is the pitch angle and $D_1(r,\theta,z)$ parametrizes its spatial variation: decreasing exponential behaviour are assumed both for increasing $r$ and $z$.   
Its general expression is reported in Eq.~(7) of~\cite{Sun2008}.
$D_2(r,\theta,z)$ includes possible reversals and asymmetries in the disk magnetic field distribution.
We consider the ASS and BSS options corresponding to the expressions in Eq.s~(8) and (9) of~\cite{Sun2008} respectively. 
The expression for the halo field, corresponding to a double-torus configuration with a scale-height of few kpc's, is given in Eq.~(10) of the same paper.

No vertical component is present in these models.

\item[\textsf{Pshirkov2011 ASS/BSS}]

The model adopts the same parametrisation as the \textsf{Sun2007} model. 
The best-fit parameters, derived from a wider set of Faraday Rotation measurements, is however different as reported in Tab.~3 of~\cite{2011ApJ...738..192P} for both the ASS and BSS configurations. 

\item[\textsf{Jansson2012}]
\par 

In \cite{2012ApJ...757...14J}, the authors use a Galactocentric ($r$, $\phi$, $z$), and a right-handed Cartesian ($x$, $y$, $z$) coordinate system, with the Sun along the negative $x$-axis, at $x_{\odot} = -8.5$ kpc. The Galactic north is in the positive $z$-axis, and $\phi = 0$ is along the positive $x$-axis. The magnetic field is set to zero for $r>20$ kpc, and in a $1$ kpc radius sphere centred on the Galactic centre\footnote{The public \texttt{C++} code has been excerpted from: \href{http://www.ccppastroparticle.com/projects/jf12/}{http://www.ccppastroparticle.com/projects/jf12/}.}. 

The model features a disk, a toroidal, and an out-of-plane component. 

\begin{itemize}
\item The disk component is a generalisation of the model derived in \cite{2007ApJ...663..258B}.
\item The toroidal component has a purely toroidal, i.e., azimuthal, component with different radial and vertical extension in the northern and southern regions. 
\item The out-of-plane component is poloidal and axisymmetric, and specified -- at any position ($r, z$) -- in terms of $r_{\rm p}$ (i.e. the radius at which the field line passing through ($r, z$) crosses the mid-plane ($z = 0$)). 
Motivated by the X-shaped field structures seen in radio observations of external, edge-on galaxies, the authors refer to this field as the "X-field" component (see also Fig. \ref{fig:B}).
\end{itemize}

The model also features a  {\it striated} random (or {\it ordered random}) field which is aligned along some particular axis over a larger scale, but whose strength and sign varies chaotically on a small scale.
In \cite{2012ApJ...757...14J} the energy density of this component is assumed to be proportional to that of the regular field. 
We refer to that paper for all the details. The relevant plots are shown in Fig.~\ref{fig:B}.

\begin{figure}[t]
\centering
\includegraphics[width=0.49\columnwidth]{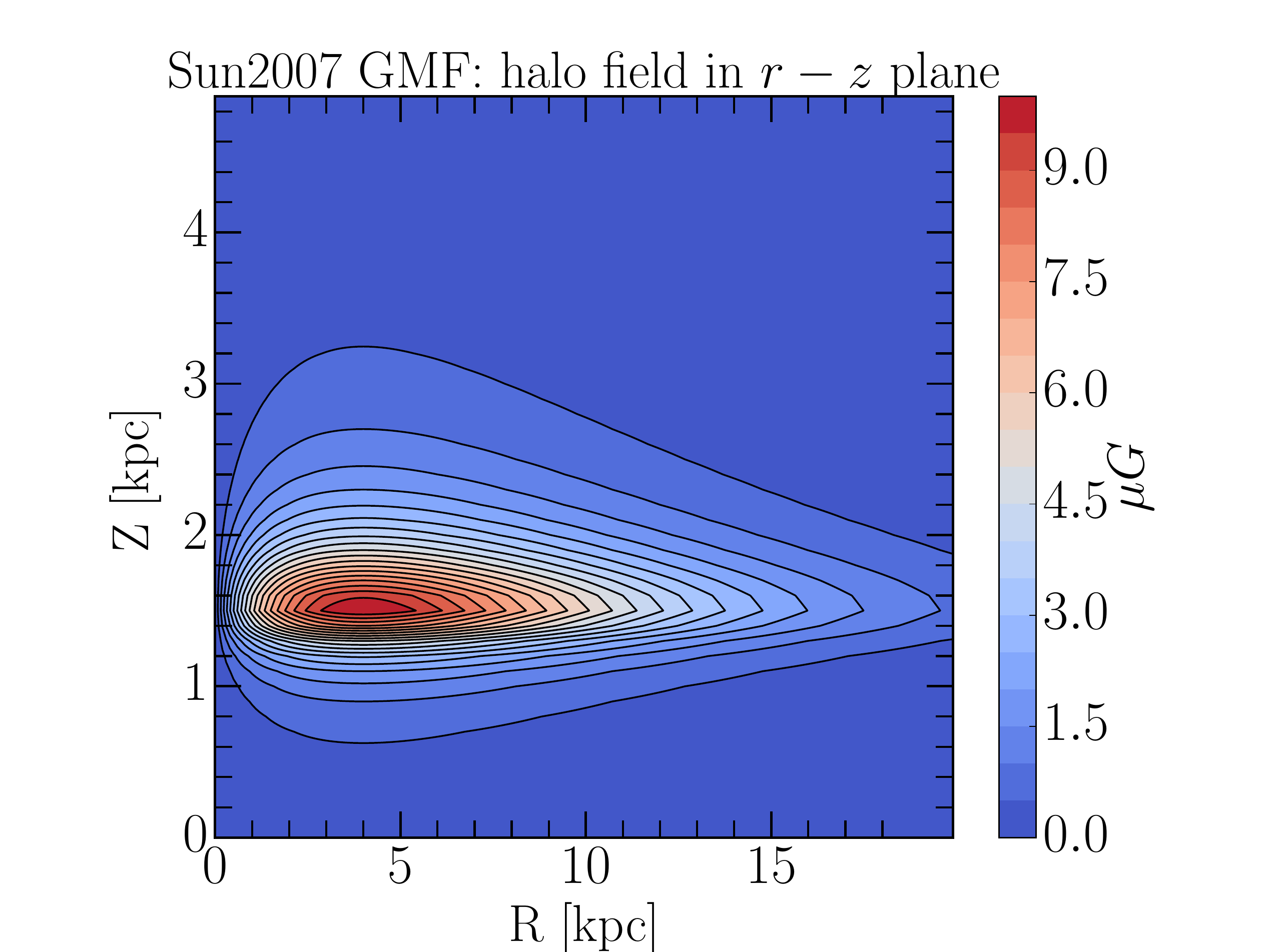}
\hspace{\stretch{1}}
\includegraphics[width=0.49\columnwidth]{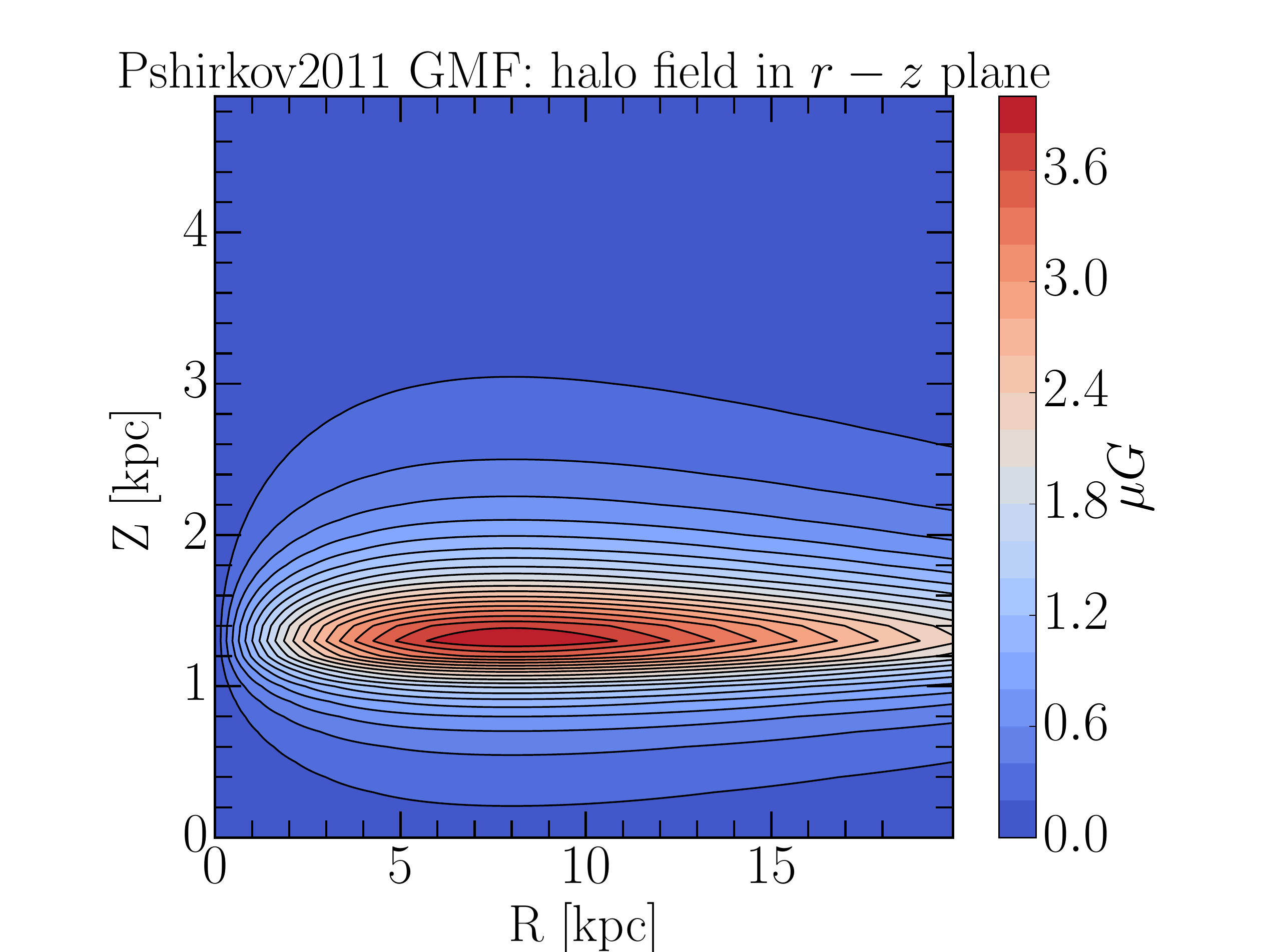} \\
\includegraphics[width=0.49\columnwidth]{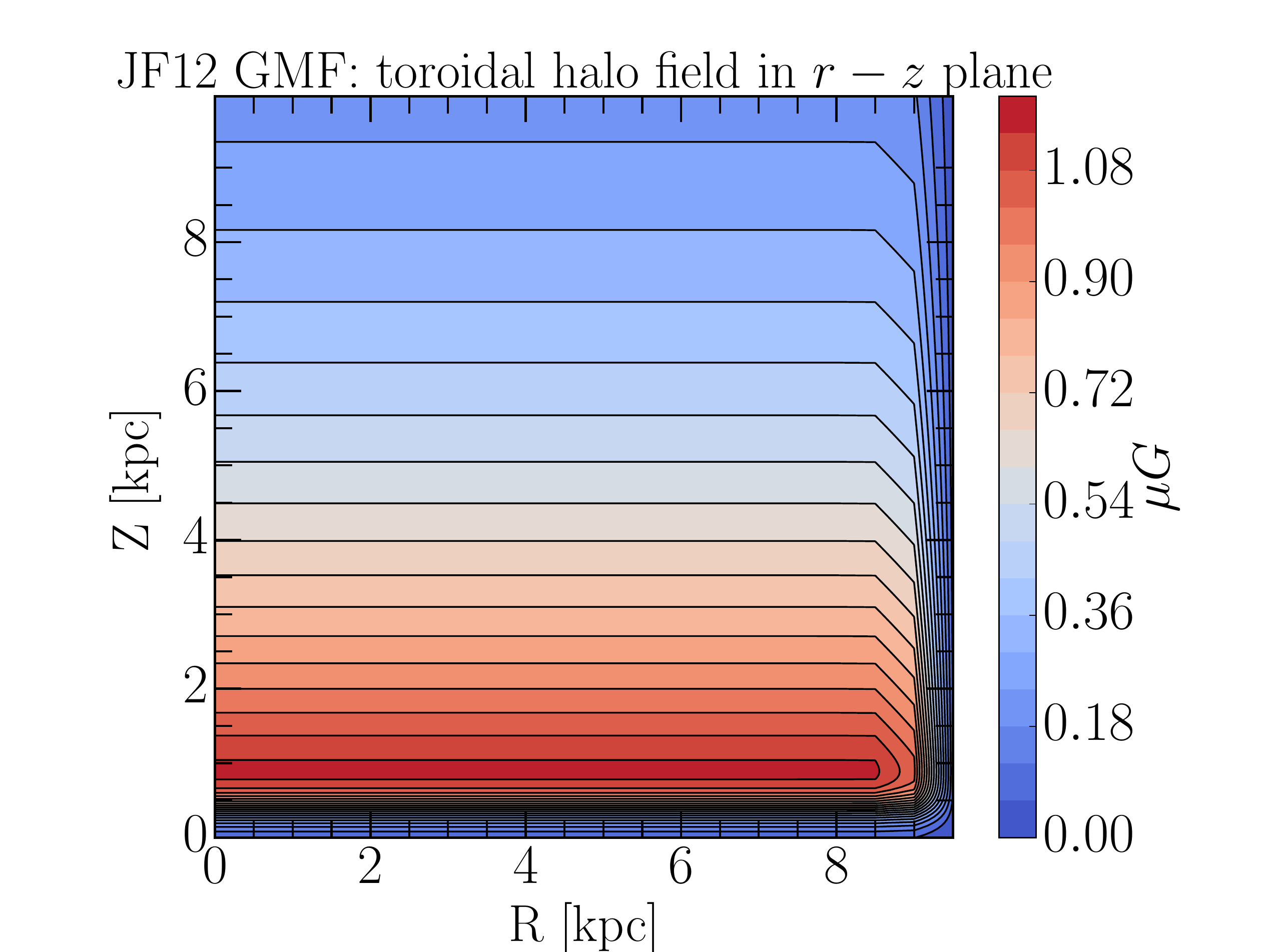}
\hspace{\stretch{1}}
\includegraphics[width=0.49\columnwidth]{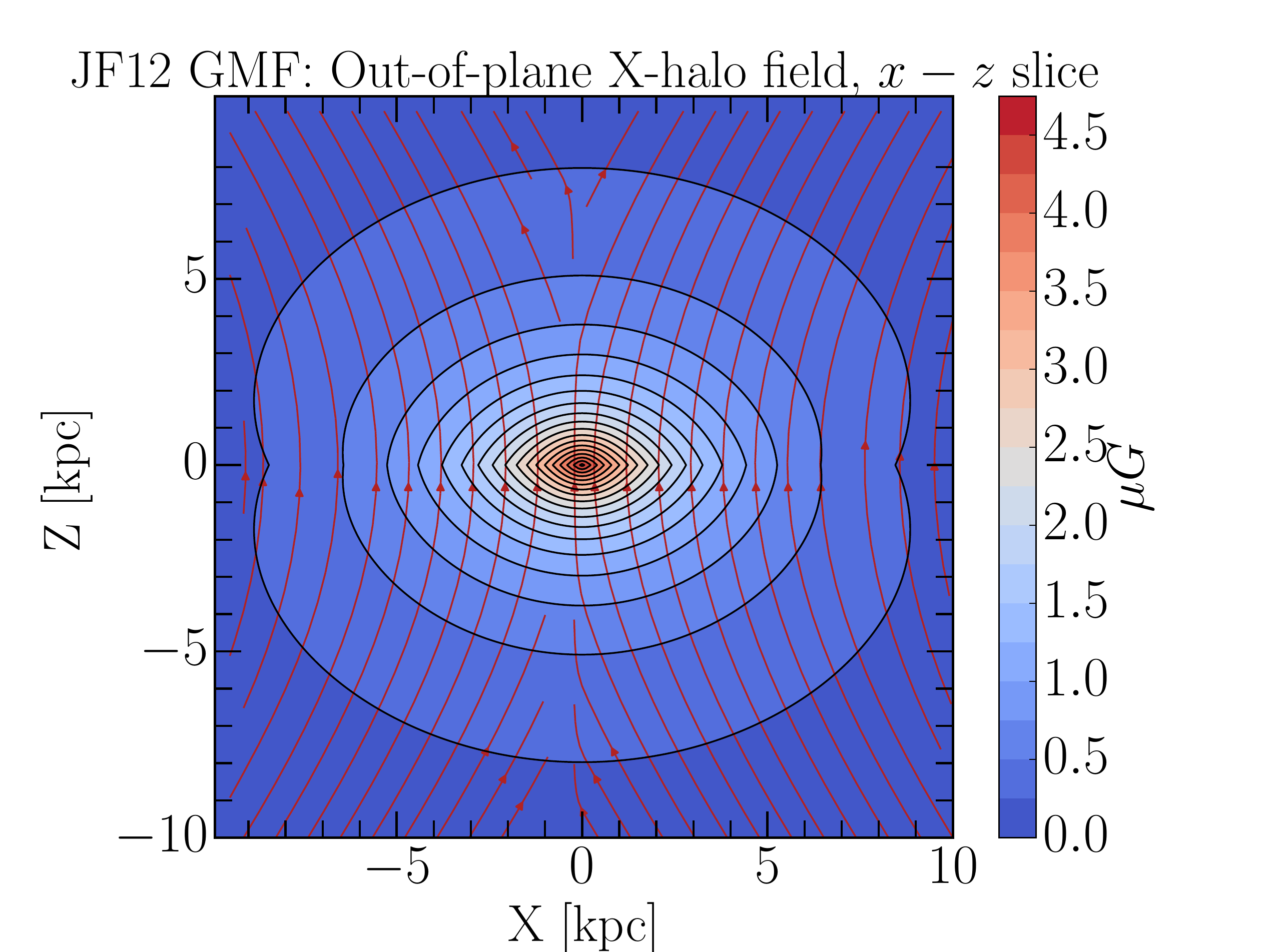}
\caption{Contour plots of the halo magnetic field for~\textsf{Sun2007} model \cite{2003A&A...410....1P} ({\it upper left panel}) for~\textsf{Pshirkov2011BSS} and~\textsf{Pshirkov2011ASS} models ({\it upper right panel}) and of the toroidal and out-of-plane components for~\textsf{Jansson2012} model ({\it lower panels}) the are displayed. Each line of the contour is associated to a different value for the halo magnetic field written in the plot in units of $\mu$G.}
\label{fig:B} 
\end{figure}

\end{description}

\subsubsection{Random component}

The random component of the Galactic magnetic field plays a crucial role in CR diffusion and is expected to determine the diffusion tensor and the size of the diffusion region.
Unfortunately, very little is known so far about its properties and its relation with the regular field. 
 
From the observational point-of-view, one of the most relevant information comes from the spatial correlation of Faraday Rotation measurements.
Using extra-Galactic sources the authors of~\cite{1996ApJ...458..194M} found a structure function compatible with a Kolmogorov power spectrum up to a scale of $\sim$few pc with a rms amplitude of $\sim 1~\mu$G.
On larger scales a flatter spectrum was found on the basis of pulsar rotation and dispersion measurements with an amplitude $6.1 \pm 0.5~\mu$G in the Galactic plane region~\cite{Han2004}.
Interestingly this value is in good agreement with that inferred from the intensity of the synchrotron emission of the Galaxy (which was found to be dominated by the effect of random field)~\cite{Strong2000,DiBernardo:2012zu}. 

Synchrotron emission can also be used to estimate the vertical extent (along $z$ coordinate) of the turbulent field, hence that of the cosmic ray diffuse halo. 
Recent studies~\cite{Bringmann:2011py,2013JCAP...03..036D}, showed that this quantity must be larger than 2 kpc favouring thicker ($L \sim 10$~kpc) halos.
The shape of the vertical profile, however, is poorly constrained: exponential, Gaussian or step-like profiles are all compatible with presently available data. 

Although disfavoured from those results a correlation between the turbulent and regular fields can not, in principle, be excluded.

For this reasons, in \dragon~we implement for the turbulent field the following three different prescriptions:

\begin{description}
\item[ProportionalToRegular] In this model, $B_{\rm rms}$ is given by an user-provided rescaling factor of the regular magnetic field intensity $|\vec{B}_{\rm reg}|$.
\item[ExponentialModel]
This amounts to a double exponential in $r$ and $z$:
\begin{equation}
B_{\rm ran}(r,z) = B_0 \, \exp{\left( -\frac{r - R_\odot}{R_B}\right)}\ \exp{\left( -\frac{|z|}{z_B} \right)}~.
\end{equation}
This structure resembles that of the simplest (azimuthally symmetric) regular magnetic fields models (see, e.g.,~\cite{2006ApJ...642..868H}).
 
It was shown that for $B_0 = 6.1~\mu$G, $R_B \simeq 6$~kpc and $z_B \sim 2$~kpc this model provides a reasonable description of the synchrotron emission of the Galaxy \cite{Strong2000}. 

\item[GaussianModel] 
A Gaussian vertical profile has been also used, e.g., in~\cite{2011APh....35..192G,2010JCAP...03..022G,2012ApJ...761L..11J}.
We consider the following parametrisation that has been derived in~\cite{2012ApJ...761L..11J} for the random magnetic field:
\begin{equation}
\label{eq:halofarrarfuncnew}
B_{\rm{ran}}(r,z) = B_0 \, \exp{\left( -\frac{z^2}{2 z^2_B} -\frac{r}{r_B} \right)},
\end{equation} 
where $r_B = (10.97 \pm 3.80)$ kpc, $z_B=(2.84\pm1.30)$ kpc and $B_0=(4.68\pm1.39)$ $\mu$G.

\end{description}

\subsection{Interstellar radiation field}
\label{sec:isrf}

\begin{figure}[!t]
\centering
\includegraphics[width=0.49\textwidth]{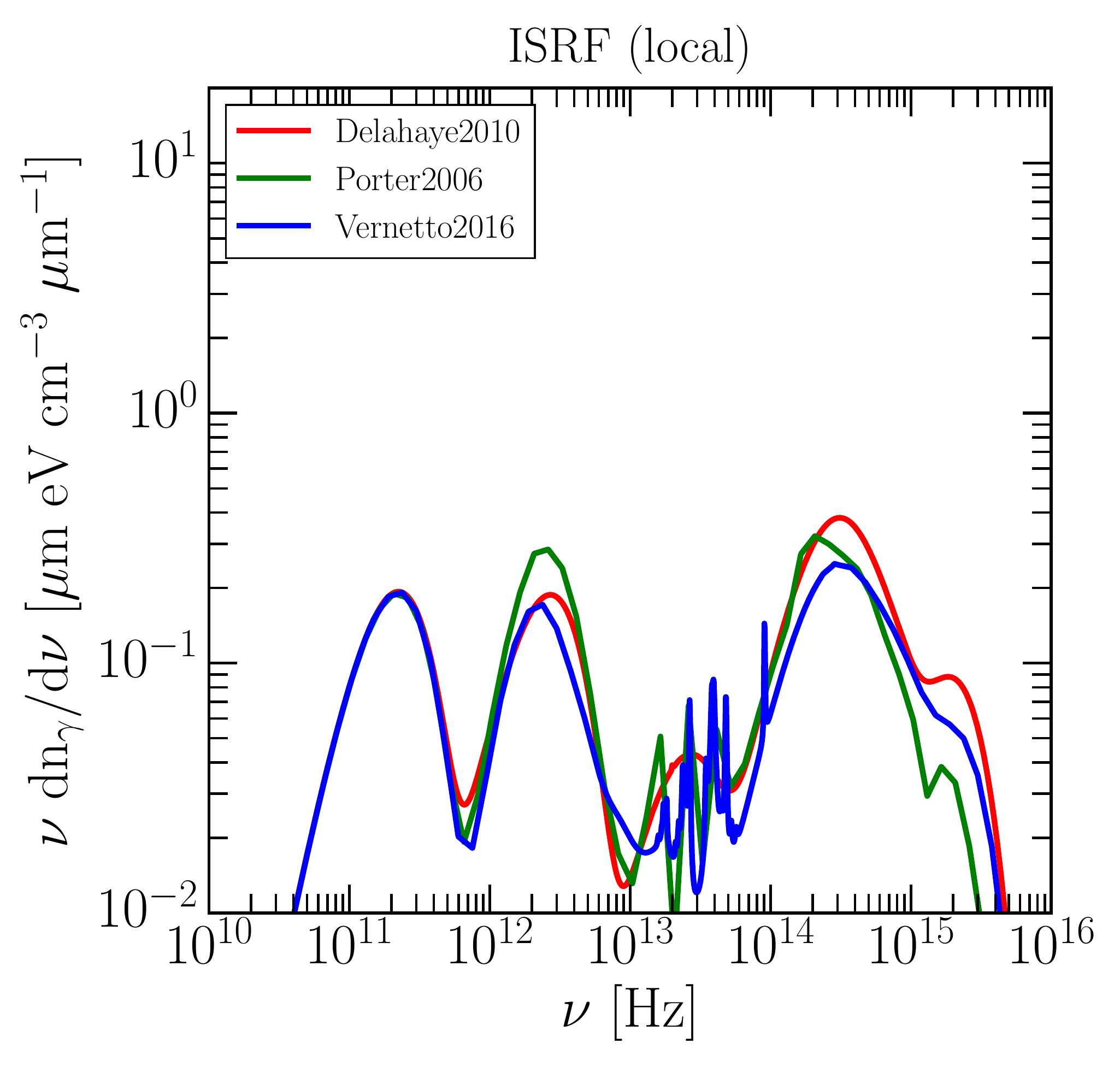}
\hspace{\stretch{1}}
\includegraphics[width=0.49\textwidth]{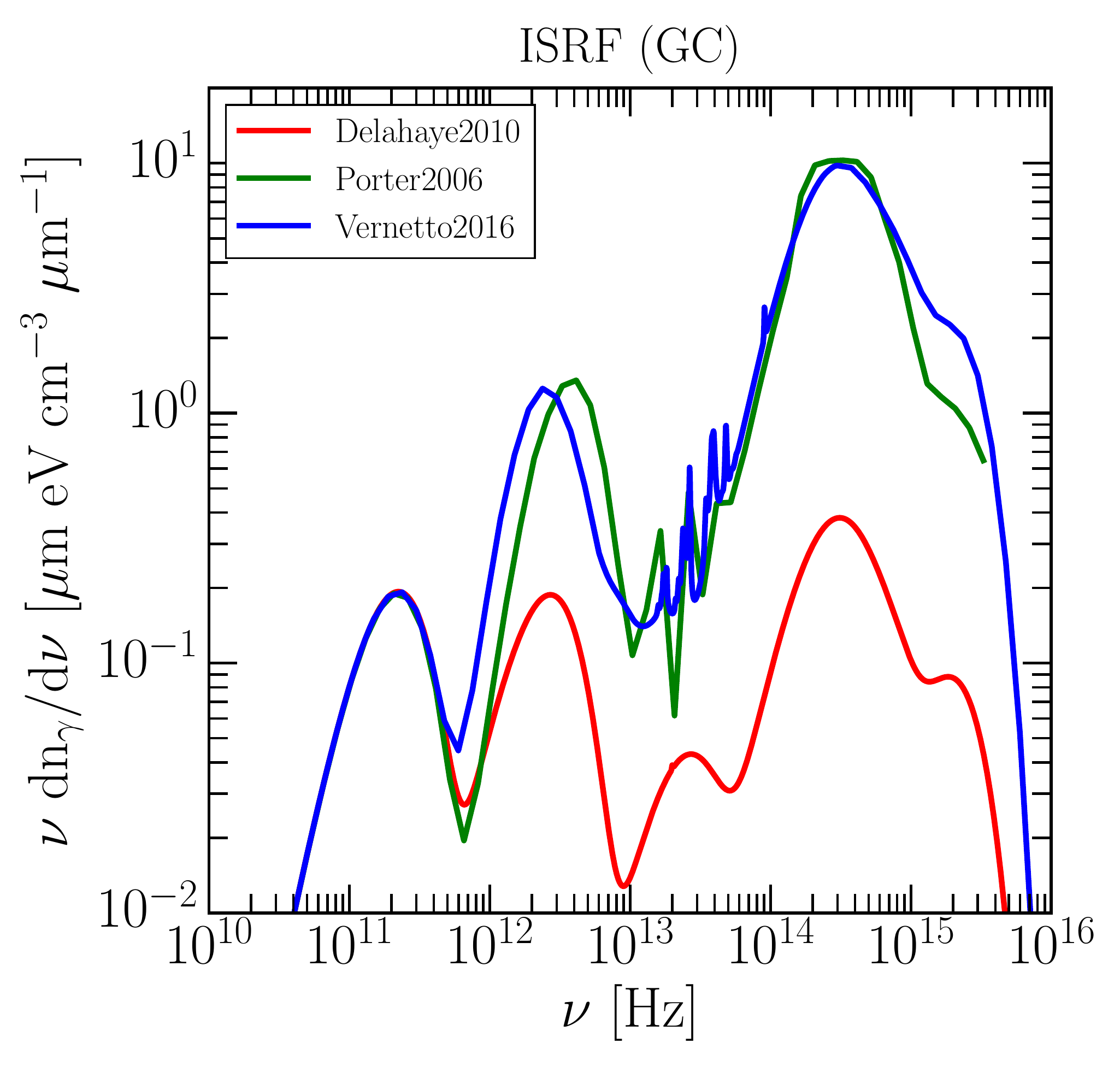}
\caption{A comparison between the ISRF spectrum evaluated at Sun ({\it left panel}) and Galactic Center ({\it right panel}) position for the three models implemented in~\dragon.}
\label{fig:isrf}
\end{figure}

The InterStellar Radiation Field (ISRF) is relevant for the leptonic energy losses via Inverse Compton scattering (ICS). 
The low-energy photons involved in this process originate from stars, and are further reprocessed by Galactic dust; CMB photons also contribute with a comparable energy density.  

We implement in \dragon~the following three models:

\begin{description}

\item[Porter2006] 
With this model, we make use of the public dataset provided with \galprop~\cite{Ackermann2012,Vladimirov2012} and available for download on the code website\footnote{\url{http://galprop.stanford.edu/FITS/MilkyWay_DR0.5_DZ0.1_DPHI10_RMAX20_ZMAX5_galprop_format.fits.gz}}. This file, distributed in FITS format, contains the ISRF spatial distribution for 56 log-spaced frequencies in the range $\sim 0.1 \div 36000$~$\mu$m.

This distribution is based on a realistic modelling of the interactions between starlight and the interstellar matter, which takes into account an accurate knowledge of both the star, gas and dust in the Galaxy, and the infrared emissivities (and spectral shape) per dust grain: a detailed calculation is presented in~\cite{Strong2000} and updated in~\cite{Porter2005}. 

\item[Delahaye2010]
We implement the ISRF model described in~\cite{Delahaye2010}.
In this model, the spectral shape is obtained as a superposition of 6 Planck distributions, with effective temperatures for the CMB, stellar, and ultraviolet components as in table~\ref{tab:isrf_parameters}. 
However, these parameters can be also modified by the user.

\item[Vernetto2016]  
In this more recent result~\cite{2016PhRvD..94f3009V}, a precise description of the Galactic dust emission is computed following the parametrisation suggested by~\cite{2006A&A...459..113M}. 
The energy density of the radiation fields in the solar neighbourhood is found to reproduce well the measurements of COBE-FIRAS and the sky maps of COBE-DIRBE in the wavelength range $60 \div 600$~$\mu$m.
The spectral density as a function of the Galactic position in cylindrical coordinates is provided by the authors as a FITS file.

\end{description}

\begin{table}[!t]
\begin{center}
\begin{center}
\begin{tabular}{| l | r | r |}
\hline
Component & $T$ [K] & Normalization \\ 
\hline
\hline
CMB & $2.7$ & $1$ \\ 
IR & $33.1$ & $4.5 \cdot 10^{-5}$ \\
Optical & $313.3$ & $1.2 \cdot 10^{-9}$ \\
UV I & $3249.3$ & $7.03 \cdot 10^{-13}$ \\
UV II & $6150.4$ & $3.39 \cdot 10^{-14}$ \\
UV III & $23209.0$ & $8.67 \cdot 10^{-17}$ \\
\hline
\end{tabular}
\end{center}
\caption{Temperatures and relative normalisations for the Planck distributions as implemented in the \textsf{Delahaye2010} ISRF (see also Fig.~\ref{fig:isrf}).}
\label{tab:isrf_parameters}   
\end{center}
\end{table}

\noindent We plot in Fig.~\ref{fig:isrf} the spectral shape of \textsf{Delahaye2010} compared to \textsf{Porter2006} and \textsf{Vernetto2016} 
evaluated at the Sun position and at the Galactic Centre. 

\subsection{Source term}
\label{sec:sourceterm}

Assuming SNR as the main CR sources and with universal injection spectrum, the source term can be written as it follows: 
\begin{equation}
Q (r, z, \rho) = Q_{0} \, q_{\rm SN} (r,z) \, \Phi (\rho) \, \exp \left( - \frac{\rho}{\rho_c} \right)
\end{equation}
where $Q_{0}$ is a normalisation factor, $q_{\rm SN} (r,z)$ is the rate per unit of volume, $\Phi (\rho)$ is the injection spectrum and $\rho_c$ the rigidity cut-off.

Moreover, the code allows the user to provid a generic function of space and momentum to be implemented as a source term.
  
\subsubsection{Source profile}

For the axisymmetric CR source distribution, we adopt the parametrisation suggested by the authors in~\cite{Yusifov2004} to model the radial distribution of the pulsar progenitors in the Milky Way, and described by the following function:
\begin{equation}
\label{eq:snr_term}
q_{\rm SN} (r, z) =  A \biggl(\frac{r + R_{1}}{R_{\odot} + R_{1}}\biggr)^{a} \exp\biggl(-b\times \frac{r - R_{\odot}}{R_{\odot} + R_{1}}  - \frac{\lvert z \rvert}{z_{0}}\biggr) \, , 
\end{equation} 
where $A$, $a$, $b$, $R_1$ and $z_0$ are model parameters. In particular, $A$ is a normalisation constant and it is computed to account for the SN galactic rate.
$R_{1}$ is included to obtain a nonzero surface density at $r = 0$, which may be not realistic (see also \cite{Kaspi2005}). 
For what concerns the vertical distribution, the $z-$dependence in Eq.~\ref{eq:snr_term} reflects instead the assumption of  confinement of the sources in the disc.

In Tab.~\ref{tab:snr_parameters} we summarize the best-fit parameters characterising the source models based on Eq.~\ref{eq:snr_term}, and included in~\dragon~code as possible source term (see also Fig.~\ref{fig:snr_distribution}).

A different parametrisation for the SNR spatial profile is given in \cite{Ferriere2001}. 
This model traces both type-II and type-I SNR distributions. In particular, the spatial distribution of type-II SNe is traced by the H II regions - produced by their OB progenitor stars - or PSR, instead the type-I SNe follow the distribution of old disc stars.

The SNR profile in this \textsf{Ferriere2001} model is then given by the sum of two contributions:
\begin{equation}
q_{\rm I} (r,z) = (7.3~{\rm kpc}^{-3}~{\rm Myr}^{-1}) \times \exp \left( -\frac{r - R_{\odot}}{4.5} - \frac{\lvert z \rvert}{0.325} \right)
\end{equation}
and
\begin{equation}
\begin{split}
q_{\rm II} (r, z) &= (177.5~{\rm kpc}^{-3}~{\rm Myr}^{-1}) \times \biggl\{0.79\times \exp \biggl[-\biggl(\frac{z}{0.212}\biggr)^{2} \biggr] + 0.21\times \exp \biggl[-\biggl(\frac{z}{0.636}\biggr)^{2} \biggr]\biggr\}\times \\
	 	& \times \exp\biggl(-\frac{r - 3.7}{2.1}\biggr)^{2} \qquad \text{for } r < 3.7~{\rm kpc}\\
q_{\rm II} (r, z) &= (50~{\rm kpc}^{-3}~{\rm Myr}^{-1}) \times \biggl\{0.79\times \exp \biggl[-\biggl(\frac{z}{0.212}\biggr)^{2} \biggr] + 0.21\times \exp \biggl[-\biggl(\frac{z}{0.636}\biggr)^{2} \biggr]\biggr\}\times \\
	 	& \times \exp\biggl(-\frac{r - R_{\odot}}{6.8}\biggr)^{2} \qquad \text{for } r > 3.7~{\rm kpc} \, .
	\end{split}
\end{equation}

\begin{figure}[!t]
\centering
\includegraphics[width=0.49\textwidth]{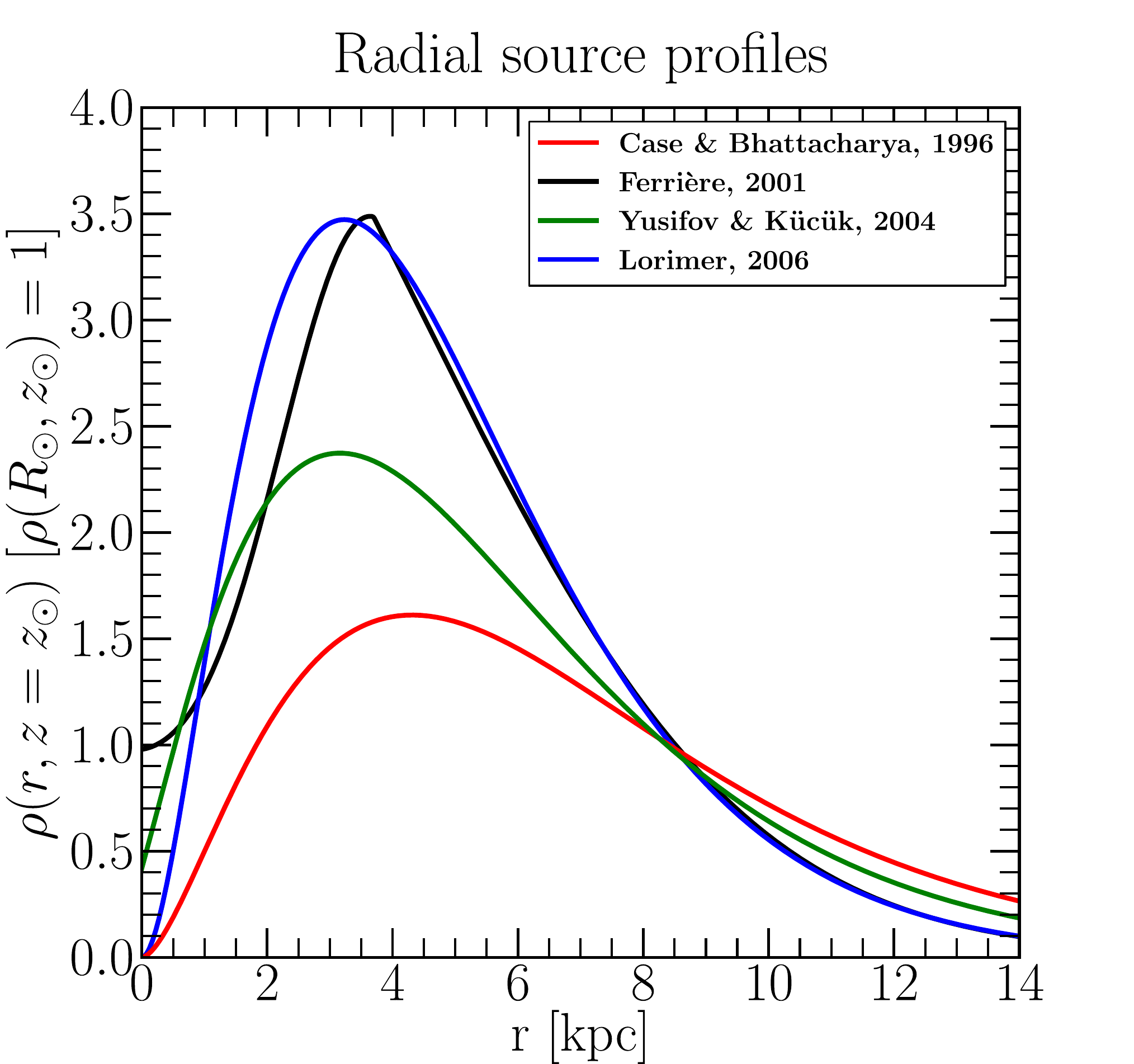}
\hspace{\stretch{1}}
\includegraphics[width=0.49\textwidth]{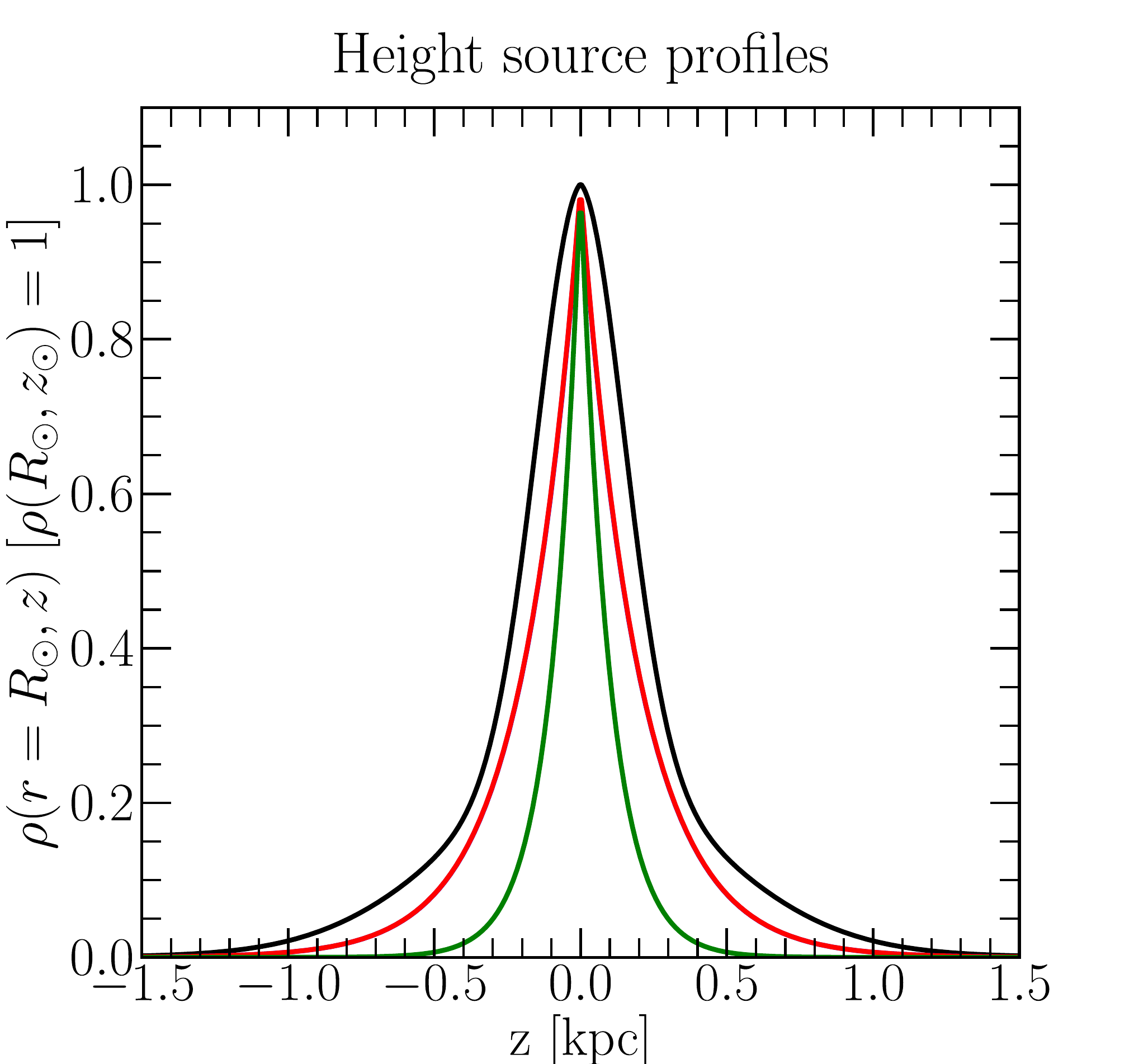}
\caption{Radial and height profile of the distribution functions assumed for the CR sources.}
\label{fig:snr_distribution}
\end{figure}

\begin{table}[!t]
\begin{center}
\begin{center}
\begin{tabular}{| l | c | c | c | c | c | r }
\hline
{\it Model name} & $a$ & $b$ & $R_{1}$ [kpc] & $z_{0}$ [kpc] & Tracer \\ 
\hline
\hline
\textsf{Case1998} \cite{Case1998} & $1.69$ & $3.33$ & $0$ & $0.2$ & SNR \\ \hline
\textsf{Yusifov2004} \cite{Yusifov2004} & $1.64$ & $4.01$ & $0.55$ & $0.1$& Pulsar \\ 
\hline
\textsf{Lorimer2006} \cite{Lorimer2006} & $1.9$ & $5.0$ & $0$ & $0.2$ & Pulsar \\ 
\hline    
\end{tabular}
\end{center}
\caption{The different implementation of the SNR distribution based on Eq.~\ref{eq:snr_term}.}
\label{tab:snr_parameters}   
\end{center}
\end{table}

\subsubsection{Source spectra}

The injection spectrum can be chosen between two different models:

\begin{itemize}
\item a multiple broken power-law with an exponential cutoff:

\begin{equation}
\Phi (\rho) = \sum_i \frac{f_i}{\beta(\rho)} \left( \frac{\rho}{\rho_{b,i}} \right)^{-\alpha_i} \Theta(\rho - \rho_{b,i}) \Theta(\rho_{b,i+1} - \rho) \exp \left( - \frac{\rho}{\rho_{c,i}} \right)
\end{equation}
where $\Theta(x)$ is the Heaviside function and 
\begin{equation}
f_i =
\begin{cases}
1 & \quad i = 0 \\
\prod_{j=0}^{i-1} \left( \frac{\rho_{b,j+1}}{\rho_{b,j}} \right)^{-\alpha_j}  & \quad i \neq 0 \\
\end{cases}
\end{equation}

In doing so, $\alpha_i$, $\rho_{b,i}$, $\rho_{c,i}$, $f_i$ are free parameters provided by the user and can be different for each species $i$.

\item {\bf a log parabolic power-law:}

\begin{equation}
\Phi (\rho) = f_i \left( \frac{\rho}{\rho_{b,i}} \right)^{-(\alpha_i + \beta_i \log(\rho/\rho_{c,i}))}
\end{equation}
with free parameters: $\alpha_i$, $\beta_i$, $\rho_{b,i}$, $\rho_{c,i}$, $f_i$.

\end{itemize}

\subsubsection{Source normalization}

We foresee that, in some applications, the source term requires to be normalised. 
The user is allowed to provide the galactic source luminosity, $L_{\rm SN}$, such that the source term is normalised as it follows:
\begin{equation}
L_{\rm SN} = \int dV \int dE \, p(E) \, Q (r,z,p)
\end{equation}

\subsection{Wind velocity}

The wind velocity is important for adiabatic energy losses and advective transport.

In the new \dragon~code we adopt three different parameterisations:

\begin{description}

\item[ConstantGradient] The wind velocity is assumed to increase linearly with distance from the plane and constant radially:
\begin{equation}
v_w (z) = v_{w,0} + \frac{d v_w}{dz} z
\end{equation}
where $v_{w,0}$ (in km/s) and $\frac{d v_w}{dz}$ (in km/s/kpc) are input parameters.
 
\item[ConstantAtLargeHeight] Cosmic-ray driven wind models predicts that convection velocity reaches a constant value (e.g.,~\cite{Everett2008}), as described by the following functional form:
\begin{equation}
v_w (z) = {\rm sgn}(z)v_{w,0} + v_{w,\infty} \, \rm{tanh} \left( \frac{z}{z_w} \right) 
\end{equation}
where $v_{w,\infty}$ and $z_w$ are additionally free parameters.

\item[RadialDependent] In order to test a radially dependent convection model, we assume that convection velocity is proportional to source profile, as expected from winds which are sustained by SN kinetic energy release~\cite{Gebauer2009,Evoli2011}.

Following this idea, the wind velocity is given by:
\begin{equation}
v_w(r,z) = v_{w,0} \, \left[ \frac{q_{\rm SN} (r)}{q_{\rm SN} (r_\odot)} \right]_{z=0} + \left(\frac{d v_w}{dz}\right) z
\end{equation}

\end{description}

\subsection{Alfv\'en velocity}

For Alfv\'en velocity, entering in momentum diffusion calculations, the user can choose to provide a spatially constant value ({\bf Constant}) or to consistently calculate $v_A$ as function of position in terms of the magnetic field and ionised gas density distribution ({\bf SpatialDependent}): 
\begin{equation}
v_A = 2.18 \left( \frac{n_{\rm HII}}{\rm cm^{-3}} \right)^{-1/2} \left( \frac{B}{\mu \rm G} \right) \, \, \, {\rm km/s} 
\end{equation}

In the latter case, we assume that $v_A$ is bounded by $100$~km$/$s even for very small gas densities.

\subsection{Spiral galactic pattern}

We introduce in the code the possibility to superimpose a spatial pattern to the distributions of different astrophysical quantities: source term, gas, ISRF and magnetic field.

In order to apply a generic pattern $f(\vec{x})$ to the distribution $c(\vec{x})$, we first compute the new distribution:
\begin{equation}
c'(\vec{x}) \,=\, c(\vec{x})\cdot f(\vec{x})
\end{equation}
Then, we rescale the altered term $c'$ imposing that the volume integral is conserved 
\begin{equation}
\int{{\rm d}^3\, c'(\vec{x})} \, = \, \int{{\rm d}^3 \, c(\vec{x})}
\end{equation}

\begin{figure}[!t]
\centering
\includegraphics[width=0.49\textwidth]{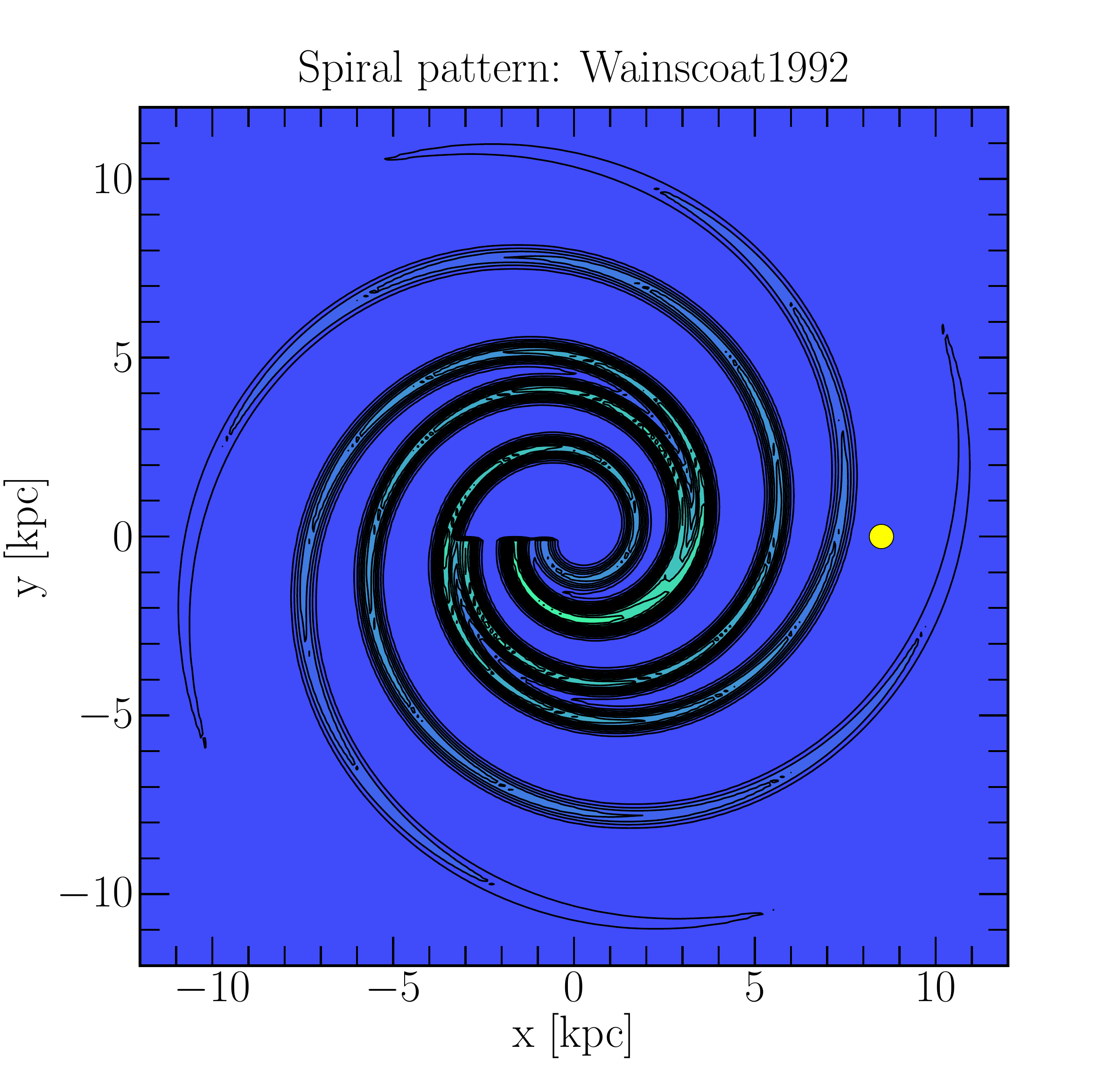}
\hspace{\stretch{1}}
\includegraphics[width=0.49\textwidth]{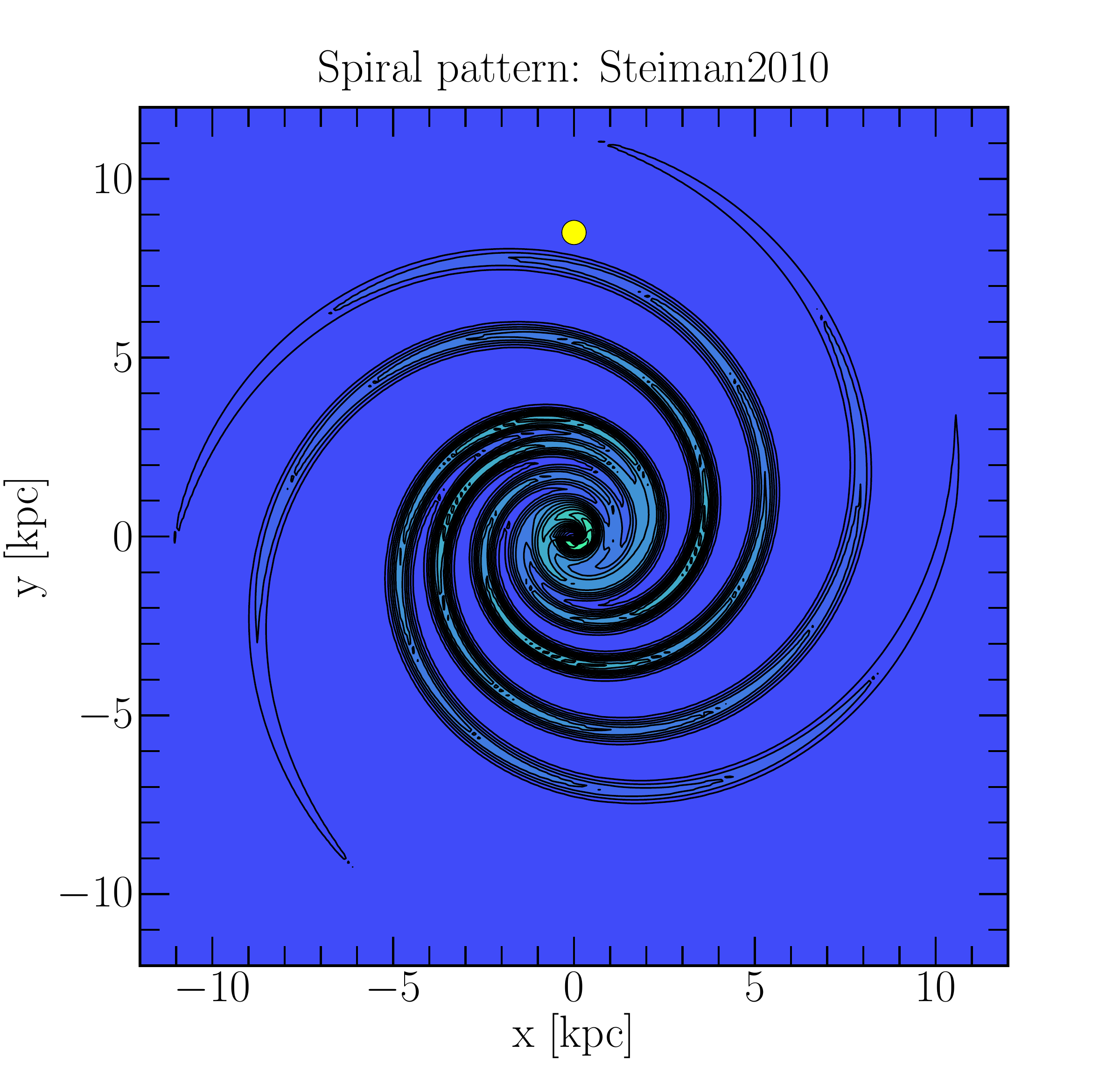}
\caption{The spiral patterns \textsf{Wainscoat1992} \cite{Wainscoat1992} and \textsf{Steiman2010}\cite{Steiman2010}. The Solar System location is identified with a yellow circle.}
\label{fig:spiral_pattern}
\end{figure}

Several astronomical observations point to the presence of a spiral arm structure in our Galaxy. There has been a general consensus on the existence of this pattern between $\sim 3$ and $\sim 10$ kpc, although some disagreement remains on the details regarding the number of spiral arms and their geometry.

Many astronomical data may be used to trace the details of the spiral structure. The arms are generally modelled as logarithmic spirals, and two-, three-, four-arms patterns can be also found in literature.
 
In particular, the spiral structure may be derived:

\begin{itemize}

\item From the catalogues of point sources: They reveal the presence of the pattern and permit to fit the arm parameters. For example, the IRAS catalogue of sources in the the 8-25 micron band (mainly red giant stars) was used in \cite{Wainscoat1992} to derive a physically realistic representations of the Galactic disk, bulge, stellar halo, spiral arms. The arm parameters derived in this work are adopted in several more recent works \cite{Kaspi2005,Blasi2012}.

\item From the study of the far-infrared cooling lines of the interstellar gas. Since the spiral arms, compared to inter-arm regions, show higher gas column densities and star formation rates, they have enhanced line radiation whose observations can be used to trace the arms.
For example in \cite{Steiman2010} the [CII] 158 $\mu$m and [NII] 205 $\mu$m lines observed by the FIRAS instrument mounted on COBE satellite are used to determine the parameters of a four-arm logarithmic spiral pattern.
\end{itemize}

In \dragon~we implement the spiral structure with the following formalism.
Each arm is modelled as a logarithmic spiral with arm width $\sigma_a$. The {\it locus} corresponding to this curve -- in polar coordinates -- is defined by:

\begin{equation}
\theta (R) \, = \, K \, \log \left( \frac{r}{R_0} \right) \, + \theta_0
\end{equation}

The parameter $K$ is related to the {\it pitch angle} $p$ though the following equation: $p \, = \, \arctan(1/K)$.

The user is allowed to implement an arbitrary number of spiral arms, characterised by the values of $K$, $R_0$, $\theta_0$, and $\sigma_a$. 
In Fig.~\ref{fig:spiral_pattern} we provide two examples of spiral patterns, with $\sigma_a = 200$~pc and arm parameters reported in Tab.~\ref{tab:spiral_parameters}.

\begin{table}[!t]
\small
\begin{center}
\begin{center}
\begin{tabular}{| c | c | c | c | c | c | c | c | c | c | c | c | c |}
\hline
{\it Model name} & $K_1$ & $R_{1}$ & $\theta_{1}$ & $K_2$ & $R_2$ & $\theta_2$ & $K_3$ & $R_3$ & $\theta_3$ & $K_4$ & $R_4$ & $\theta_4$ \\ 
\hline
\hline
\textsf{Wainscoat1992} & 4.25 & 3.48 & 0 & 4.25 & 3.48 & 3.14 & 4.89 & 4.90 & 2.52 & 4.89 & 4.90 & -0.62 \\
\textsf{Steiman2010} & 4.17 & 0.38 & 0 & 4.13 & 0.25 & 0 & 4.02 & 0.45 & 0 & 3.58 & 0.61 & 0 \\ 
\hline
\end{tabular}
\end{center}
\caption{Parameters for the spiral pattern models in Fig.~\ref{fig:spiral_pattern}. Both models depict four-arms structures. The spiral width is assumed to be $0.2$~kpc.}
\label{tab:spiral_parameters}   
\end{center}
\end{table}

\subsection{Spatial diffusion coefficient}
\label{sec:spatialcoefficient}

As mentioned in Section \ref{sec:propagation}, the interaction between charged CRs and the Alfv\'en waves travelling through
the turbulent Galactic plasma is responsible for the CR random walk, which is well described in terms of a position-dependent diffusion tensor. 

We remark that accounting for a spatially inhomogeneous diffusion coefficient is a innovative feature of {\tt DRAGON} since its first release. Moreover, as detailed below, this feature is essential to interpret recent measurements of CR fluxes and gamma-ray diffuse emissions.

\begin{description}

\item[OneZoneModel] A constant diffusion coefficient can be adopted to reproduce local observables. It can be parametrised as:  
\begin{equation}\label{Eq:dcost}
D_{\rm cst}(\rho) = D_0 \beta^\eta \, \left(\frac{\rho}{\rho_0}\right)^{\delta}
\end{equation}
where $D_0$ is the normalisation of the diffusion coefficient, $\rho_0$ is a reference rigidity, $\delta$ is  the slope and $\eta$ allows us to take into account physics taking place at low energy, including some non-linear phenomena such as the dissipation of magneto-hydrodynamics waves by their resonant interaction with CRs \cite{Ptuskin2006}.

\item[ExponentialHalo] The assumption of diffusion inversely proportional to the magnetic field turbulent component turns into a decrease of $D$ in the halo.

To account for this, we model $D$ as:
\begin{equation}
D(\rho,z) = D_{\rm cst}(\rho) \, {\rm exp}\left(\frac{|z|}{z_t}\right)
\end{equation}
where $z_t$ is a caracteristic scale, and the halo size is $\sim 3 z_t$.

\item[PropToSourceTerm] This model was introduced in \cite{Evoli2012} to provide a solution to the long-standing {\it gradient problem} in the outer Galaxy. The diffusion coefficient follows the radial source profile as:
\begin{equation}
D(\rho,r) = D_{\rm cst}(\rho) \left[ \frac{q_{\rm SN}(r)}{q_{\rm SN}(r_\odot)} \right]^{\tau}
\end{equation}
where $\tau$ is a free parameter.

\item[VariableSlope] This model was introduced in \cite{2015PhRvD..91h3012G} in order to reproduce $\gamma$-ray emissivity in various Galactic regions. 
The slope in Eq.~\ref{Eq:dcost} is replaced by a function of the Galactic radius:
\begin{equation}
\delta(r) = a r + b
\end{equation}
with local normalization $\delta(r_{\odot}) = 0.5$.
$A$ is the free parameter which can be fitted against $\gamma$-ray data. Authors in \cite{2015PhRvD..91h3012G} provide as best fit parameter $A = 0.035$  kpc$^{-1}$. 

\end{description}

\subsection{Momentum diffusion coefficient}

The turbulent fields affect particles trajectory essentially in two ways: particles experience parallel and perpendicular scattering with respect to the background field (mainly due to the turbulent magnetic fields), but they also experience momentum diffusion or stochastic acceleration.
In particular, the nonlinear interplay between particles and turbulent waves/modes is a stochastic process that drains energy from plasma turbulence to particles.
In this section, we will focus on stochastic reacceleration of CRs due to resonant interaction with turbulence, in particular with low-frequency waves: we refer to two different kind of waves, namely Alfv\'enic and Magnetosonic waves.

\begin{description}
\item [Alfv\'enic] With this model we adopt the momentum diffusion coefficient for the ISM described in~\cite{Berezinskii1990}. In doing so: 
\begin{equation}\label{Eq:DppBerezinskii}
D_{pp} = \frac{4}{3 \delta (4-\delta^2) (4-\delta) w} \frac{p^2 v_A^2}{\langle D \rangle}
\end{equation}

where $w$ is the turbulence level (usually assumed to be $\sim 1$), $v_A$ is the Alfv\'en speed and $\langle D \rangle$ is the direction averaged spatial diffusion coefficient at each position. 

\item[Magnetosonic] The diffusion coefficient in momentum for scattering by fast magnetosonic waves is~\cite{1988SvAL...14..255P}:
\begin{equation}
D_{pp} = p^2 \frac{8 \pi D_{xx}}{9} \int_{1/L}^{k_d} \, dk \, \frac{W(k) k^4}{v_F^2 + D^2 k^2} 
\end{equation}
where $L$ is the scale of turbulence injection, $v_F$ is the magnetosonic phase velocity, $1/k_d$ is the dissipation scale, and $W(k) = W_0 L^{-2/3} k^{-11/3}$ is the energy density power spectrum.

This model could be relevant in highly-turbulent environments, as exploited in~\cite{Mertsch2011} to study stochastic reacceleration inside the Fermi bubbles.

\end{description}

\subsection{Energy losses}
\label{Sec:elosses}

Nucleons lose energy via ionisation, Coulomb interactions with the interstellar gas and pion production, while electrons and positrons lose energy via inverse Compton, synchrotron emission and bremsstrahlung as well.

In the following, we review how these processes are implemented in \dragon. Figure~\ref{fig:alltau} 
 present a comparison of the energy loss time for these mechanisms.

\subsubsection{Synchrotron radiation}
\label{synchrotron_losses}

Synchrotron radiation is emitted by an ultra-relativistic particle interacting with a magnetic field. The emitted power is $\propto m^{-4}$, therefore this process is relevant only for leptons, due to their relatively small mass. 
A detailed description of the synchrotron radiation theory can be found in~\cite{Pacholczyk1970,Pacholczyk1977,Blumenthal1970}.

We implement in \dragon~the average loss rate for relativistic electrons in an isotropic magnetic field, as derived in~\cite{Ginzburg1979} and~\cite{Schlickeiser2002}:

\begin{equation}
-\left(\frac{dE}{dt}\right) =  \frac {4}{3} \sigma_T c U_{B} \gamma^2 \beta^2 \sim 2.53 \times 10^{-18} \left( \frac{B}{\rm \mu G} \right)^2 \left( \frac{E}{\rm GeV} \right)^2 \text{   GeV s$^{-1}$}
\label{Synchro_def}
\end{equation}

where $\sigma_T$ is the Thomson cross section and $U_{\rm B} = \left( |\vec{B}_r|^2 + |\vec{B}_t|^2 \right) / 8\pi$ is the total energy density of the magnetic field. 

\subsubsection {Inverse Compton Scattering}

Inverse Compton scattering (ICS) takes place when highly relativistic electrons and positrons scatter with low energy photon fields. This process can be neglected when treating hadronic CRs.

Because of this process leptons lose a part of their energy and photons are up-scattered into high-energetic $\gamma$ rays. 

We take into account highly energetic $e^{\pm}$ CRs which scatter against the "sea" of photons constituted by the Galactic ISRF.
The energy-loss rate for an electron with an initial energy$\epsilon_i$ and a final energy $\epsilon_f$ (after ICS) is expressed in the following way \cite{Blumenthal1970,Longair1992}:

\begin{equation} 
-\frac{dE}{dt} = \int^{\infty}_0 d\epsilon_i \int^{\infty}_{\epsilon_i} d\epsilon_f (\epsilon_f -\epsilon_i) \times \frac{dN_{coll}}{dt  d\epsilon_i d\epsilon_f }.
\label{IC1}
\end{equation}

The term $\frac{dN_{coll}}{dt  d\epsilon_i d\epsilon_f }$ is the collision rate, given by

\begin{equation} 
\frac{dN_{coll}}{dt  d\epsilon_i d\epsilon_f } = \frac{3 \sigma_T c}{4 {\gamma_e}^2 \epsilon_i} \frac{d \eta}{d\epsilon_i}  \left\{ 1 + 2q \left(\ln{q} - q + \frac{1}{2}\right) + \frac{1- q}{2} \frac{(\Gamma q)^2}{1 + \Gamma q}  \right\}
\label{IC2}
\end{equation} 

where $d \eta/d\epsilon_i$ is the energy distribution of ISRF field in the photon gas frame (see Section~\ref{sec:isrf}), $\gamma_e=E/m$ is the Lorentz factor for the electron, $q \equiv \frac{\hat{\epsilon}_f}{\Gamma(1-\hat{\epsilon}_f)}$, $\hat{\epsilon}_f\equiv \frac{\epsilon_f}{\gamma_e m c^2}$ and $\Gamma \equiv \frac{4 \gamma_e \epsilon_i}{m c^2}$. 
Taking into account kinematic rules of the process, $\hat{\epsilon}_f$ can vary in the range $[\hat{\epsilon}_i, \Gamma/(1+\Gamma)]$, which translates into $q  = [\frac{1}{4\gamma_e^2}, 1]$. It is so possible to rewrite Eq.~\ref{IC1} as an integral over $q$

\begin{equation}
\begin{aligned}
\label{eq:icdef}
-\frac{dE}{dt} = 3 \sigma_T c \int^{\infty}_0 d\epsilon_i  \epsilon_i \int^{1}_{1/(4 {\gamma_e}^2)} dq \;  \frac{(4 \gamma^2_e - \Gamma)q - 1}{(1+\Gamma q)^3}  \frac{d \eta}{d\epsilon_i}  \times \\ 
 \left\{ 1 + 2q \left(\ln{q} - q + \frac{1}{2}\right) + \frac{1- q}{2} \frac{(\Gamma q)^2}{1 + \Gamma q}  \right\}.
\end{aligned}
\end{equation}

We show in Fig.~\ref{fig:icsync} the energy loss (both for the Galactic centre and local position) due to ICS with the approximation of the ISRF as composed by a superposition of black body distributions $d \eta_a/d\epsilon_i$ given by

\begin{equation}
 \frac{d \eta_a}{d\epsilon_i} = \mathcal{N}_a \frac{8\pi{\epsilon_i}^2}{(2\pi\hbar c)^3} \left(e^{\frac{\epsilon_i}{k_b T_a}} - 1\right)^{-1} .
\end{equation}

where the normalizations $\mathcal{N}_a$ and temperatures $T_a$ of the ISRF components for the local position in the Galaxy are reported in Tab.~\ref{tab:isrf_parameters}.

Fig.~\ref{fig:icsync} shows also a comparison between synchrotron and ICS energy losses: the contributions to ICS losses due to different components of ISRF are shown. 

\begin{figure*}
\begin{centering}
\includegraphics[width=0.49\textwidth]{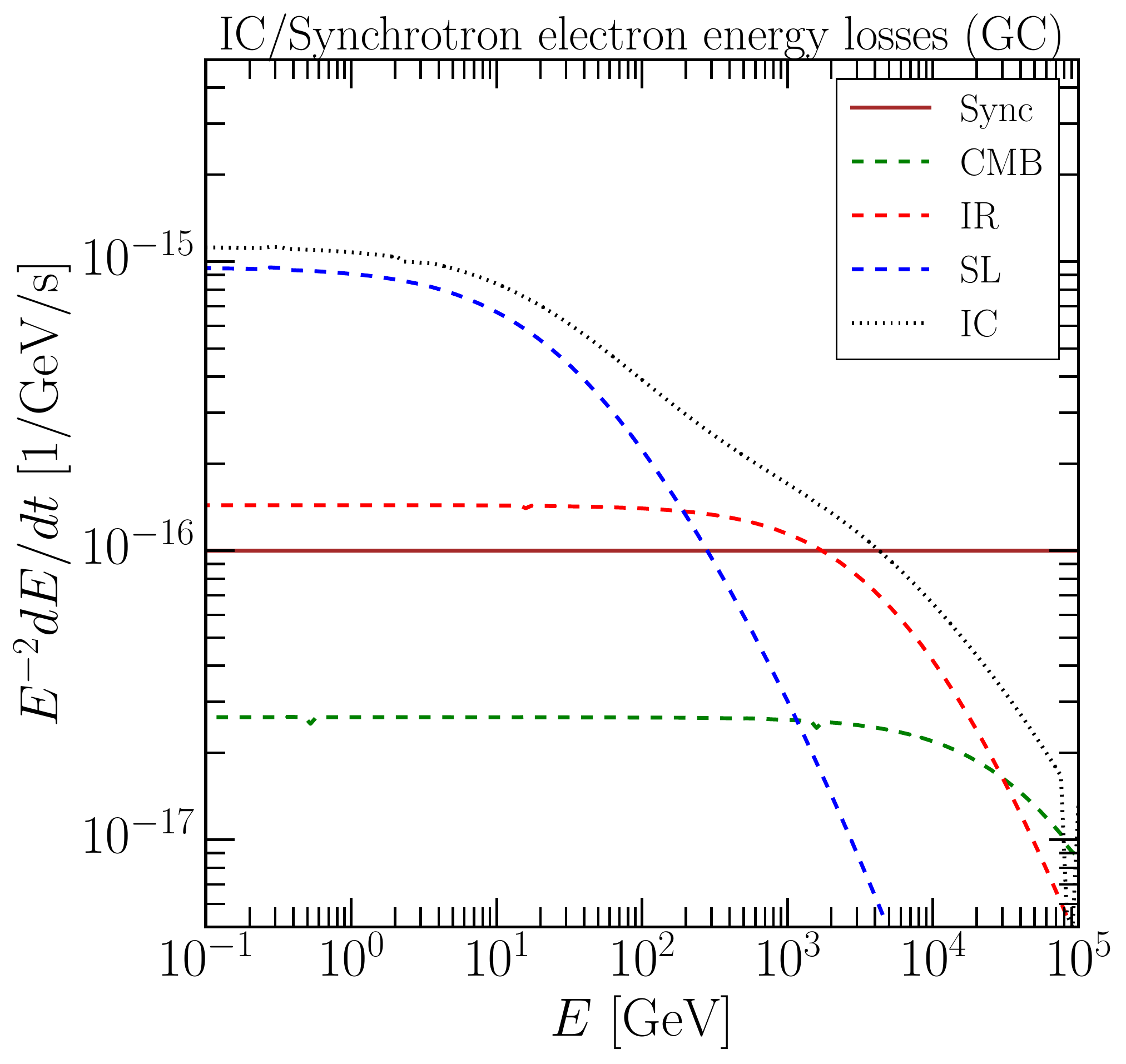}
\includegraphics[width=0.49\textwidth]{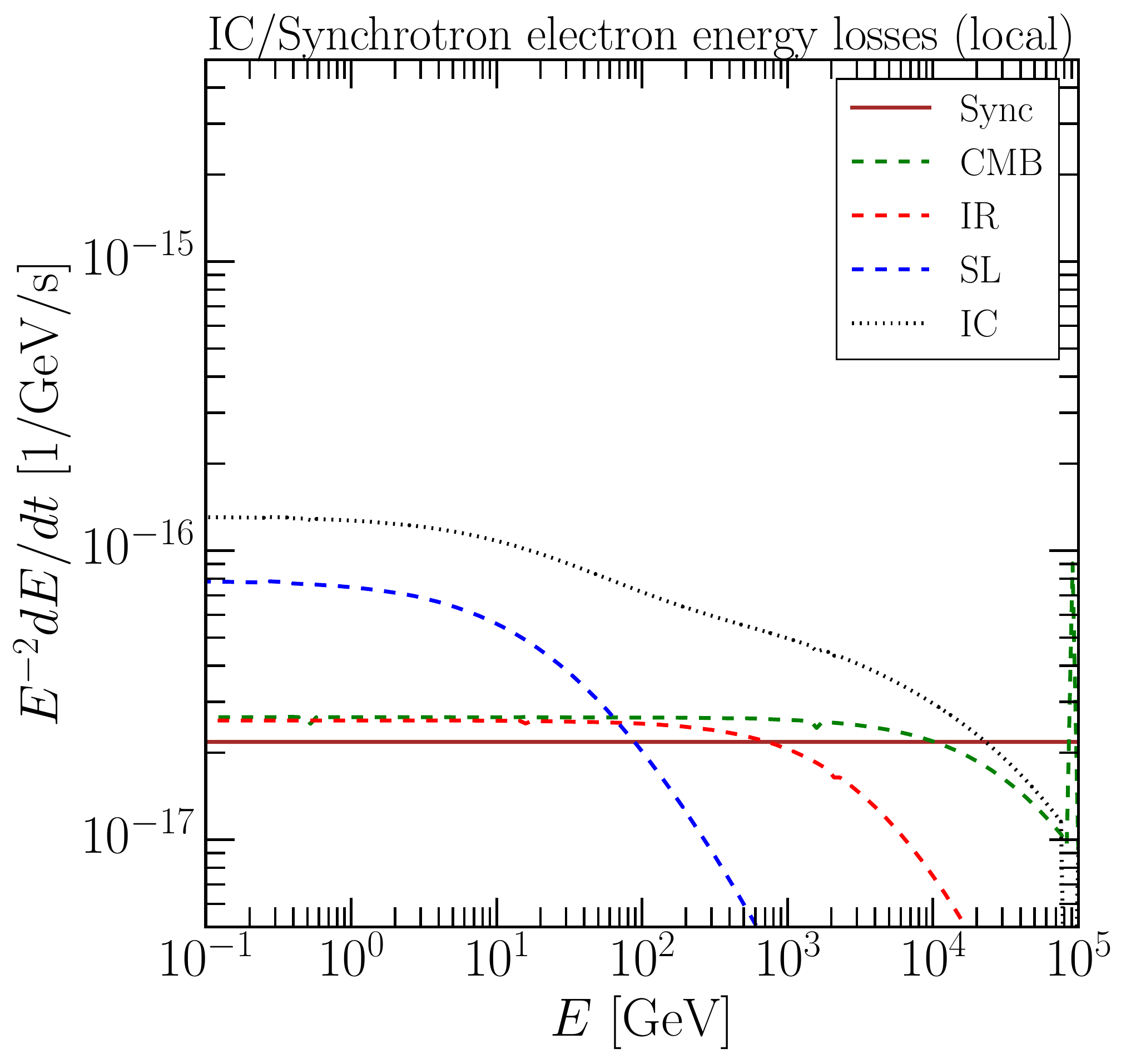}
\caption{In the {\it left (right) panel} the electron energy loss in the Galact centre (local region) due to ICS on the ISRF and to synchrotron radiation with the Galactic magnetic field. In each figure we display the energy loss for ICS on CMB (dashed green dashed line), IR (dashed red line), SL (dashed blue line), for synchrotron radiation (solid brown line) and the total ICS energy loss (dotted black line).}
\label{fig:icsync} 
\end{centering}
\end{figure*}

\subsubsection {Bremsstrahlung}

This process, also called \textit{braking radiation}, occurs when an electron or a positron is accelerated by the electric fields associated with interstellar ions or nuclei. Bremsstrahlung due to hadronic CRs is then neglected, since because of the heavy masses involved  accelerations are very small.

In \dragon ~we implement the following expressions. For a complete derivation, see  \cite{Ginzburg1979} and~\cite{Schlickeiser2002}.

\begin{itemize}

\item Ionised gas and weakly-shielded (WS) neutral gas.

\begin{equation}
\begin{aligned}
-\left(\frac{dE}{dt}\right)_{\text{WS}} = \frac{3\alpha c \sigma_{T}}{2\pi}~m_{e}\gamma c^{2}~\left[\ln{(2 \gamma)} - \frac{1}{3}\right]  \sum_{i= \rm H,He} Z_i (Z_i + 1) \, n_{\rm i} \sim \\
3.55\cdot 10^{-20} \left(\frac{E}{\text{GeV}}\right) \left[\ln{(2 \gamma)} - \frac{1}{3}\right]\left[2\left(\frac{n_{\rm H}}{\text{cm$^{-3}$}}\right) + 6 \left(\frac{n_{\rm He}}{\text{cm$^{-3}$}}\right) \right] \, \text{GeV s$^{-1}$},
\end{aligned}
\label{eq:weaks}
\end{equation}

This expression applies to the interaction with ionised gas at all energies.
In the case of interaction with neutral gas, the formula only applies for $\gamma<100$, also denominated weak-shielded regime on neutral gas. 

\item Strong-shielded (SS) neutral gas~\cite{Ginzburg1979}.

\begin{equation}
\begin{aligned}
-\left(\frac{dE}{dt}\right)_{\text{SS}}= c E \sum_{i=H,He} \frac{n_i M_i}{T_i} \sim 8 \times 10^{-16} \text{ GeV s$^{-1}$} \left( \frac{E}{\rm{GeV}} \right) \left( \frac{n_{\rm H}}{\text{cm$^{-3}$}} + 2.7 \, \frac{n_{\rm He}}{\text{cm$^{-3}$}} \right) 
\end{aligned}
\label{eq:strongs}
\end{equation}
where $M_i$ is the atomic mass, and $T_i$ is the radiation length with $T_{H} \approx 62.8$ g/cm$^2$ for H and $T_{\rm{He}} \approx 93.1$ g/cm$^2$ for He.
This expression holds for neutral gas in strong shielding case ($\gamma \geq 800$).

\item Intermediate-shielded neutral gas.

This regime holds for $100<\gamma<800$ and it's given by a linear interpolation of the strong and weak shielding in Eqs.~\ref{eq:strongs} and \ref{eq:weaks}~\cite{Ginzburg1979}.

\end{itemize}

\subsubsection{Ionisation Losses}

Relativistic charged particles moving through a material medium interact with electrons belonging to atoms in that same material:  the interaction thus excites or ionises the atoms. 

This process applies to both hadrons and leptons.

\begin{itemize}

\item
For ionisation losses suffered by hadrons, a general formula derived from \cite{Mannheim1994} was implemented in \dragon: 

\begin{equation}
-\left(\frac{dE}{dt}\right) = -\frac{3 \sigma_T \, c \, m c^2}{4 \beta} Z^2 \sum_{\text{s = H, He}} [n_i A_i] \sim 7.64\cdot 10^{-18} Z^2 \sum_{\text{s = H, He}} \left[\frac{n_i}{\rm cm^{-3}} \frac{A_i}{\rm eV} \right]  \text{ GeV s$^{-1}$},
\label{Ioniz_hadr}
\end{equation}
where 
\begin{equation}
A_i = \left[ \ln{\left( \frac{2 m_e c^2 \beta^2 \gamma^2 q_{max}}{\tilde{I}_i^2}\right)} - 2\beta^2 \right].
\end{equation} 

Here, $\tilde{I}_s$ denotes the geometric mean of all ionisation and excitation potentials of the atom: in \dragon, we used $\tilde{I}_{H}$ = 19 eV and $\tilde{I}_{He}$ = 44 eV, as derived in \cite{Mannheim1994}.
Also, $q_{max}$ is the largest possible energy transfer from the incident particle to the atomic electron, as defined by kinematics: 
 \begin{equation}
 q_{max} \sim \frac{2m_e c^2 \beta^2 \gamma^2}{1+ [2\gamma m_e/ M]}
 \end{equation}
where $M \gg m_e$ is the nucleon mass.

\item
If the incoming relativistic particle is an electron Eq.~\ref{Ioniz_hadr} is no longer valid and the ionisation losses is given by \cite{Longair1992}:

\begin{equation}
	\begin{aligned}
-\left(\frac{dE}{dt}\right)  = &  \frac{3 \sigma_T \, c \, m c^2}{4 \beta}\sum_{\text{i = H, He}}  Z_{i} n_i \left[\ln\left(\frac{\gamma^2 m_e c^2 E_{max}}{2I_i^2}\right)  \right. \\ 
& \left. - \left( \frac{2}{\gamma} - \frac{1}{\gamma^2}\right) \ln{2}+ \frac{1}{\gamma^2} +\frac{1}{8}\left(1-\frac{1}{\gamma}\right)^2 \right] 
 	\end{aligned}
\label{Ioniz_ele}
\end{equation}
where $I_i$ is the ionisation potential, for which we use $I_H$ = 13.6 eV, $I_{He}$ = 24.59 eV, and $E_{max}=\frac{\gamma^2 m_2 c^2}{1 + \gamma}$ is the maximum kinetic energy which can be transferred to the stationary electron.
A simplified expression of Eq.~\ref{Ioniz_ele} is:

\begin{eqnarray}
	\begin{aligned}
&-\left(\frac{dE}{dt}\right) \sim    7.64\cdot 10^{-18} \text{ GeV s$^{-1}$} \times \\
&  \sum_{\text{i = H, He}}  Z_{i} \frac{n_i}{\rm cm^{-3}}\left[\ln\left(\frac{\gamma^2 E}{2I_i^2 (1+\gamma)}\right)  \right. \left. - \left( \frac{2}{\gamma} - \frac{1}{\gamma^2}\right) \ln{2}+ \frac{1}{\gamma^2} +\frac{1}{8}\left(1-\frac{1}{\gamma}\right)^2 \right] 
 	\end{aligned}
\label{Ioniz_ele_simpl}
\end{eqnarray} 

Eq. \ref{Ioniz_ele_simpl} is different with respect to what is included in \galprop, where the Bethe-Bloch formula as taken from~\cite{Ginzburg1979} is considered. 
However, the differences between the two implementations are always smaller than $1\%$.

\end{itemize}

\begin{figure}[t!]
\begin{centering}
\includegraphics[width=0.49\textwidth]{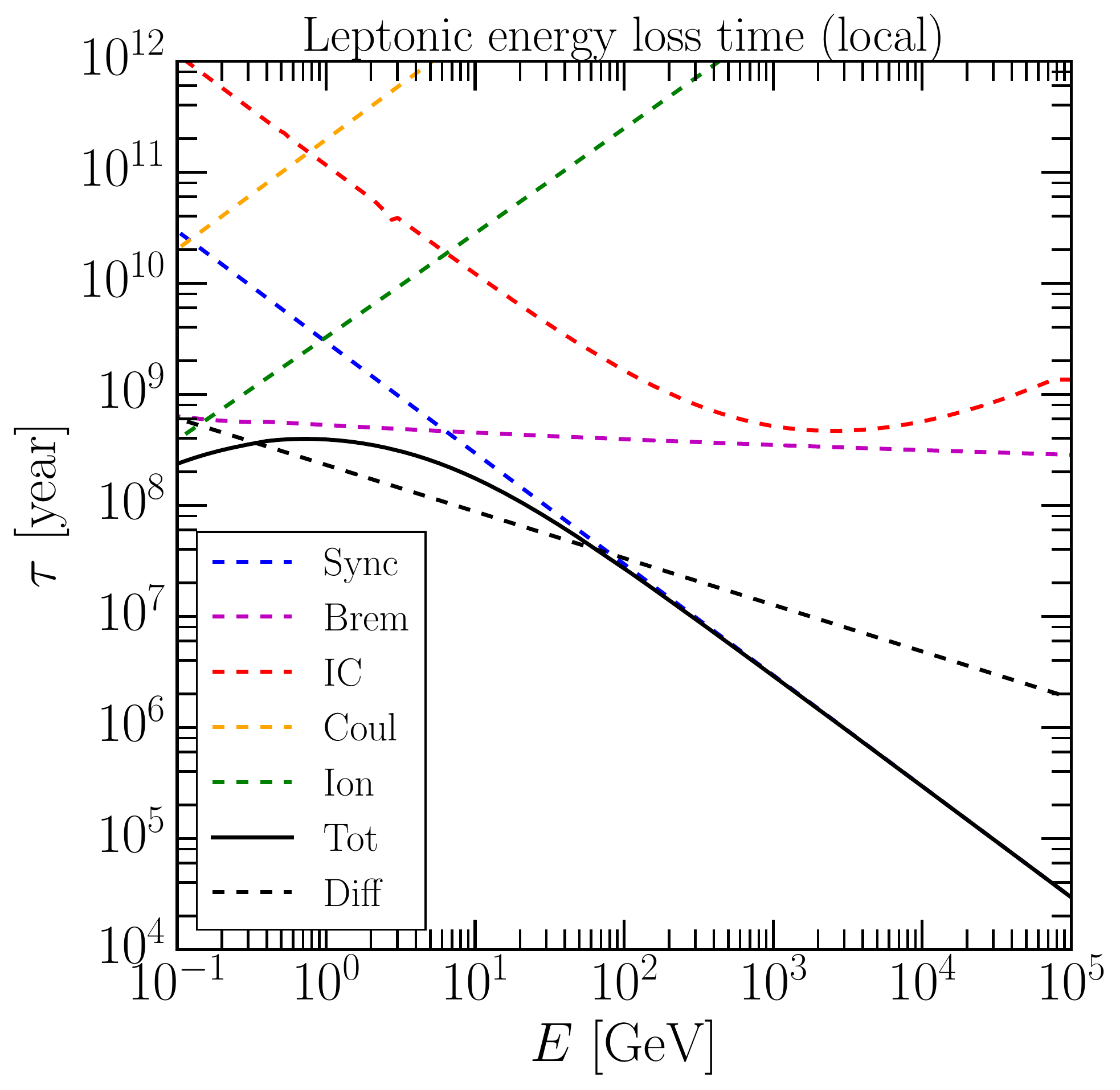}
\includegraphics[width=0.49\textwidth]{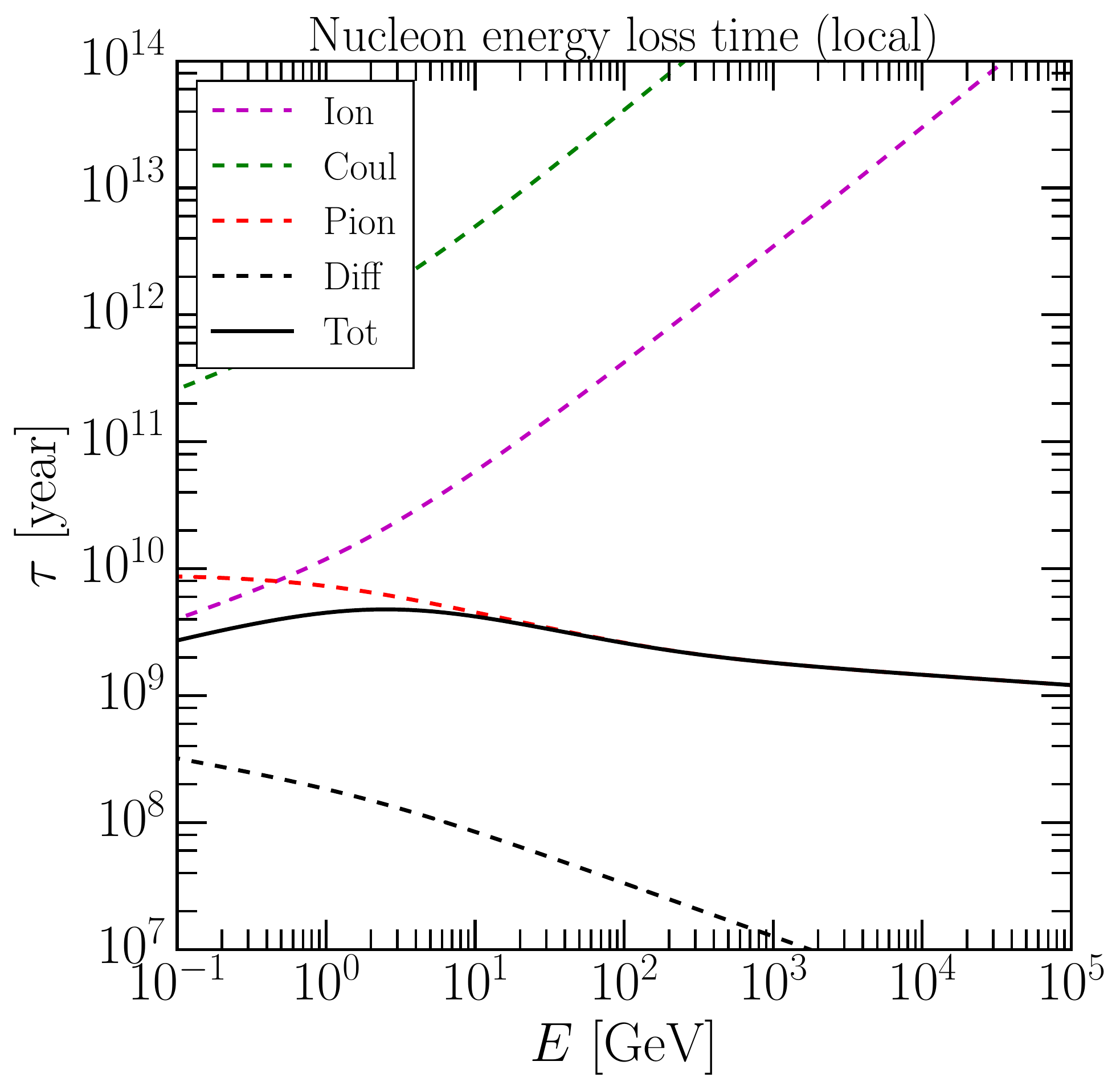}
\includegraphics[width=0.49\textwidth]{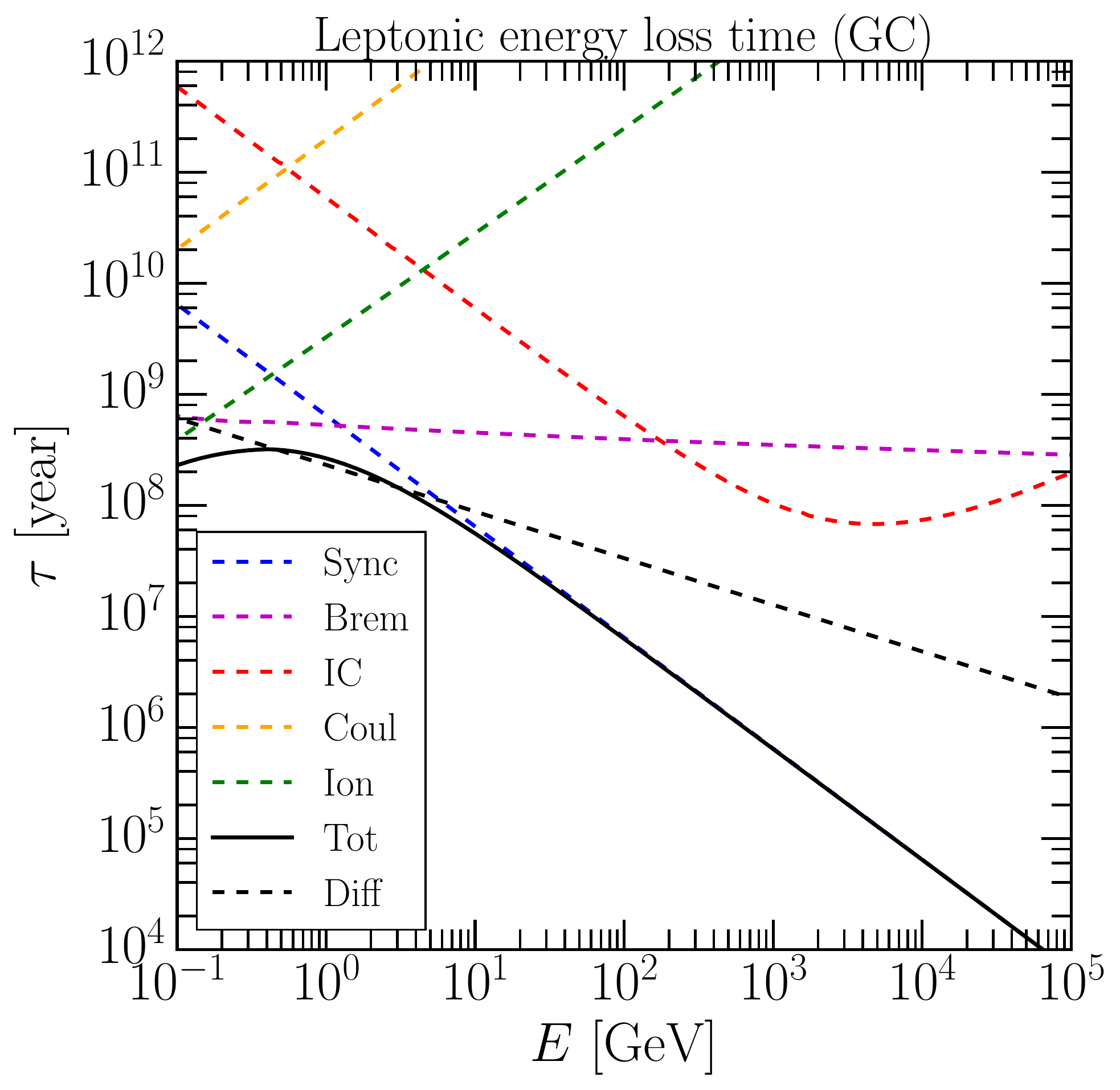}
\includegraphics[width=0.49\textwidth]{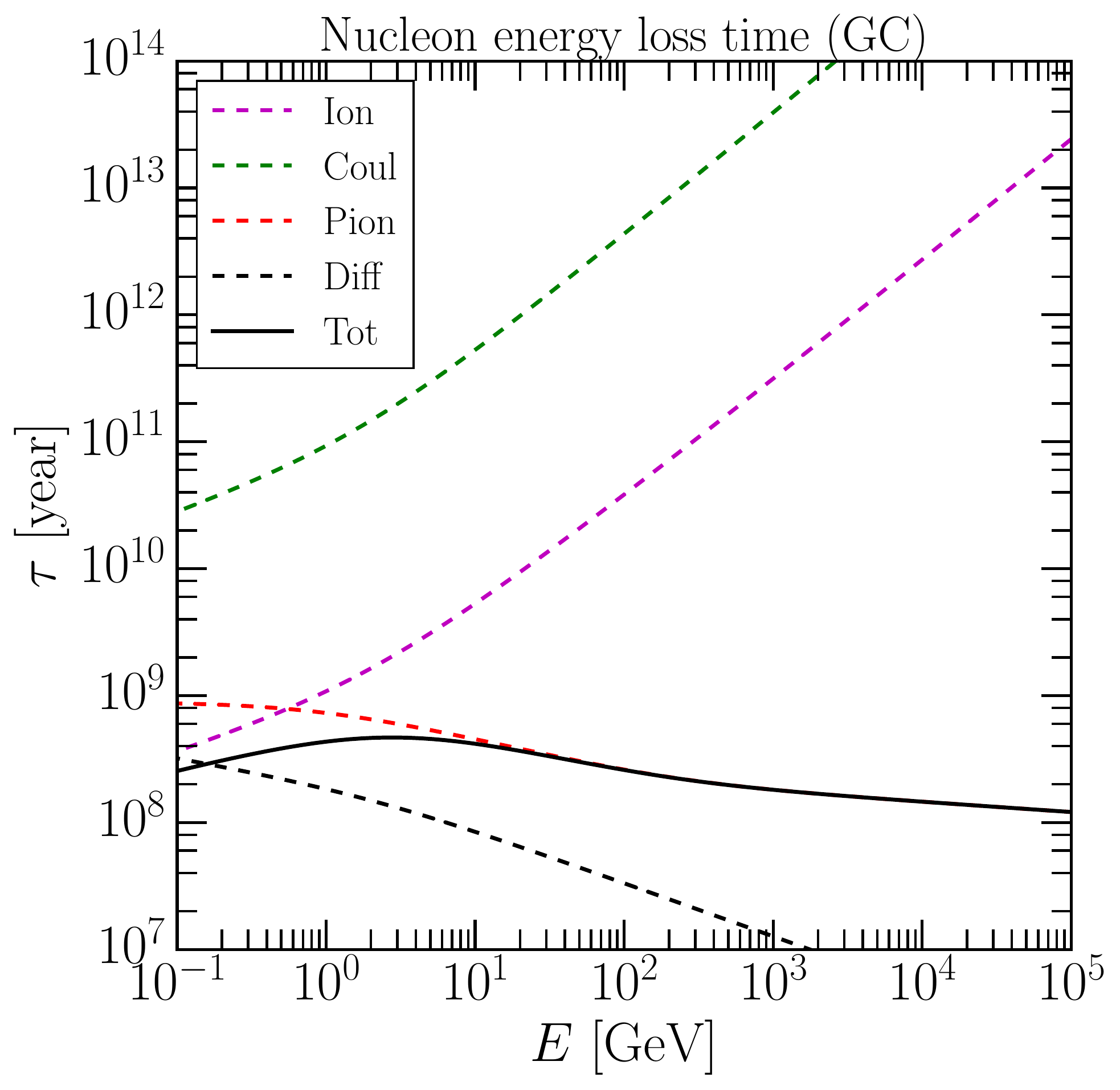}

\caption{The energy loss timescales for electrons or positrons (left panels) and protons (right panels) are shown for the mechanisms reported in Section~\ref{Sec:elosses}. For the local gas density we assume $n_{\rm H} = 0.9$~cm$^{-3}$ (upper panels) and $n_{\rm H} = 10$~cm$^{-3}$ for the GC (lower panels). To compute the leptonic losses we assume the constant ISRF from~\textsf{Delahaye2010} and the magnetic field model \textsf{Sun2007ASS}. The total energy loss timescale (black solid line) is compared with the diffusion timescale (black dashed line) in a halo with $H=4$~kpc and diffusion coefficient with $D_0 = 10^{28}$~cm$^2$/s and $\delta = 0.4$.}

\label{fig:alltau} 
\end{centering}
\end{figure}

\subsubsection{Coulomb scattering}

Coulomb collisions in a completely ionised plasma are dominated by scattering off the thermal electrons. 

\begin{itemize}

\item

The corresponding expression for hadronic particles is given by \cite{Mannheim1994}:

\begin {equation}
-\left(\frac{dE}{dt}\right)  = \frac{3}{2} \sigma_T \, c \, m_e c^2 Z^2 n_e \ln{\Lambda} 
\frac{1}{\beta} W_e \left(\frac{\beta}{\beta_e} \right)
\label{Coul_hadr}
\end{equation}

where we define:

\begin{equation}
\beta_e \equiv \sqrt{\frac{2 k_b T_e}{m_e c^2}}
\end{equation}

Here, $T_e$ is the electron temperature, and the functional form of $W_e$ is given by:
\begin{equation}
W_e(x) =  {\rm erf}(x)  - \frac{2}{\sqrt \pi} \left( 1+\frac{m_e}{A m_p} \right) x {\rm e}^{-x^2}
\end{equation}

The Coulomb logarithm, $\ln{\Lambda}$  in the cold plasma limit has been derived, e.g., in \cite{Dermer1985}:
 
\begin {equation}
\ln{\Lambda} \sim \frac{1}{2}\ln{\left(\frac{m_e^2 c^2}{\pi r_e \hbar^2 n_e} \frac{A m_p \gamma^2 \beta^4}{A m_p + 2\gamma m_e c^2} \right)}, 
\end{equation}

A numerical approximation of Eq.~\ref{Coul_hadr} is provided by~\cite{Mannheim1994}:

 \begin{equation}
-\left(\frac{dE}{dt}\right) \sim 3.1 \times 10^{-16} Z^2 \left(\frac{n_e}{\rm cm^{-3}} \right) 
\left(\frac{\beta^2}{x_m^3 + \beta^3}\right) \, \text{ GeV s$^{-1}$}
\end{equation}

where

\begin{equation}
x_m \sim 0.0286 \left( \frac{T_e}{2 \times 10^6 \, {\rm K}} \right)^{1/2}
\end{equation}

\item

Concerning electrons and positrons, the Coulomb energy loss rate in the fully ionised medium (with electron density $n_e$), can be written in the following way:~\cite{Ginzburg1979}:

\begin {equation}
-\left(\frac{dE}{dt}\right)  = \frac 3 4 \sigma_T \, c \, m_e c^2 n_e \left[ \ln{\left( \frac{E m_e c^2}{4\pi r_e\hbar^2 c^2 n_e}\right) - \frac{3}{4}}   \right]. 
\label{Coul_ele}
\end{equation}

or, in numerical form,

\begin{equation}
-\left(\frac{dE}{dt}\right)  \sim 7.64 \times 10^{-18} \left(\frac{n_e}{\rm cm^{-3}} \right) \left[ \ln \left(\frac{E}{m_e c^2} \right) - \ln \left(\frac{n_e}{\rm cm^{-3}} \right) + 73.57 \right]  \, \text{ GeV s$^{-1}$}\end{equation}
\end{itemize}

\subsubsection{Pion production}
\label{sec:pion_pion}

CR hadrons can lose energy through the production of pions that might follow a collision against a nucleus of the interstellar medium. 

This energy loss has been described for the first time in~\cite{Mannheim1994} and it has recently been revisited in~\cite{2015ApJ...802..114K}, where the impact of the new parameterisations for the pion production cross section derived in~\cite{2000PhRvD..62i4030B,2006PhRvD..74c4018K} has been discussed. 

In our implementation of the energy loss due to pion production, we use the analytical formula provided by~\cite{2015ApJ...802..114K} which reads: 
\begin{equation}
-\left( \frac{dE}{dt} \right)_p = 3.85 \times 10^{-16} \left( \frac{n_{\mathrm{gas}}}{\mathrm{cm}^{-3}}\right)\left(\frac{E}{\mathrm{GeV}}\right)^{1.28} \times \left( \frac{E}{\mathrm{GeV}} + 200\right)^{-0.2} \,\,\,\text{ GeV s$^{-1}$}
\end{equation}
where $E$ is the energy of the proton, while $n_{\mathrm gas}=n_{\mathrm {HI}}+2 n_{\mathrm {H}_2}$ denotes the interstellar gas density. 
Analogously to~\cite{2015ApJ...802..114K}, to model the energy loss by heavier nuclei we assume that the loss rate increases by a factor $A^{0.79}$.

We notice that the formula above can be applied only for protons with~$E \gg 1$~GeV. 
At lower energies however this loss mechanism is usually subdominant with respect to the other terms.

\section{Notations in this paper}

\begin{center}
\begin{tabular}{| c  c |}
\hline
$\alpha$ &  fine structure constant \\
$A$ & atomic mass \\
$B$ & magnetic field intensity \\
$b_i$ & magnetic field versor \\
$\beta$ & particle speed in units of c \\
$c$ & light speed \\
$\delta$ & diffusion coefficient slope index \\ 
$\delta_{ij}$ &  Kronecker delta symbol \\
$D_{\rm xx}$ & spatial diffusion coefficient along $x$ direction\\
$D_{\rm pp}$ & momentum diffusion coefficient \\
$E$ & particle kinetic energy per nucleon \\ 
$\epsilon_{ijk}$ & the complete antisymmetric Levi-Civita tensor \\
$\eta$ & low energy dependence index  of the diffusion coefficient \\
$\gamma$ & Lorentz factor ($\equiv E/m$) \\
$H$ & galaxy halo size \\
$\hbar$ & reduced Plank constant \\
$k_b$ & Boltzmann constant \\
$\Lambda$ & Coulomb logarithm\\
$m_e$ & electron mass \\
$m_p$ & proton mass\\
$n_{\rm e}$ & free electron number density \\
$n_{\rm H}$ & Hydrogen number density\\
$n_{\rm He}$ & Helium number density, assumed to be $0.11 n_{\rm H}$ \\
$p$ & particle momentum per nucleon \\
$\Phi$ & CR injection spectrum by SNR\\
$q_{\rm SN}$ & rate of CR injected by SNR per unit of volume\\
$r$ & radial coordinate in cylindrical coordinate system\\
$r_e$ & classical electron radius \\
$R_\odot$ & Sun position with respect to the galaxy centre \\
$R$ & galaxy radius \\
$\rho$ & particle rigidity ($\equiv p/Z$) per nucleon \\ 
$\sigma_T$ & Thomson cross section \\
$U_B$ & magnetic energy density \\
$v_A$ & Alfv\'en velocity \\
$v_w$ & Galactic wind velocity\\
$w$ & turbulence level\\
$x$ & coordinate in Cartesian system\\
$y$ & coordinate in Cartesian system\\
$z$ & vertical coordinate in Cartesian system or in cylindrical system \\
$z_t$ & vertical scale-length of diffusion coefficient \\
$Z$ & atomic number \\
\hline
\end{tabular}
\end{center}

\newpage

\bibliographystyle{JHEP}
\bibliography{dragon_paper_I}

\end{document}